\documentclass[a4paper, 10pt]{article}
\usepackage[T1]{fontenc}			
\usepackage[english]{babel}
\usepackage{array}						
\usepackage{graphics}					
\usepackage{indentfirst} 			
\usepackage{vmargin}					
\usepackage{amsmath,amsthm,amssymb,stmaryrd,wasysym}
\usepackage{color}
\usepackage{stackrel}
\usepackage{mathrsfs}
\usepackage{fancyref}
\usepackage{graphicx}
\usepackage{enumerate}
\usepackage{mdwlist}
\usepackage{caption}
\usepackage{fancybox}
\usepackage{pifont}
\usepackage[all]{xy}
\usepackage{makecell}
\usepackage{epigraph}
\usepackage{pifont}
\usepackage{dsfont}
\usepackage{relsize}
\usepackage{rotating}
\usepackage{multirow}
\newcolumntype{?}{!{\vrule width 1pt}}
\usepackage{csquotes}
\usepackage{cite}
\input amssym.def

\usepackage{fancyhdr}
\makeindex

\def\bqn{\begin{eqnarray}}
\def\eqn{\end{eqnarray}}
\newcommand*\xbar[1]{
  \hbox{
    \vbox{
      \hrule height 0.5pt 
      \kern0.5ex         
      \hbox{
        \kern-0.25em      
        \ensuremath{#1}
        \kern-0.25em
      }
    }
  }
} 

\definecolor{linkcolor}{rgb}{0,0,0.6}
\usepackage[	pdftex,colorlinks=true,	
			pdfstartview=FitV,
			linkcolor=linkcolor,
			citecolor=linkcolor,
			urlcolor=linkcolor,
			hyperindex=true,
			hyperfigures=false]{}

\newcommand{\bea}{\begin{eqnarray}}
\newcommand{\eea}{\end{eqnarray}}
\newcommand{\p}{\partial}

\newcommand{\mR}{\mathbb R}

\newcommand{\bi}{\begin{itemize}}
\newcommand{\ei}{\end{itemize}}
\newcommand{\be}{\begin{enumerate}}
\newcommand{\ee}{\end{enumerate}}
\newcommand{\nn}{\nonumber}

\newcommand{\pl}{\left(}
\newcommand{\pr}{\right)}

\newcommand{\crl}{\left[}
\newcommand{\crr}{\right]}

\newcommand{\w}{\wedge}

\newcommand{\Lag}{\mathcal{L}}

\newcommand{\alg}{\mathfrak g}

\newcommand\fonc[1]{{\mathscr C}^\infty\pl#1\pr}

\newcommand{\vf}{\field{T\M}}

\newcommand{\M}{\mathscr{M}}

\newcommand{\dx}{\dot{x}}

\newcommand{\Om}{\Omega}
\newcommand{\om}{\omega}

\newcommand{\un}{^{-1}}

\newcommand{\ie}{\textit{i.e.}\, }

\newcommand{\cf}{\textit{cf.}\, }
\newcommand{\Cf}{\textit{Cf.}\, }
\newcommand{\eg}{\textit{e.g.}\, }

\newcommand{\etal}{\textit{et al.}\, }
\newcommand{\ala}{\textit{\`a la}\, }

\newcommand{\br}[2]{\crl #1,#2 \crr}

\newcommand\pset[1]{\left \lbrace #1\right \rbrace}
\newcommand\bprop[1]{\begin{prop}#1\end{prop}}
\newcommand\bpropp[2]{\begin{prop}[#1]#2\end{prop}}
\newcommand\bdefi[2]{\begin{defi}[#1]#2\end{defi}}

\newcommand\blem[2]{\begin{lem}[#1]#2\end{lem}}

\newcommand\bcor[2]{\begin{cor}[#1]#2\end{cor}}
\newcommand\bthm[2]{\begin{thm}[#1]#2\end{thm}}
\newcommand\bexa[2]{\begin{exa}[#1]#2\end{exa}}

\newcommand{\Ker}{\text{\rm Ker }}

\newcommand{\Rad}{\text{\rm Rad }}
\newcommand{\Ann}{\text{\rm Ann }}
\newcommand{\End}{\text{\rm End}}

\newcommand{\gal}{\mathfrak{gal}}

\newcommand{\barg}{\mathfrak{bar}}

\newcommand\N[1]{\overset{N}{#1}}

\newcommand\ovA[1]{\overset{A}{#1}}

\newcommand\Riem[4]{R^{\hspace{0.3mm}#1}_{\hspace{1.5mm}{#2}\hspace{0.2mm}{#3}\hspace{0.2mm}{#4}}}

\newcommand\form[1]{\Omega^1\pl #1\pr}

\newcommand\field[1]{\Gamma\pl #1\pr}
\newcommand{\ff}{\form{\M}}

\newcommand\Prop[1]{Proposition \ref{#1}}
\newcommand\Defi[1]{Definition \ref{#1}}

\newcommand\Span[1]{\text{\rm Span}\hspace{1.2mm} #1}

\newcommand{\Milneg}{\fpsi}
\newcommand{\Milnegh}{\fhpsi}

\newcommand{\lmn}{{}^\lambda_{\mu\nu}}

\newcommand{\Lagxi}{\Lag_\xi}
\newcommand{\Lagxih}{\Lag_{\hat \xi}}

\newcommand{\FO}{FO\pl\M,\psi\pr}

\newcommand{\EC}{EC\pl\M,\xi\pr}

\newcommand\inv{\text{\rm inv}}

\newcommand\foncinvn[1]{{\mathscr C}^\infty_{\inv|\neq0}\pl#1\pr}

\newcommand{\bproof}[1]{\begin{proof}#1\end{proof}}

\newcommand{\mbP}{\mathbb P}
\newcommand{\vfh}{\Gamma\big(T\hat\M\big)}
\newcommand{\ffh}{\Om\big(\hat\M\big)}
\newcommand{\foncm}{{\mathscr C}^\infty(\M)}
\newcommand{\foncmh}{{\mathscr C}^\infty\big(\hat\M\big)}
\newcommand\foncmhn{{\mathscr C}^\infty_{\neq0}\big(\hat\M\big)}
\newcommand\foncmn{{\mathscr C}^\infty_{\neq0}\big(\M\big)}
\newcommand\foncmhinv{{\mathscr C}^\infty_\inv\big(\hat\M\big)}
\newcommand\foncmhinvn{{\mathscr C}^\infty_{\inv|\neq0}\big(\hat\M\big)}

\newcolumntype{M}[1]{>{\centering}m{#1}}

\renewcommand{\theequation}{\thesection.\arabic{equation}}

\theoremstyle{plain}
\newtheorem{thm}{Theorem}[section]
\newtheorem{lem}[thm]{Lemma}

\newtheorem{cor}[thm]{Corollary}
\newtheorem{prop}[thm]{Proposition}

\theoremstyle{definition}
\newtheorem{defi}[thm]{Definition}
\newtheorem{notation}[thm]{Notation}

\newtheorem{exa}[thm]{Example}
\newtheorem{rem}[thm]{Remark}
\newtheorem{fact}[thm]{Fact}
\newtheorem{term}[thm]{Terminology}

\usepackage{ulem}

\definecolor{rougef}{rgb}{0.56,0,0}		
\definecolor{vertf}{rgb}{0,0.5,0}		
\definecolor{bleuf}{rgb}{0,0,0.8}

\newlength\foo

\newcommand{\bremi}[1]{\begin{rem}[]~\\
\vspace{-8mm}\bi#1 \ei\end{rem}}

\newcommand{\bprope}[1]{\begin{prop}[]~\\
\vspace{-8mm}\be#1 \ee\end{prop}}

\newcommand{\bexai}[2]{\bexa{#1}{~\\
\vspace{-8mm}\bi#2 \ei}}

\newcommand{\bproofe}[1]{\bproof{~\\
\vspace{-8mm}\be#1 \ee}}

\newcommand{\half}{{{\textstyle\frac{1}{2}}}}

\newcommand{\beqa}{\begin{eqnarray} }
\newcommand{\eeqa}{\end{eqnarray} }
\newcommand{\ba}{\begin{array}}
\newcommand{\ea}{\end{array}}
\newcommand{\bpm}{\begin{pmatrix}}
\newcommand{\epm}{\end{pmatrix}}

\newcommand\wnab{{\overset{w}{\nabla}}}

\newcommand\fpsi{\Gamma(\Ker\psi)}
\newcommand\fhpsi{\Gamma(\Ker\hat\psi)}

\usepackage{color}

\renewcommand{\title}[1]{\vbox{\center\LARGE{#1}}\vspace{5mm}}
\renewcommand{\author}[1]{\vbox{\center\large#1}\vspace{5mm}}

\newcommand{\itemnum}{\hfill\refstepcounter{equation}\textup{(\theequation)}}

\newcommand\btitle[1]{\newblock ``\textit{{#1}}''}

\newcommand\barxiv[1]{\newblock \href{http://arXiv.org/abs/#1}{\texttt{arXiv:#1}}}

\usepackage{chngcntr}
\counterwithin*{equation}{section}

\begin{document}

\thispagestyle{empty}

\vspace{1cm}

 \begin{centering}

{\large {\bfseries 
Embedding Galilean and Carrollian geometries I.\\\vspace{1mm}
Gravitational waves
}
}
\\\vspace{0.5cm}
Kevin Morand
\\
\vspace{0.5cm}
Department of Physics, Sogang University, 35 Baekbeom-ro, Mapo-gu,  Seoul  04107, South Korea\\

\vspace{0.5cm}
{\tt morand@sogang.ac.kr}

\vspace{0.5cm}

\end{centering}
\begin{flushright} \textit{To the memory of Christian Duval} \end{flushright}
\begin{abstract}
\centering\begin{minipage}{\dimexpr\paperwidth-6.2cm}
\noindent 
\noindent The aim of this series of papers is to generalise the ambient approach of Duval \etal regarding the embedding of Galilean and Carrollian geometries inside gravitational waves with parallel rays. 
\noindent In this first part, we propose a generalisation of the embedding of {\it torsionfree} Galilean and Carrollian manifolds inside larger classes of gravitational waves. 
\noindent On the Galilean side, the quotient procedure of Duval \etal is extended to gravitational waves endowed with a lightlike hypersurface-orthogonal Killing vector field. 
This extension is shown to provide the natural geometric framework underlying the generalisation by Lichnerowicz of the Eisenhart lift.
\noindent On the Carrollian side, a new class of gravitational waves -- dubbed Dodgson waves -- is introduced and geometrically characterised. 
Dodgson waves are shown to admit a lightlike foliation by Carrollian manifolds and furthermore to be the largest subclass of gravitational waves satisfying this property.
This extended class allows to generalise the embedding procedure to a larger class of Carrollian manifolds that we explicitly identify. 
As an application of the general formalism, (Anti) de Sitter spacetime is shown to admit a lightlike foliation by codimension one  (A)dS Carroll manifolds. 
\end{minipage}
\end{abstract}
\pagebreak
\pagenumbering{gobble}
\tableofcontents

\newpage
\setcounter{secnumdepth}{0}
\section{Introduction}

\setcounter{secnumdepth}{3}

\pagenumbering{arabic}

Recognition that the motion of point particles -- \ie {\it kinematics} -- is underlain by 
a four dimensional space dates back as early as 1873, when P. L. Chebyshev, in a letter to J. J. Sylvester, addressed to the English mathematician the following advice\footnote{As quoted in \cite{Chebyshev}.}: ``Take to kinematics, it will repay you; it is more fecund than geometry; it adds a fourth dimension to space.''
The importance of four dimensional geometry became even more manifest 
with the advent of special relativity, more precisely its reformulation\footnote{In a famous essay with telling title \cite{Mink}.} by H. Minkowski as a theory of four dimensional spacetime endowed with a flat (pseudo)-Riemannian geometry. The crowning achievement of the present line of thought was conducted by A. Einstein whose theory of general relativity imparted four dimensional spacetime geometry the origin of gravitational phenomena \ie gravitation is kinematical in nature. 

Soon after the inception of general relativity, E. Cartan realised that classical mechanical forces (for a one-particle system) are as kinematical as the relativistic gravitational interaction\footnote{Namely, in the sense that trajectories of particles submitted to such forces are geodesics with respect to a (suitable) connection. }.  Nonrelativistic classical mechanics can thus equally be described within a four dimensional framework, though the underlying geometry -- referred to as Newton-Cartan (or \textit{Galilean}) geometry -- differs from the (pseudo)-Riemannian (or \textit{Lorentzian}) geometry  of general relativity. The origin of the discrepancy between the geometries underlying relativistic and nonrelativistic physics can be traced back at the algebraic level to a different  choice of kinematical group, \ie the group of automorphisms of the flat spacetime geometry. In other words, the kinematical group dictates the spacetime geometry which in turn prescribes the particle dynamics\footnote{Or in the words of the French philosopher J.S. Partre: ``Le cin\'ematisme est un dynamisme.''
}. 

A classification of these ``possible kinematics'' has been performed by H. Bacry and J. M. L\'evy--Leblond in a seminal paper \cite{Bacry:1968zf} (\cf also the recent classifications \cite{Figueroa-OFarrill:2017sfs,Andrzejewski:2018gmz}). Their classification divides kinematical groups into three families\footnote{We restrict here to the subset of \textit{geometric} kinematical groups \ie kinematical groups whose corresponding Lie algebra admits a faithful representation on the space of vector fields on a manifold of same dimension as the subspace of transvections, \cf \cite{Morand2018b} for details.} and features, along with the known relativistic (Poincar\'e and (Anti) de Sitter groups) and nonrelativistic (Galilei and Newton-Hooke groups) families, a novel family of ``ultrarelativistic'' groups composed of the (flat) Carroll group\footnote{The Carroll group was originally introduced -- mostly for pedagogical purposes -- by L\'evy--Leblond in \cite{LevyLeblond1965} as a ``degenerate cousin of the Poincar\'e group''. The rationale behind the reference to L. Carroll is justified in \cite{LevyLeblond1965} as originating from the lack of causality in a Carrollian universe (or flat Carroll spacetime) as  well as for the arbitrariness of time intervals (\cf CHAPTER VII - {\it A Mad Tea-Party} in \cite{Carroll1865}). Later, F. Dyson \cite{Dyson:1972sd} further justified the reference to Carroll by appealing to the immobility of (``timelike'') Carrollian observers as reminiscent of the following dialog between Alice and the Red Queen (CHAPTER II - {\it The Garden of Live Flowers} \cite{Carroll1871}):

\begin{displayquote}
``Well, in our country,'' said Alice, still panting a little, ``you'd generally get to somewhere else-if you ran very fast for a long time as we've been doing.''

``A slow sort of country!'' said the Queen. ``Now, here, you see, it takes all the running you can do, to keep in the same place.'' 
\end{displayquote}
} together with its curved avatars \big(the (A)dS Carroll groups\big).  
Consistently with the previous leitmotiv, this new algebraic person\ae\  carries its own geometry, referred to as \textit{Carrollian geometry} \cite{Duval:2014uoa} (\cf also \cite{Henneaux1979} for an early study).
Recent years have seen a surge of interest regarding Galilean and Carrollian geometries (collectively referred to as \textit{non-Riemannian}) due to their applications in a variety of contexts such as condensed matter \cite{Son:2013rqa,Geracie:2014zha,Jensen:2014aia,Geracie:2014mta,condmatt}, effective field theories \cite{Son:2013rqa,Jensen:2014aia}, (fractional) Quantum Hall effect \cite{Son:2013rqa,Hall,Banerjee:2015rca}, Hall viscosity \cite{Geracie:2014mta}, hydrodynamics \cite{Geracie:2014zha,fluid}, flat holography \cite{flatholography}, Lifshitz and Schr\"odinger holography \cite{Christensen:2013lma,Christensen:2013rfa,Bergshoeff:2014uea,lifshitz}, Ho{\v{r}}ava-Lifshitz gravity \cite{horava,Banerjee:2015rca}, Galilean string \cite{galstring}, Carrollian string \cite{Cardona:2016ytk} or Stringy gravity \cite{stringygravity}.

On the relativistic side of the story, yet another thread was added by T. Kaluza who first contemplated the merits of adding a {fifth} dimension to space as a means to geometrise both gravitational and electromagnetic interactions. Kaluza's idea was then incorporated within the nonrelativistic realm by L. P. Eisenhart \cite{Eisenhart1928} who established a correspondence between dynamical trajectories associated with a given nonrelativistic classical mechanical system on the one hand, and geodesics of a specific five dimensional relativistic spacetime on the other. 
\noindent This bridge between nonrelativistic and relativistic physics has since been independently rediscovered and considerably generalised, both at the algebraic (\cf \cite{ambientgroup}) and geometric level, most notably through the important work of Duval and collaborators \cite{Duval:1984cj,Duval:1990hj} (\cf also \cite{Balachandran:1986hv}). This so-called \textit{ambient approach} to nonrelativistic physics has recently been generalised to include ultrarelativistic structures. Specifically, Duval \etal  showed in \cite{Duval:2014uoa} that both Galilean and Carrollian geometries could be embedded inside higher dimensional gravitational waves with parallel rays, the former as quotient and the latter as lightlike hypersurface.

Since its inception, the ambient approach to non-Riemannian structures allowed to shed new light on various classical and quantum mechanical systems \cite{nullclassicalandquantummechanicalsystems} via their embedding into relativistic manifolds and found a number of applications in various contexts such as hydrodynamics \cite{nullfluids}, condensed matter \cite{nullcondensedmatter}, effective field theories \cite{Jensen:2014aia}, cosmology \cite{nullcosmology}, 3D spin 1 and spin 2 theories \cite{null3D} and holography \cite{Christensen:2013lma,Christensen:2013rfa,nullholography,Cariglia:2018hyr}. \\

The embedding of non-Riemannian structures performed in \cite{Duval:1984cj,Duval:2014uoa} relies crucially on a particular class of Lorentzian spacetimes, called {\it Bargmann--Eisenhart waves} in the following\footnote{\label{footnoteBarg}Note that the recognition of the r\^ole played by the  Bargmann group -- the central extension of the Galilei group -- in connection with nonrelativistic structures dates back to \cite{Duval1977} where Newtonian connections were geometrically characterised as connections on the affine extension of the frame bundle by the Bargmann group (whose associated curvature only takes values in the homogeneous Galilei algebra). The connection between the Bargmann group and Galilean geometry was further explored in \cite{Duval:1976ht} and recently readdressed in \cite{Andringa:2010it}.}. This particular class has been proposed as string vacua in \cite{stringvacua} and notably includes Minkowski spacetime as well as the renowned subclass of pp-waves, \cf \cite{Amati:1988sa,Horowitz:1989bv}. Geometrically, Bargmann--Eisenhart waves lie at the intersection of two interesting categories of structures:
\be
\item \textbf{Gravitational waves} \ie Lorentzian manifolds endowed with a (class of) lightlike and hypersurface-orthogonal vector field(s). Bargmann--Eisenhart waves are characterised among gravitational waves by the existence of a parallel lightlike vector field\footnote{Here and throughout the rest of this first part, the notion of parallelism on Lorentzian manifolds is provided by the Levi--Civita connection for the associated metric. }.
\item \textbf{Bargmannian manifolds} \ie Cartan geometries for the Bargmann algebra. Explicitly, Bargmannian manifolds are Lorentzian manifolds endowed with a lightlike vector field together with a (possibly torsionful) connection preserving both the metric and the lightlike vector field. In this context, Bargmann--Eisenhart waves identify with {\it torsionfree} Bargmannian manifolds\footnote{\label{footnotestructuremanifold}In the following, we will maintain a terminological distinction between manifolds endowed with metric structures (referred to as \textbf{structures}) and structures supplemented with a compatible connection (hereafter referred to as \textbf{manifolds}). For example, a {\it Lorentzian structure} will refer to a pair $(\M,g)$ consisting of a manifold $\M$ endowed with a Lorentzian metric $g$ while a {\it Lorentzian manifold} will denote a triplet $(\M,g,\nabla)$ where the Lorentzian structure is supplemented with a compatible connection $\nabla$. A {\it torsionfree manifold} will therefore refer to a manifold for which the connection has vanishing torsion. 
}.
\ee
This two-fold characterisation suggests two possible (and mutually exclusive) directions suitable for generalisation. The aim of the present series of papers is to systematically explore these two avenues. 
In this first part, we will focus on gravitational waves and thus only retain the torsionfree condition while relaxing the parallel condition for the lightlike vector field. 
{\it Contrariwise}, the second part \cite{Morand2018b} will address Bargmannian manifolds (as well as their generalisation, called {\it Leibnizian manifolds}) which will imply relaxing the torsionfree condition on the connection while retaining compatibility with the lightlike vector field. 

A first motivation for the present work consists in enlarging the class of Lorentzian spacetimes allowing the ambient approach. 
This was the main leitmotiv of the work \cite{Bekaert:2013fta} which already analysed the extension -- at the level of metric structures -- of the embedding procedure of Duval \etal to a larger class of gravitational waves, dubbed {\it Platonic waves} therein. This larger class was shown to contain both Anti de Sitter and Schr\"odinger spacetimes, thus hinting to possible applications of the ambient approach to -- relativistic and nonrelativistic -- holography. In the present work, we will extend the approach of \cite{Bekaert:2013fta} at the level of parallelism and further generalise it in order to include ultrarelativistic structures.

Another motivation comes from the fact that the class of non-Riemannian spacetimes that can be embedded inside Bargmann--Eisenhart waves is restricted to:
\be
\item \textbf{Newtonian spacetimes} on the quotient manifold \ie torsionfree Galilean manifolds whose Riemann curvature tensor satisfies a linear constraint\footnote{Referred to hereafter as the {\it Duval--K\"unzle condition}. }.
\item \textbf{Invariant Carrollian spacetimes} on lightlike hypersurfaces \ie torsionfree Carrollian manifolds whose connection is preserved by the Carrollian vector field.
\ee
{\it Alas}, known important examples of non-Riemannian spacetimes fall outside these two categories. 
On the torsionfree side, the most prominent examples are perhaps the maximally symmetric (A)dS Carroll spacetimes\footnote{Whose respective isometry group is the (A)dS Carroll kinematical group, \cf \cite{Bergshoeff:2015wma}. } which, as we will show, are non-invariant and as such cannot be embedded inside a Bargmann--Eisenhart wave\footnote{Note that (A)dS Carroll spacetimes are the only maximally symmetric non-Riemannian manifolds that do not admit an embedding inside a Bargmann--Eisenhart wave. In particular, their nonrelativistic counterparts (Newton-Hooke spacetimes) are known to be embedded inside a Hpp wave \cite{Gibbons:2003rv}, as reviewed below.}. This calls for a generalisation of the embedding procedure to account for a  larger class of torsionfree Carrollian manifolds, as performed in the present paper.
The second important restriction has to do with torsion. Bargmann--Eisenhart waves are  by definition torsionfree and thus can only induce torsionfree non-Riemannian geometries. 
Considering the importance of torsionful non-Riemannian geometries in the recent literature, 
it will be the ambition of the second paper \cite{Morand2018b} of the present series --  building on the earlier work \cite{Bekaert:2015xua} -- to provide a consistent ambient description of these torsionful (Galilean and Carrollian) geometries.

\paragraph{Summary and main results}
~\\

Section \ref{Intrinsic geometries} {\it begins at the beginning
}by providing a review of Galilean and Carollian geometries from an intrinsic perspective. We focus on the torsionfree case (\cf the companion paper \cite{Morand2018b} for the torsionful case) and recall classification results of torsionfree Galilean and Carrollian connections. The particular examples of maximally symmetric Galilean ({\it flat Galilei} and {\it Newton-Hooke}) and Carrollian ({\it flat Carroll} and ({\it A}){\it dS Carroll}) manifolds are discussed as an illustration of the general formalism. 
The (hierarchised) notions  of {\it invariant} and {\it pseudo-invariant} Carrollian manifolds are introduced (Definitions \ref{deficarrinvar} and \ref{propinvKoszulCarrpseudo}, respectively) in order to account for the flat and (A)dS Carroll manifolds, respectively. 
~\\

\vspace{-5mm}
Section \ref{Gravitational waves} is dedicated to the geometry of gravitational waves.
We focus our attention on the class of Kundt waves whose relevance regarding the embedding of nonrelativistic manifolds has been emphasised in \cite{Bekaert:2013fta}. 
A new subclass of Kundt waves, dubbed {\it Dodgson waves} is introduced (\Defi{defiDodgsonwave}) and geometrically characterised. The latter is shown to contain as subclasses some interesting families of waves previously discussed in the literature such as Walker, Platonic and Bargmann--Eisenhart waves. We conclude the section by displaying some distinguished examples of Dodgson waves, among which (Anti) de Sitter spacetime. In order to make the geometrical definitions more concrete, we make use of two privileged coordinate systems, adapted for the embedding of Galilean and Carrollian manifolds, respectively. We conclude by locally reinterpreting some of the previous classification results using Brinkmann coordinates \cite{Brinkmann} (\cf also the lecture notes \cite{Blau2004}). 
~\\

\vspace{-5mm}
In Section \ref{secDuval}, we review the seminal works of Duval \etal regarding the embedding of Galilean and Carrollian manifolds inside Bargmann--Eisenhart waves. Section \ref{secnewtmanquot} reviews the projection procedure of the geometry of a Bargmann--Eisenhart wave onto a Newtonian geometry on the associated quotient manifold \cite{Duval:1984cj}. This projection procedure is then shown to provide the geometric background underpinning the Eisenhart lift \cite{Eisenhart1928}. In Section \ref{secDuvalCarr}, we review the dual procedure by showing that Bargmann--Eisenhart waves admit a natural foliation by torsionfree Carrollian manifolds \cite{Duval:2014uoa}. 
We further characterise the space of torsionfree Carrollian manifolds that can be embedded in this way as endowed with an {\it invariant} connection. 
~\\

\vspace{-5mm}
The results of Sections \ref{secnewtmanquot} and \ref{secDuvalCarr} are then generalised in Section \ref{secELl} and \ref{secCarrembed}, respectively. Section \ref{secELl} can be seen as a follow up of the work \cite{Bekaert:2013fta} in which Platonic waves were introduced and characterised as conformal Bargmann--Eisenhart. The projection of the Platonic structure onto the quotient manifold was then worked out geometrically at the level of the metric structures.  Section \ref{secELl} of the present work reviews and completes these previous results by addressing the connection side. Explicitly, the main result of this section consists in the introduction of a new projection procedure generalising the one described in Section \ref{secnewtmanquot}. This new projection is shown to be suitable to account for the larger class of Platonic waves, for which the former procedure cannot be applied. The projected connection is non-canonical but depends on a constant referred to as the {\it weight}. The induced Newtonian connection on the quotient manifold differs from the one induced by the conformally related Bargmann--Eisenhart wave via a shift in the Newtonian potential depending on both the weight and the conformal factor. When focusing on the projection of geodesics, this procedure is shown to recover the generalisation of the Eisenhart lift by Lichnerowicz \cite{Lichnerowicz1955}, thus providing a new geometric understanding of the latter.
~\\

\vspace{-5mm}
Section \ref{secCarrembed} extends the results of Section \ref{secDuvalCarr} to the larger class of Dodgson waves. Explicitly, we show that any Dodgson wave admits a natural lightlike foliation by \textit{pseudo-invariant} Carrollian manifolds and conversely that any pseudo-invariant Carrollian manifold can be embedded inside a (class of) Dodgson wave(s). This result thus allows to generalise the embedding procedure to a larger class of Carrollian manifolds that can not be embedded inside a Bargmann--Eisenhart wave. In particular, the maximally symmetric (A)dS Carroll manifold  \cite{Bergshoeff:2015wma} is shown to admit a natural embedding into the (Anti) de Sitter spacetime, considered as a Dodgson wave. Conversely, (Anti) de Sitter spacetime is shown to admit a natural foliation by (A)dS Carroll manifolds of equal radii. We conclude the section by displaying an ultrarelativistic avatar of the Eisenhart lift -- referred to as \textit{Carroll train}\footnote{Apart from the ``horizontal'' character of the embedding of Carrollian manifolds -- as opposed to the vertical Eisenhart lift -- , the terminology ``Carroll train'' refers to the short-lived comic journal \textit{The Train}, edited by Edmund Yates, who published in 1856 the first piece of work -- the romantic poem \textit{Solitude} -- of the Oxford college mathematics lecturer Charles Lutwidge Dodgson to be signed under his more well-known pen name Lewis Carroll. In other words, \textit{The Train} welcomed the transition from Dodgson to Carroll, hence our choice to refer to it in the present context. } -- applicable to the class of Dodgson waves. 
~\\

\textbf{Note}: Most of the results presented in Section \ref{secELl} already appeared in an earlier form in the thesis work \cite{Morand:2016rrt} of the author as a result of a collaboration with Xavier Bekaert. 

\pagebreak
\section{Intrinsic geometries}
\label{Intrinsic geometries}
We start by reviewing Galilean and Carrollian geometries from an intrinsic perspective. After recalling the definitions of Galilean (resp. Carrollian) metric structures, particular attention is given to the respective notions of connection in both geometries, focusing on the torsionfree case. Unlike in the Riemannian case, torsionfree connections compatible with a given metric structure are not unique. We recall classification results of the spaces of compatible connections as well as explicit component expressions for the most general connection in each case. In view of subsequent discussions in the ambient context, we review distinguished subclasses  of Galilean and Carrollian manifolds. On the Galilean side, we recall the definition of \textit{Newtonian} manifolds, a subclass of Galilean manifolds containing the maximally symmetric flat Galilei and Newton-Hooke spacetimes. On the Carrollian side, the class of Carrollian manifolds with {\it invariant} connection is discussed and argued to provide the dual counterpart of Newtonian manifolds. Both dual geometries will be shown to admit an embedding inside Bargmann--Eisenhart waves through the ambient approach of Duval \etal in Section \ref{secDuval}. The class of invariant Carrollian manifolds contains the maximally symmetric flat Carroll spacetime but fails to account for the (A)dS Carroll spacetime. A  larger subclass, dubbed {\it pseudo-invariant} Carrollian manifolds, is then introduced and shown to contain (A)dS Carroll spacetime. The class of pseudo-invariant Carrollian manifolds will be shown in Section \ref{secCarrembed} to be the largest class of torsionfree Carrollian manifolds admitting an embedding inside gravitational waves. 
\subsection{Galilean}
\label{Galilean}
In the present section, we review Galilean (or Newton-Cartan) geometry (\cf \eg \cite{Kuenzle:1972zw,Bernal:2002ph,Bekaert:2014bwa}), starting with the Galilean notion of metric structure\footnote{Galilean structures were referred to as {\it Leibnizian} structures in \cite{Bernal:2002ph,Bekaert:2014bwa}. In order to avoid confusion with the ambient notion of Leibnizian manifolds \cite{Bekaert:2015xua,Morand2018b}, we did not retain this terminology in the present work. }:
\bdefi{Galilean structure}{\label{defiGalstruc}A Galilean structure is a triplet $(\M,\psi,h)$ where 
\bi
\item $\M$ is a manifold of dimension $d+1$.
\item $\psi\in\ff$ is a non-vanishing 1-form.
\item $h\in\field{\vee^2T\M}$ is a contravariant metric satisfying the following properties:
\bi
\item $h$ is of rank $d$ and of signature $(0,\underbrace{+1,\ldots,+1}_d)$.
\vspace{-1.2mm}
\item The radical of $h$ is spanned by $\psi$ \ie $h(\alpha,\cdot)=0\Leftrightarrow \alpha\sim\psi$, where $\alpha\in\ff$.
\ei
\ei
}
\bremi{\label{remgalfrob}
\item Alternatively, a Galilean structure can be defined as a triplet $(\M,\psi,\gamma)$ where $\gamma\in\Gamma\big(\vee^2(\Ker\psi)^*\big)$ is a rank $d$ covariant Riemannian metric on $\Ker\psi$ of signature $(\underbrace{+1,\ldots,+1}_d)$. 

Since the data of $h$ and $\gamma$ are equivalent, we will denote the associated Galilean structure interchangeably as $\big(\M,\psi,h\big)$ or $\big(\M,\psi,\gamma\big)$.
\item The 1-form $\psi$ is sometimes referred to as an (absolute) \textbf{clock} while the metric $h$ (or $\gamma$) is called a collection of (absolute) \textbf{rulers}.
\item The clock $\psi$ allows to distinguish between two classes of vector fields $X\in\vf$:
\bi
\item \textbf{Spacelike}: $\psi(X)=0$
\item \textbf{Timelike}: $\psi(X)\neq0$.
\ei
A normalised timelike vector field  \big(\ie $N\in\vf$ | $\psi(N)=1$\big) will be called a \textbf{field of observers}. 
\item Additional constraints can be imposed on the absolute clock $\psi$ in order to make the notion of absolute time and space more manifest:
\bi
\item $d\psi\w\psi=0$\, \ie $\psi$ is \textbf{Frobenius} so that the distribution $\Ker\psi$ of spacelike vectors is involutive, hence integrable. The Frobenius condition ensures that the spacetime $\M$ admits a natural foliation by codimension one hypersurfaces corresponding to \textbf{absolute spaces} (or simultaneity slices). Locally, one can always make use of the Frobenius theorem to express $\psi=\Om\, dt$, with $t$ the absolute time function (labelling the absolute spaces) and $\Om$ the ``time unit'' function\footnote{Although two timelike observers sharing the same starting and ending points will generically measure different proper times, the time unit allows them to synchronize their proper time with the absolute time, \cf \eg \cite{Bekaert:2014bwa} for details.}.

\item $d\psi=0$\, \ie $\psi$ is \textbf{closed} and can thus always be expressed locally as $\psi=dt$\footnote{In this case, any timelike observer is automatically synchronized with the absolute time.}.
\item Whenever $d\psi\w\psi\neq0$, the spacetime is called \textbf{acausal}\footnote{This terminology is justified by the following fact: in a spacetime endowed with a non-Frobenius clock, all points in a given neighbourhood are simultaneous to each other, in the sense that any pair of points can be joined by a spacelike curve, \cf \eg \cite{Bekaert:2014bwa}.}. 
\ei
\item In view of the previous remark, Galilean structures endowed with a Frobenius clock are privileged in view of their causal properties. The space of Galilean structures with Frobenius clock is preserved by the following action of the abelian multiplicative group of nowhere vanishing functions $\foncmn$:
\bea
\psi\overset{\Lambda}{\mapsto}\Lambda^2\, \psi\quad,\quad h\overset{\Lambda}{\mapsto}\Lambda^{-2} h \quad,\quad \gamma\overset{\Lambda}{\mapsto}\Lambda^2\, \gamma\quad\quad \text{with }\Lambda\in\foncmn.\label{groupactiongalconf}
\eea
Transformation \eqref{groupactiongalconf} allows to define orbits\footnote{Or equivalently $[\psi,\gamma]$.} $[\psi,h]$ of Galilean structures with Frobenius clocks. Note that any such orbit contains a distinguished representative $(\bar\psi,\bar h)$ such that $\bar\psi$ is closed. The latter will be referred to as the \textbf{special} structure of $[\psi,h]$.
}
Galilean structures with closed absolute clock can be upgraded to torsionfree Galilean manifolds via the introduction of a compatible connection:
\bdefi{Torsionfree Galilean manifold}{\label{deftorgal}A torsionfree Galilean manifold is a quadruplet $(\M,\psi,h,\nabla)$ where:
\bi
\item $(\M,\psi,h)$ is a Galilean structure with closed $\psi$ \ie $d\psi=0$.
\item $\nabla:\Gamma\pl T\M\pr\to \End\big(\field{T\M}\big)$ is a torsionfree Koszul connection on $\M$ compatible with both $\psi$ and $h$ \ie satisfying
\be
\item $\nabla \psi=0$
\item $\nabla h=0$
\ee
and referred to as the \textbf{Galilean connection}. 
\ei
}
As shown in \cite{Kuenzle:1972zw}, the closedness condition $d\psi=0$ is necessary in order to ensure the existence of torsionfree Galilean connections. The latter are classified as follows \cite{Kuenzle:1972zw}:
\bpropp{Classification of torsionfree Galilean connections}{\label{proptorsionfreeGal}Let $(\M,\psi,h)$ be a Galilean structure with closed clock $\psi$.
\bi
\item 
The space of torsionfree connections compatible with $(\M,\psi,h)$ is an affine space modelled on the vector space $\Om^2(\M)$ of differential 2-forms on $\M$. 
\item The most general expression\footnote{An index free ``Koszul-like'' formula is displayed in \cite{Bernal:2002ph,Bekaert:2014bwa}. } for torsionfree connections compatible with $(\M,\psi,h)$ is given by:
\bea
\Gamma^{\lambda}_{\mu\nu}&=&N^\lambda\p_{(\mu}\psi_{\nu)}+\half h{}^{\lambda\rho}\big( \p_\mu\N\gamma_{\rho\nu}+\p_\nu\N\gamma_{\rho\mu}-\p_\rho\N\gamma_{\mu\nu}\big)+h^{\lambda \rho}\psi_{(\mu}\N F_{\nu)\rho}\label{Galconn}
\eea
where 
\bi
\item $N\in\vf$ is a \textbf{field of observers } \ie  $\psi(N)=1$. \itemnum\label{FOdef}
\item $\N \gamma\in\field{\vee^2T^*\M}$ is the unique solution to:
\bea
\quad \N\gamma{}_{\mu\nu}N^\nu=0\quad,\quad h{}^{\mu\rho}\N\gamma_{\rho\nu}=\delta^{\mu}_\nu-N^\mu \psi_\nu\label{eqdefiNg}
\eea 
and referred to as the \textbf{transverse metric} associated with $N$.
\item $\N F\in\Om^2(\M)$ is an arbitrary 2-form called the \textbf{gravitational fieldstrength}.
\ei
\ei
}
\bremi{\label{remspacelikeabsspace}
\item When acting on spacelike vector fields, any Galilean connection $\nabla$ compatible with $(\M,\psi,h)$ can be shown\footnote{The proof follows straightforwardly from the Koszul-like formula for Galilean connections \cite{Bernal:2002ph,Bekaert:2014bwa}.} to reduce to the Levi--Civita connection $\nabla_\gamma$ associated with the Riemannian metric $\gamma$, that is $\nabla|_{\Ker\psi}=\nabla_\gamma$. Two distinct Galilean connections will thus differ by their action on timelike vector fields. 
\item The gravitational fieldstrength $\N F$ encodes the arbitrariness in the torsionfree Galilean connection $\nabla$ \ie the part of the connection that is not fixed by the compatibility relations with $\psi$ and $h$. It can be intrinsically defined as $\N F_{\mu \nu}=-2\, \N\gamma_{\lambda[\mu}\nabla_{\nu]}N^\lambda$. 
\item The space of fields of observers \big(\ie vector fields on $\M$ satisfying \eqref{FOdef}\big) is denoted $\FO$. The latter is an affine space modelled on $\Milneg$ \ie for any two fields of observers $N,N'\in\FO$, there exists $V\in\vf$ such that $\psi(V)=0$ and $N'-N=V$. Seen as an abelian group, the vector space $\Milneg$ is referred to as the \textbf{Milne group} \cite{Duval:1993pe}.  
\item Expressions \eqref{Galconn}-\eqref{eqdefiNg} are left invariant by a \textbf{Milne-boost} transformation:
\bea
N\overset{V}{\mapsto} N+V\quad,\quad
\overset{N}{\gamma}\overset{V}{\mapsto}\overset{N}{\gamma}-\Phi\otimes \psi-\psi\otimes \Phi\quad,\quad
\N F\overset{V}{\mapsto} \N F+d\Phi\label{Milneboost}
\eea
where $V\in\Milneg$ \ie $\psi(V)=0$, $V^\flat_\mu:=\N\gamma_{\mu\nu}V^\nu$ and $\Phi\in\ff$ is defined as $\Phi:=V^\flat-\half \,{\gamma}\pl V,V\pr\psi$.
}
The action \eqref{Milneboost} on $\FO\times\Om^2(\M)$ allows to articulate the following classification result for torsionfree Galilean connections \cite{Kuenzle:1972zw}, \cf also \cite{Bekaert:2014bwa}:
\bprop{\label{corGalileantorless}The space 
of torsionfree Galilean connections compatible with $(\M,\psi,h)$ possesses the structure of an affine space canonically isomorphic to the affine space $\frac{\FO\times\Om^2(\M)}{\Milneg}$.
}
Before addressing some examples of Galilean manifolds, we introduce an interesting subclass thereof dubbed {\it Newtonian manifolds}:
\bdefi{Newtonian manifold}{\label{defiNewtonman}
A Newtonian manifold is a torsionfree Galilean manifold $(\M,\psi,h,\nabla)$  such that the Galilean connection $\nabla$ is Newtonian, \ie $\nabla$ satisfies the \textbf{Duval--K\"unzle condition}
\bea
R\, {}^{\mu\, \, \, \, \nu}_{\, \, \, \rho\, \, \, \, \sigma}=R\, {}^{\nu\, \, \, \, \mu}_{\, \, \, \sigma\, \, \, \, \rho}\label{eqDK}
\eea
with  $R\, {}^{\mu\, \, \, \, \nu}_{\, \, \, \rho\, \, \, \, \sigma}:= h^{\nu \lambda}\, \Riem{\mu}{\rho}{|\lambda}{\sigma}$ and $\Riem{\mu}{\rho}{|\lambda}{\sigma}:=dx^\mu\crl R\pl \p_\lambda,\p_\sigma;\p_\rho\pr\crr$.

}
\bremi{\label{remNewtconTrump}
\item A non-trivial result \cite{Kuenzle:1972zw} states that the subspace of Newtonian connections possesses the structure of an affine space\footnote{Despite the seemingly non-linear nature of the Duval--K\"unzle condition.} modelled on the vector space of {\it closed} 2-forms. Explicitly, the torsionfree Galilean connection \eqref{Galconn} satisfies the Duval--K\"unzle condition if and only if the 2-form $\N F$ is closed \ie
\bea
\text{For all torsionfree Galilean connections }\quad\quad R\, {}^{\mu\, \, \, \, \nu}_{\, \, \, \rho\, \, \, \, \sigma}=R\, {}^{\nu\, \, \, \, \mu}_{\, \, \, \sigma\, \, \, \, \rho}\Leftrightarrow d\N F=0.\label{appliedDK}
\eea

 Note that condition \eqref{appliedDK} is Milne invariant since a Milne boost \eqref{Milneboost} rescales $\N F$ by an exact 2-form. 
 
 \item Locally, and given a field of observers $N\in\FO$, the arbitrariness in a Newtonian connection is encoded in an equivalence class of 1-forms $[\N A]$, called the \textbf{gravitational potential}, and satisfying the two following properties:

 \bi
\item For any representative $\N A\in[\N A]$, we have $\N F=d\N A$.
  \item Any two representatives $\N A,\N A{}'\in[\N A]$ differ by an exact 1-form \ie $\N A{}'-\N A=df$. 
 \ei

 \item The components of a Newtonian connection can be written in a manifestly Milne-invariant form as:
 \bea
\Gamma\lmn=Z^\lambda\p_{(\mu}\psi_{\nu)}+\half h^{\lambda\rho}\big( \p_\mu g_{\rho\nu}+\p_\nu g_{\rho\mu}-\p_\rho g_{\mu\nu}\big)\label{NewtMilne}
\eea
through the use of the following variables, (\cf \cite{Trumper1983,Duval:1993pe} and more recently \cite{Bergshoeff:2014uea,Bekaert:2014bwa}):
\bea
Z^\mu=N^\mu-h^{\mu\nu} A_{\nu}\quad,\quad \phi=2\, {A}_\mu N^\mu-h^{\mu \nu} A_\mu A_\nu\quad,\quad g_{\mu \nu}=\gamma_{\mu \nu}+\psi_\mu  A_{\nu}+  A_{\mu}\psi_\nu\label{Trumpervariables}
\eea
where the field of observers $Z$ is referred to as \textbf{Coriolis-free}, the metric $g$ as \textbf{Lagrangian} and the scalar $\phi$ as the \textbf{Newtonian potential}. 
The latter satisfy the set of relations:
\bea
\psi_\mu Z^\mu=1\quad,\quad g_{\mu\nu}Z^\nu=\phi\, \psi_\mu\quad,\quad h^{\mu\nu}\psi_\nu=0 \quad,\quad h^{\mu\rho}g_{\rho\nu}=\delta^{\mu}_\nu-Z^\mu \psi_\nu.\label{reltrump}
\eea 
\item Note that expression \eqref{NewtMilne} is invariant under a so-called \textbf{Maxwell-gauge transformation}:
\bea
Z^\mu\overset{f}{\mapsto}Z^\mu-h^{\mu\nu}\p_\nu f\quad,\quad\phi\overset{f}{\mapsto}\phi+2\, Z^\mu\p_\mu f-h^{\mu\nu}\p_\mu f\p_\nu f\quad,\quad g_{\mu\nu}\overset{f}{\mapsto}g_{\mu\nu}+2\, \psi _{(\mu}\p_{\nu)}f
\label{eqtransU1}
\eea
parameterised by $f\in\fonc{\M}$.
\item The set of relations \eqref{reltrump} is preserved by a shift of the Newton potential:
\bea
Z\overset{{\bar\phi}}{\mapsto}Z\quad,\quad\phi\overset{{\bar\phi}}{\mapsto}\phi+{\bar\phi}\quad,\quad g\overset{{\bar\phi}}{\mapsto}g+{\bar\phi}\, \psi\otimes\psi\quad,\quad\text{where }\bar\phi\in\fonc{\M}.\label{eqshifttrumpvar}
\eea
The latter transformation induces an action of the abelian multiplicative group $\fonc{\M}$ on the set of Newtonian connections (compatible with a given closed Galilean structure) as:
\bea
\Gamma\lmn\overset{{\bar\phi}}{\mapsto}\Gamma\lmn-\half h^{\lambda \rho}\p_\rho{\bar\phi}\, \psi_\mu\, \psi_\nu.\label{eqgroupactionNewtpotshift}
\eea
\item Newtonian manifolds are distinguished among torsionfree Galilean manifolds by the fact that the geodesic equations associated with a Newtonian connection are Lagrangian. Explicitly, given a timelike observer $\dot x$ (\ie $\psi(\dot x)\neq0$\footnote{Note that this constraint is holonomic if $\psi$ is closed, \cf \eg \cite{Bekaert:2014bwa}. }), Euler-Lagrange equations of motion associated with the Lagrangian density $\Lag=\frac{1}{2}\frac{g(\dot x,\dot x)}{\psi(\dot x)}$ identify with the geodesic equations associated with the Newtonian connection \eqref{NewtMilne}. Under a Maxwell-gauge transformation \eqref{eqtransU1}, the Lagrangian density is shifted by a boundary term $\Lag\overset{f}{\mapsto} \Lag+\frac{df}{d\tau}$ so that the associated equations of motion are Maxwell-invariant. 
}
We conclude this review of Galilean manifolds by displaying two distinguished examples:
\bexai{Galilean manifolds}{\label{exaGal}
\item \textbf{Flat Galilei manifold}: The flat Galilei manifold is defined as the quadruplet $(\M,\psi,h,\nabla)$ where:
\bi
\item $\M$ is a $d+1$-dimensional spacetime coordinatised by $(t,x^i)$ where $i\in\pset{1,\ldots,d}$.
\item $h$ stands for the Galilean metric $h=\delta^{i j}\p_i\vee \p_j$.
\item $\psi=dt$ spans the radical of $h$.
\item $\nabla$ is the flat connection (with vanishing coefficients $\Gamma=0$) preserving both $\psi$ and $h$.
\ei

\item \textbf{Newton-Hooke manifold:} The Newton-Hooke manifold differs from the flat Galilei manifold by the connection $\nabla$. The latter is defined in this case as the {\it non-flat} connection preserving both $\psi$ and $h$ whose only non-vanishing coefficients read $\Gamma^i_{tt}=-\frac{x^i}{R^2}$ where $R$ stands for  the \textbf{Newton-Hooke radius}.  

Two cases are to be distinguished according to the sign of the square of the \textbf{Newton-Hooke radius} $R$:
\bi
\item $R^2>0$: expanding Newton-Hooke 
\item $R^2<0$: oscillating Newton-Hooke. 
\ei
The flat Galilei manifold is recovered in the limit $R\to \infty$.

}
\bremi{\label{remGalNH}
\item Note that the two examples displayed in Example \ref{exaGal} share the same Galilean structure $(\M,h,\psi)$ but differ by the choice of Galilean connection, thus illustrating the arbitrariness in the latter (\cf \Prop{proptorsionfreeGal}). 
\item Both the flat Galilei and Newton-Hooke manifolds are examples of Newtonian manifolds, \ie the associated Galilean connection is torsionfree and satisfies the Duval--K\"unzle condition \eqref{eqDK}. 
\item In particular, the Newton-Hooke connection can be put in the form \eqref{Galconn} with:
\bea
N=\p_t\quad,\quad\N\gamma=\delta_{ij}\,  dx^i\vee  dx^j\quad,\quad\N F=d\N A\quad{\text{with}}\quad\N A=\frac{|x|^2}{2R^2}dt
\eea
or alternatively in the form \eqref{NewtMilne} through the identification:
\bea
Z=\p_t\quad,\quad g=\frac{|x|^2}{R^2}dt\vee dt+\delta_{ij}\,  dx^i\vee  dx^j\quad,\quad\phi=\frac{|x|^2}{R^2}.
\eea
\item The Newton-Hooke connection thus differs from the flat Galilei connection by a shift \eqref{eqshifttrumpvar}-\eqref{eqgroupactionNewtpotshift} of the Newtonian potential parameterised by $\bar\phi=\frac{|x|^2}{R^2}$.
\item Both the flat Galilei and Newton-Hooke manifolds are maximally symmetric Galilean manifolds \ie their isometry algebras\footnote{Let us emphasise that the connection is part of the structure and, as such, should be preserved by an isometric transformation. Technically, the affine Killing condition $\Lag_X\nabla=0$ is necessary in order to ensure that the isometry algebra is finite-dimensional, \cf footnote \ref{footnoteaff}.} 
\bea\alg:=\pset{X\in\vf\ |\ \Lag_X h=0, \Lag_X \psi=0 \text{ and }\Lag_X\nabla=0}
\eea
 are both of dimension $\frac{(d+2)(d+1)}{2}$. The latter are isomorphic to the Galilei algebra $\mathfrak{gal}(d,1)$ and the Newton-Hooke algebras $\mathfrak{nh}_\pm(d,1)$, respectively.
 \item The parameterised geodesic equation associated with the Newton-Hooke connection for an observer $x(\tau)=\pset{t(\tau),x^i(\tau)}$ reads:
 \be
 \item $\ddot t=0$
 \item $\ddot x^i-\frac{x^i}{R^2}\dot t^2=0$.
 \ee
Solving the first equation as $t=\tau$ leads to the harmonic (resp. expanding) oscillator equation $\ddot x^i=\frac{x^i}{R^2}$ for $R^2<0$ (resp. $R^2>0$).
}
\subsection{Carrollian}
\label{subsecCarr}
The present section aims at providing a self-contained review of torsionfree Carrollian geometries, following \cite{Duval:2014uoa,Bekaert:2015xua} (\cf also \cite{Hartong:2015xda}). As such, it can be considered as the dual of Section \ref{Galilean}, following the leitmotiv of \cite{Duval:2014uoa} describing Carrollian structures as dual of Galilean ones\footnote{\Cf \eg Table I. of \cite{Bekaert:2015xua} for a summary of this duality at the level of geometric structures. }. 
\bdefi{Carrollian structure}{A Carrollian structure is a triplet $(\M,\xi,\gamma)$ where 
\bi
\item $\M$ is a manifold of dimension $d+1$.
\item $\xi\in\vf$ is a non-vanishing vector field.
\item $\gamma\in\field{\vee^2T^*\M}$ is a covariant metric satisfying the following properties:
\bi
\item $\gamma$ is of rank $d$ and of signature $(0,\underbrace{+1,\ldots,+1}_d)$.
\vspace{-1.2mm}
\item The radical of $\gamma$ is spanned by $\xi$ \ie $\gamma(X,\cdot)=0\Leftrightarrow X\sim\xi$, where $X\in\vf$.
\ei
\ei
}
\bremi{\item Alternatively, a Carrollian structure can be defined as a triplet $(\M,\xi,h)$ where $h\in\Gamma\big(\vee^2(\Ann\xi)^*\big)$ is a rank $d$ contravariant Riemannian metric on\footnote{Recall that sections of $\Ann \xi$ are 1-forms $\alpha\in\ff$ satisfying $\alpha(\xi)=0$.} $\Ann \xi$ of signature $(\underbrace{+1,\ldots,+1}_d)$.
\item Recall that in the Galilean case, the 1-form $\psi$ endowed our spacetime structure with a notion of time by allowing to distinguish between spacelike and timelike vector fields $X\in\vf$. In the Carrollian case, this r\^ole is played by the Carrollian vector field $\xi$ -- although in a slightly trivial way -- as follows:
\bi
\item \textbf{Spacelike}: $X\nsim\xi$
\item \textbf{Timelike}: $X\sim\xi$.
\ei
In other words, in a Carrollian spacetime, timelike observers worldlines identify with curves of the congruence defined by the non-vanishing Carrollian vector field $\xi$. Furthermore, such observers are geodesic with respect to (any) Carrollian connection, as follows from the defining conditions of the latter. 
}
As in the Galilean case, Carrollian spacetimes can be endowed with a notion of parallelism via the introduction of a compatible connection:
\bdefi{Torsionfree Carrollian manifold}{A torsionfree Carrollian manifold is a quadruplet $(\M,\xi,\gamma,\nabla)$ where:
\bi
\item $(\M,\xi,\gamma)$ is a Carrollian structure such that $\gamma$ is \textbf{invariant} \ie $\Lagxi\gamma=0$.
\item $\nabla:\Gamma\pl T\M\pr\to \End\big(\field{T\M}\big)$ is a torsionfree Koszul connection on $\M$ compatible with both $\xi$ and $\gamma$ \ie satisfying
\be
\item $\nabla \xi=0$
\item $\nabla \gamma=0$
\ee
and referred to as the \textbf{Carrollian connection}. 
\ei
}

As shown in \cite{Bekaert:2015xua}, the Killing condition $\Lagxi\gamma=0$ is necessary in order to ensure the existence of torsionfree Carrollian connections. The latter are classified as follows \cite{Bekaert:2015xua}:

\bpropp{Classification of torsionfree Carrollian connections}{\label{proptorfreeCarr}Let $(\M,\xi,\gamma)$ be a Carrollian structure with invariant metric $\gamma$.
\bi
\item 
The space of torsionfree connections compatible with $(\M,\xi,\gamma)$ is an affine space modelled on the vector space $\field{\vee^2\Ann\xi}$. 
\item The most general expression\footnote{An index free ``Koszul-like'' formula for Carrollian connections can be found  in \cite{Bekaert:2015xua}. } for torsionfree connections compatible with $(\M,\xi,\gamma)$ is given by:
\bea
\Gamma^{\lambda}_{\mu\nu}&=&\xi^\lambda\p_{(\mu}A_{\nu)}+\half \ovA h{}^{\lambda\alpha}\big( \p_\mu\gamma_{\alpha\nu}+\p_\nu\gamma_{\alpha\mu}-\p_\alpha\gamma_{\mu\nu}\big)-\xi^\lambda  A_{(\mu}\Lag_{\xi} A_{\nu)}+ \xi^\lambda\overset{ A}{\Sigma}_{\mu\nu}\label{Carrconn}
\eea
where 
\bi
\item $A\in\ff$ is an \textbf{Ehresmann connection} \ie $A(\xi)=1$.\itemnum\label{eqdefA}
\item $\ovA h\in\field{\vee^2T\M}$ is the unique solution to:
\bea
\quad \ovA h{}^{\mu\nu}A_\nu=0\quad,\quad \ovA h{}^{\mu\rho}\gamma_{\rho\nu}=\delta^{\mu}_\nu-\xi^\mu A_\nu\label{eqdefhA}
\eea 
and referred to as the \textbf{transverse cometric} associated with $A$.
\item $\overset{ A}{\Sigma}\in\field{\vee^2\Ann\xi}$ \ie $\overset{ A}{\Sigma}_{[\mu\nu]}=0\quad,\quad\overset{ A}{\Sigma}_{\mu\nu}\, \xi^\nu=0$.\itemnum\label{eqdefSigma}
\ei
\ei
}
\bremi{\label{reminvcarroll}
\item The tensor $\ovA\Sigma$ encodes the arbitrariness in the torsionfree Carrollian connection $\nabla$ \ie the part of the connection that is not fixed by the compatibility relations with $\xi$ and $\gamma$. It can be intrinsically defined as $\ovA\Sigma_{\mu \nu}=-\nabla_{(\mu}A_{\nu)}+A_{(\mu}\, \Lag_\xi A_{\nu)}$. 
 \item The space of Ehresmann connections \big(\ie 1-forms on $\M$ satisfying \eqref{eqdefA}\big) is denoted $\EC$. The latter is an affine space modelled on $\field{\Ann\xi}$ \ie for any two Ehresmann connections $A,A'\in\EC$, there exists $\alpha\in\ff$ such that $\alpha(\xi)=0$ and $A'-A=\alpha$. 
\item Expressions \eqref{Carrconn}-\eqref{eqdefSigma} are left invariant by a \textbf{Carroll-boost} transformation:
\bea
\begin{split}
&A\overset{\alpha}{\mapsto} A+\alpha\quad,\quad
\ovA h{}\overset{\alpha}{\mapsto} \ovA h{}-\alpha^\#\otimes \xi-\xi\otimes\alpha^\#+\ovA h(\alpha,\alpha)\, \xi\otimes\xi\label{Carrboosteq1}\\
&\overset{ A}{\Sigma}_{\mu\nu}\overset{\alpha}{\mapsto} \overset{ A}{\Sigma}_{\mu\nu}-\half \Lag_{\alpha^\#}\gamma_{\mu \nu}+\alpha_{(\mu}\Lag_\xi A_{\nu)}+A_{(\mu}\Lag_\xi \alpha_{\nu)}+\alpha_{(\mu}\Lag_\xi \alpha_{\nu)}
\end{split}
\eea
where $\alpha\in\field{\Ann\xi}$ \ie $\alpha(\xi)=0$ and 
$\alpha^\#{}^\mu:=\ovA h{}^{\mu\nu}\alpha_\nu$. 
}
The action \eqref{Carrboosteq1} on $\EC\times\field{\vee^2\Ann\xi}$ allows to precisely classify Carrollian connections \cite{Bekaert:2015xua}:

\bprop{\label{corCarrtorfree}The space 
of torsionfree Carrollian connections compatible with the Carrollian structure $(\M,\psi,\gamma)$ possesses the structure of an affine space canonically isomorphic to the affine space $\frac{\EC\times\field{\vee^2\Ann\xi}}{\field{\Ann \xi}}$.
}

We now introduce a subclass of torsionfree Carrollian manifolds, dubbed {\it invariant} Carrollian manifolds:
\bdefi{Invariant Carrollian manifold}{\label{deficarrinvar}
A torsionfree Carrollian manifold $(\M,\xi,\gamma,\nabla)$ for which the Carrollian vector field $\xi$ is affine Killing\footnote{The affine Killing condition is tensorial and reads more explicitly as $(\Lag_\xi\nabla)_XY=0$ for all $X,Y\in\vf$ where $(\Lag_\xi\nabla)_XY:=\br{\xi}{\nabla_XY}-\nabla_{\br{\xi}{X}}Y-\nabla_X\br{\xi}{Y}$  or in components $\pl\Lag_\xi\nabla\pr\lmn:=\xi^\rho\p_\rho\Gamma\lmn+\Gamma^\lambda_{\rho\nu}\p_\mu\xi^\rho+\Gamma^\lambda_{\mu\rho}\p_\nu\xi^\rho-\Gamma^\rho_{\mu\nu}\p_\rho\xi^\lambda+\p_\mu\p_\nu\xi^\lambda$. 

\noindent Note that, in the torsionfree case, the latter expression can be recast as $\pl\Lag_\xi\nabla\pr\lmn=\nabla_\mu\nabla_\nu\xi^\lambda-R^\lambda{}_{\nu|\mu\rho}\xi^\rho$.} -- \ie $\Lag_\xi\nabla=0$ -- will be said {invariant}\footnote{\label{footnoteaff}Recall that any Killing vector field $\xi$ for a non-degenerate metric $g$ is automatically affine Killing for the associated Levi--Civita connection $\nabla$ \ie $\Lag_\xi g=0\Rightarrow \Lagxi\nabla=0$. However, there is no such implication in the Carrollian case, so that $\xi$ is not necessarily affine Killing for $\nabla$ despite being Killing for $\gamma$, \ie $\Lagxi\gamma=0\not \Rightarrow\Lagxi\nabla=0$. This is related to the fact that, in contradistinction with the non-degenerate case, the Carrollian metric structure $(\M,\xi,\gamma)$ does not entirely determine torsionfree compatible connections, \cf \Prop{proptorfreeCarr}.
}.
}
\bremi{
\item The torsionfree Carrollian connection \eqref{Carrconn} is invariant if and only if the following relation holds:
\bea
\Lag_\xi \ovA\Sigma_{\mu\nu}+\nabla_{(\mu}\Lag_\xi A_{\nu)}-\Lag_\xi(A_{(\mu}\Lag_\xi A_{\nu)})=0\label{invcoordcond}.
\eea

It can be checked that condition \eqref{invcoordcond} is invariant under a Carroll boost \eqref{Carrboosteq1}. 

\item As will be further advocated in Section \ref{secDuvalCarr}, invariant Carrollian manifolds can be seen as the natural Carrollian counterpart to Newtonian manifolds. Similarly to the Newtonian case, the relevant condition can be formulated in terms of a linear condition on the Riemann curvature as:
\bea
\Riem{\mu}{\nu}{|\alpha}{\beta}\, \xi^\beta=0
\eea
which is indeed equivalent to the invariant condition $\Lag_\xi\nabla=0$ since for any torsionfree Carrollian manifold, the following relation holds: 
\bea
R(\xi,X)Y=(\Lag_\xi\nabla)_XY\text{ for all }X,Y\in\vf.
\eea

}
We now discuss two distinguished examples of torsionfree Carrollian manifolds:
\bexai{Carrollian manifolds}{\label{exaAdSCarr}
\item \textbf{Flat Carroll manifold {\rm\cite{Duval:2014uoa}}:} The flat Carroll manifold is defined as the quadruplet $(\M,\xi,\gamma,\nabla)$ where:
\bi
\item $\M$ is a $d+1$-dimensional spacetime coordinatised by $(u,x^i)$ where $i\in\pset{1,\ldots,d}$.
\item $\gamma$ stands for the Carrollian metric $\gamma=\delta_{ij}\, dx^i\vee  dx^j$ with $\delta$ the $d$-dimensional flat Euclidean metric.
\item The Carrollian vector field $\xi=\p_u$ spans the radical of $\gamma$ and satisfies $\Lagxi\gamma=0$.
\item $\nabla$ is the flat connection preserving both $\xi$ and $\gamma$ with vanishing coefficients $\Gamma=0$. 
\ei

\item \textbf{(A)dS Carroll manifold {\rm\cite{Bergshoeff:2015wma}}:} The (A)dS Carroll manifold is defined as the quadruplet $(\M,\xi,\gamma,\nabla)$ where:
\bi
\item $\M$ is a $d+1$-dimensional spacetime coordinatised by $(u,x^i)$ where $i\in\pset{1,\ldots,d}$.
\item $\gamma$ stands for the Carrollian metric $\gamma=\gamma_{ij}\, dx^i\vee  dx^j$ where 
\bea
\gamma_{ij}=\delta_{ij}+\Big(\frac{R^2}{|x|^2}\sinh^2\frac{|x|}{R}-1\Big)\mbP_{ij}\text{ with }|x|:=\sqrt{\delta_{ij}x^ix^j}\text{ and }\mathbb P^i_{j}:=\delta^i_{j}-\frac{x^ix_j}{|x|^2}\label{hypmet}
\eea
 is the $d$-dimensional (hyperbolic) spherical metric whenever ($R^2>0$) $R^2<0$.
\item The Carrollian vector field $\xi=\p_u$ spans the radical of $\gamma$ and satisfies $\Lagxi\gamma=0$.
\item $\nabla$ is the non-flat connection preserving both $\xi$ and $\gamma$ whose non-vanishing coefficients read: 
\bea
\Gamma^u_{i j}=-\frac{u}{R^2}\, \gamma_{ij}\quad,\quad\Gamma^i_{j k}=\half \gamma^{il}\pl\p_j\gamma_{lk}+\p_k\gamma_{l j}-\p_l\gamma_{jk}\pr.\label{eqconnadscarr}
\eea
\ei
}
\bremi{\label{remCarrinv}
\item Being torsionfree and Carrollian, the connection \eqref{eqconnadscarr} can be put in the form \eqref{Carrconn} upon the identification 
\bea
A=du\quad,\quad\ovA h=\gamma^{ij}\, \p_i\vee  \p_j\quad,\quad\ovA\Sigma=-\frac{u}{R^2}\, \gamma
\eea
or equivalently, upon performing a Carroll boost:
\bea
\hspace{-0.5cm}A=du-\frac{u}{R} \frac{x_i}{|x|}\tanh\frac{|x|}{R}dx^i\quad,\quad\ovA h=\frac{u^2}{R^2}\tanh^2 \frac{|x|}{R}\p_u\vee \p_u+2\, \frac{u}{R} \frac{x^i}{|x|}\tanh\frac{|x|}{R}\, \p_u\vee \p_i+\gamma^{ij}\, \p_i\vee \p_j\quad,\quad\ovA\Sigma=0.
\quad\quad
\label{eqadscarrdesc}
\eea
\item The flat Carroll connection is obviously invariant, \cf \Defi{deficarrinvar}. However, the (A)dS Carroll connection \eqref{eqconnadscarr} is {\it not} invariant, rather $\Lag_\xi\nabla=-\frac{1}{R^2}\xi\otimes\gamma$. 
\item Both the flat and (A)dS Carroll manifolds are maximally symmetric \ie their isometry algebras
\bea\alg:=\pset{X\in\vf\ |\ \Lag_X\xi=0, \Lag_X \gamma=0 \text{ and }\Lag_X\nabla=0}
\eea
 are both of dimension $\frac{(d+2)(d+1)}{2}$. The latter are isomorphic to the Carroll algebra $\mathfrak{carr}(d,1)$ and the (A)dS Carroll algebra $\mathfrak{carr}_\pm(d,1)$, respectively.
 \item Contrarily to the Galilean case where maximally symmetric spacetimes share the same metric structure (differing only by the choice of compatible connection), Carrollian maximally symmetric spacetimes differ already at the metric level.

}

As we will show in Section \ref{sectionConnectionDuval}, the non-invariance of the (A)dS Carroll connection prevents the possibility of its embedding within the framework of \cite{Duval:2014uoa}. We now introduce a weaker notion of invariance for Carrollian manifolds that will prove to be relevant in order to account for the (A)dS Carroll case:
\bdefi{Pseudo-invariant Carrollian manifold}{\label{propinvKoszulCarrpseudo}Let $\pl \M,\xi,\gamma,\nabla\pr$ be a torsionfree Carrollian manifold. 

The connection $\nabla$ will be said pseudo-invariant if there exists a nowhere vanishing invariant function $\Om\in\foncinvn{\M}$, referred to as the \textbf{scaling factor}, 
such that the vector field $\bar \xi:=\Om\, \xi$ is affine Killing for $\nabla$ \ie $\Lag_{\bar\xi}\nabla=0$.
}
\bremi{\label{rempseudocar}
\item Generically, the vector field $\bar \xi$ is not parallelised\footnote{In other words $(\M,\bar\xi,\gamma,\nabla)$ is {\it not} a Carrollian manifold. } by $\nabla$ but rather recurrent \cite{Gibbons:2007zu} with respect to the latter since $\nabla\bar\xi=d\ln\Om\otimes  \bar \xi$. 
\item The affine Killing condition for $\bar\xi$ can be reformulated in terms of $\xi$ as:
\bea
\Lag_\xi\nabla\lmn=-\xi^\lambda\, \Om\un\nabla_\mu\nabla_\nu\, \Om.\label{pseudonabxicond}
\eea
\item The torsionfree Carrollian connection \eqref{Carrconn} is pseudo-invariant if and only if:
\bea
\Lag_\xi \ovA\Sigma_{\mu\nu}+\nabla_{(\mu}\Lag_\xi A_{\nu)}-\Lag_\xi(A_{(\mu}\Lag_\xi A_{\nu)})=-\Om\un\nabla_\mu\nabla_\nu\, \Om.\label{pseudoSigmacond}
\eea
for some nowhere vanishing invariant function $\Om\in\foncinvn{\M}$.
\item From \eqref{pseudoSigmacond}, it follows that the tensor $\ovA\Sigma$ associated with a pseudo-invariant Carrollian connection is invariant (\ie $\Lag_\xi\ovA\Sigma$=0) if $\Lag_\xi A=-d\ln\Om$.
}
We conclude this section by showing that:
\bprop{\label{propAdSCarr}The (A)dS Carroll connection is pseudo-invariant, with scaling factor $\Om=\cosh\frac{|x|}{R}$.}
\bproof{The proof follows straightforwardly from the following equalities:
\bi
\item $\p_i\Om=\frac{x_i}{|x|}\frac{1}{R}\sinh\frac{|x|}{R}$
\item $\p_i\ln\Om=\frac{x_i}{|x|}\frac{1}{R}\tanh\frac{|x|}{R}$
\item $\p_i\p_j\Om=\frac{1}{R|x|}\mathbb P_{ij}\sinh\frac{|x|}{R}+\frac{1}{R^2}\frac{x_ix_j}{|x|^2}\cosh\frac{|x|}{R}$
\item $\Om\un\p_i\p_j\Om=\frac{1}{R|x|}\mathbb P_{ij}\tanh\frac{|x|}{R}+\frac{1}{R^2}\frac{x_ix_j}{|x|^2}$
\item $\Gamma^k_{ij}=-\frac{2}{|x|^2}\pl1-\frac{|x|}{R}\coth\frac{|x|}{R}\pr\mbP^{k}_{(i}\, x_{j)}+\pl 1-\frac{R}{|x|}\cosh\frac{|x|}{R}\sinh\frac{|x|}{R}\pr \frac{x^k}{|x|^2}\, \mbP_{ij}$
\item $\Gamma^k_{ij}\p_k\ln\Om=\frac{1}{|x|R}\tanh\frac{|x|}{R} \mbP_{ij}-\frac{1}{|x|^2}\sinh^2\frac{|x|}{R} \mbP_{ij}$
\item $\Om\un\nabla_i\nabla_j\Om=\frac{1}{R^2}\gamma_{ij}$.
\ei
}
Beyond the fact that it accounts for the case of the (A)dS Carroll spacetime, the notion of pseudo-invariant Carrollian manifold will prove relevant within the ambient context. In particular, it will be shown in Section \ref{secCarrembed} that pseudo-invariant Carrollian manifolds constitute the most general class of torsionfree Carrollian spacetimes arising as leaves of the foliation of a gravitational wave. 

\pagebreak\section{Geometry of gravitational waves}
\label{Gravitational waves}
This section is devoted to the geometry of various Lorentzian spacetimes that fall under the label of gravitational waves\footnote{We restrict ourselves to considerations at the kinematical level \ie none of the definitions below involve equations of motion.}.
After reviewing some classical definitions, we introduce a new subclass -- dubbed {\it Dodgson waves} -- which will prove relevant regarding the ambient approach to Carrollian manifolds, \cf Section \ref{secCarrembed}. A summary of the hierarchy of spacetimes discussed throughout the present section is provided in Table \ref{diagboxes2} and Figure \ref{diagboxes} while local expressions can be found in Section \ref{appBrink}.
\begin{term}
We will denote {\it ambient} ($d+2$-dimensional) structures by topping them with a hat, in order to distinguish them from {\it intrinsic} ($d+1$-dimensional) structures, as discussed in Section \ref{Intrinsic geometries}.
\end{term}
\subsection{Bargmannian structures}
\label{secBargstruc}
Before focusing on gravitational waves, we start by recalling the notion of Bargmannian structures\footnote{We pursue with the terminology used in Section \ref{Intrinsic geometries} which distinguishes between manifolds endowed with metric structures (referred to as {\it structures}) and structures supplemented with a compatible connection (referred to as {\it manifolds}), \cf footnote \ref{footnotestructuremanifold}.} -- \ie Lorentzian spacetimes endowed with a lightlike vector field -- in order to introduce various related objects and notions as well as to fix some terminology:
\bdefi{Bargmannian structure}{\label{defifibLor}A Bargmannian structure is a triplet $\big(\hat\M,\hat \xi,\hat g\big)$ where:
\bi
\item $\hat\M$ is a manifold of dimension $d+2$.
\item $\hat\xi\in\Gamma\big(T\hat\M\big)$ is a nowhere vanishing vector field. 
\item $\hat g\in\field{\vee^2T^*\hat\M}$ is a covariant metric such that:
\bi
\item $\hat g$ is of rank $d+2$ and signature $(-1,\underbrace{+1,\ldots,+1}_{d+1})$.
\vspace{-1.2mm}
\item $\hat\xi$ is lightlike with respect to $\hat g$ \ie $\hat g(\hat\xi,\hat\xi)=0$.
\ei
\ei
}
\bremi{\label{remBM}
\item A Bargmannian structure $\big(\hat\M,\hat \xi,\hat g\big)$ supplemented with a compatible connection $\hat \nabla$ (\ie satisfying $\hat\nabla\hat\xi=0$ and $\hat\nabla\hat g=0$) will be referred to as a \textbf{Bargmannian manifold} \cite{Duval:1984cj} (\cf also the companion paper \cite{Morand2018b}). 
\item In the present section, the notion of parallelism associated with a given Bargmannian structure $\big(\hat\M,\hat \xi,\hat g\big)$ will be provided by the Levi--Civita connection $\hat\nabla$ associated with the Lorentzian metric $\hat g$. Note that, in general, the quadruplet $\big(\hat\M,\hat \xi,\hat g,\hat\nabla\big)$ is {\it not} a Bargmannian manifold since the Levi--Civita connection in general fails to preserve the vector field $\hat \xi$ (\ie $\hat\nabla\hat\xi\neq0$ in general). Among Bargmannian structures, only Bargmann--Eisenhart waves (to be introduced in \Defi{defBEW}) are Bargmannian manifolds. 
\item The nowhere vanishing 1-form dual to $\hat \xi$ with respect to $\hat g$ will be denoted $\hat \psi\in\ffh$  \big(\ie $\hat \psi:=\hat g(\hat \xi)$\big) while $\fhpsi:=\pset{V\in\vfh\,|\,\hat \psi(V)=0}$ will denote the distribution of vector fields spanning the kernel of $\hat \psi$. 

\item Note that the fact that $\hat \xi$ is assumed to be lightlike ensures that $\hat \psi(\hat \xi)=0$ \ie $\hat \xi\in\fhpsi$. 

\item Given a Bargmannian structure $\big(\hat\M,\hat \xi,\hat g\big)$, any function $\Omega\in\foncmh$ satisfying $\Lag_{\hat\xi}\, \Om=0$ will be called \textbf{invariant} and the space of invariant functions will be denoted $\foncmhinv$. 
\item The space of Bargmannian structures is invariant under the following free 
action of the abelian multiplicative group defined as the direct product of nowhere vanishing 
functions $\foncmhn\times\foncmhn$:
\bea
\hat\xi&\overset{\Om}{\mapsto}&\Om\, \hat\xi\label{groupaction1}\\
\hat g&\overset{\Lambda}{\mapsto}&\Lambda^2\, \hat g\label{groupaction2}
\eea
where $\Om,\Lambda\in\foncmhn$.\footnote{The square in \eqref{groupaction2} is introduced for later convenience. }
\item An orbit of Bargmannian structures under the group action \eqref{groupaction1} \big(resp. \eqref{groupaction2}\big) will be denoted $(\hat\M,[\hat \xi],\hat g)$ \big(resp. $(\hat\M,\hat \xi,[\hat g])$\big). 
\item Given an orbit $(\hat\M,[\hat \xi],\hat g)$ \big(resp. $(\hat\M,\hat \xi,[\hat g])$\big), the distribution $\Ker \hat \psi$, where $\hat \psi=\hat g(\hat\xi)$ with $\hat \xi$ (resp. $\hat g$) any representative of the orbit, is canonical \ie independent of the choice of representative. 
\item The covariant metric $\hat\gamma\in\Gamma\big(\vee^2(\Ker\hat\psi)^*\big)$ induced by the Bargmannian metric $\hat g$ on $\Ker\hat\psi$ \big( \ie $\hat\gamma:=\hat g|_{\Ker\hat\psi}$\big) is called the \textbf{Leibnizian metric} induced by $\big(\hat\M,\hat \xi,\hat g\big)$. One can check that $\Rad \hat\gamma=\Span\hat\xi$ so that the quadruplet $\big(\hat\M,\hat \xi,\hat\psi,\hat \gamma\big)$ forms a Leibnizian structure, as defined in \cite{Bekaert:2015xua,Morand2018b}.
}
\subsection{Gravitational waves}
\label{secGW}
\bdefi{Gravitational wave}{A gravitational wave is an orbit $(\hat\M,[\hat \xi],\hat g)$ of Bargmannian structures such that, for any representative $\hat \xi\in[\hat \xi]$, the following equivalent conditions hold:
 \begin{itemize}
\item $\hat \xi$ is hypersurface-orthogonal.  
\item $\hat \psi:=\hat g(\hat \xi)$ satisfies the Frobenius integrability condition, \ie $\hat d\hat\psi\w\hat\psi=0$.\footnote{Where $\hat d:\Om^\bullet(\hat\M)\to\Om^{\bullet+1}(\hat\M)$ denotes the de Rham differential on $\hat \M$.} 
\item The canonical distribution $\Ker\hat \psi$ is involutive. 
\end{itemize}
}

\begin{figure}[ht]
\centering
   \includegraphics[width=0.8\textwidth]{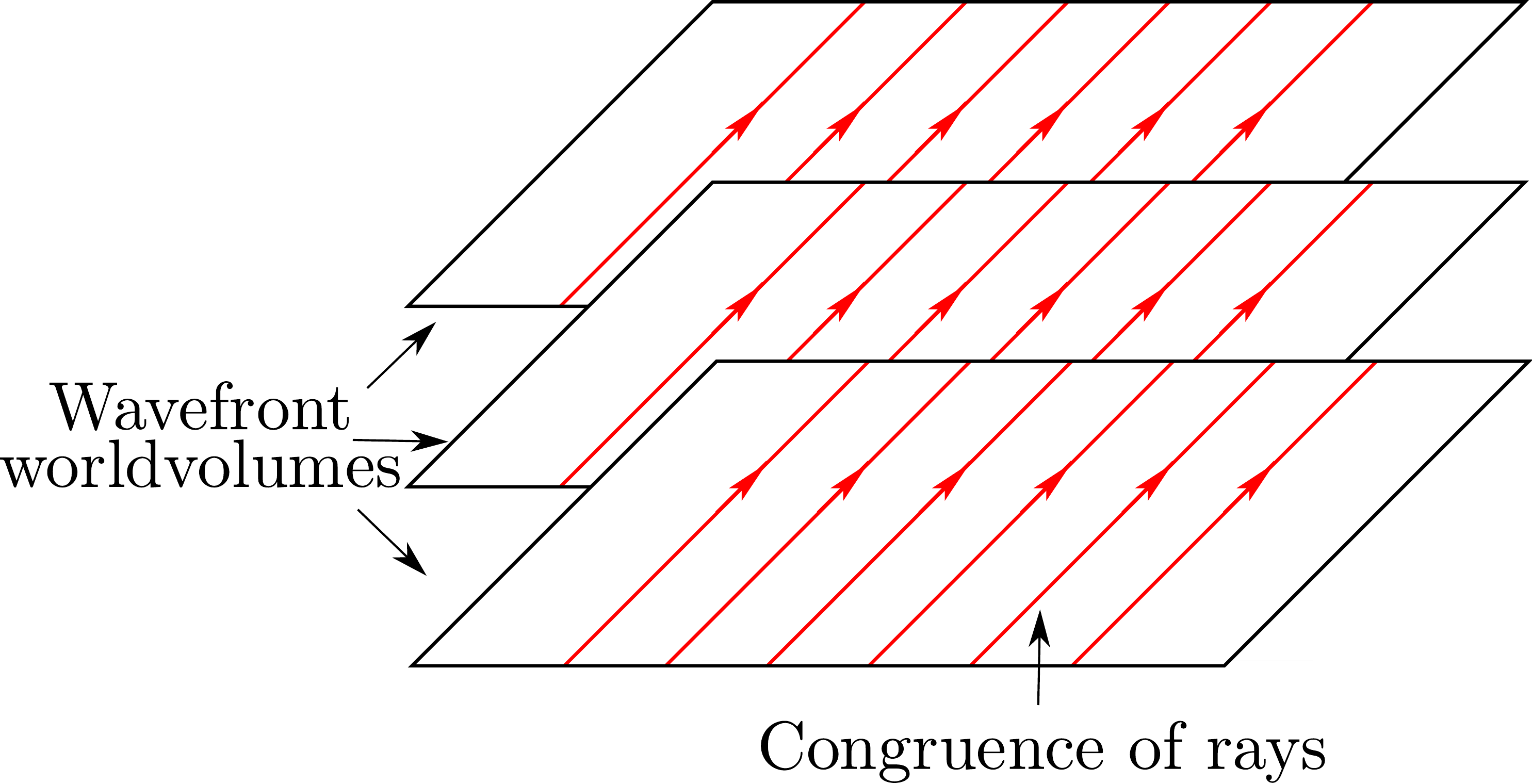}
  \caption{Foliation of a gravitational wave by wavefront worldvolumes\label{fig1}}
\end{figure}

The integrability condition ensures, via Frobenius theorem, that the $(d+2)$-dimensional spacetime $\hat\M$ of a gravitational wave admits a canonical foliation by a family of $(d+1)$-dimensional lightlike hypersurfaces $\M_t$ whose tangent space $T_x\M_t$ at each point $x\in\M_t$ is isomorphic to $\Ker\hat \psi_x$. Such hypersurfaces will be referred to as \textbf{wavefront worldvolumes}, consistently with the terminology adopted in \cite{Bekaert:2013fta,Bekaert:2014bwa,Bekaert:2015xua}, while the integral curves of $\hat\xi$ will be called \textbf{rays} (\cf Figure \ref{fig1}).

The following proposition follows straightforwardly from Frobenius Theorem:
\bprop{\label{propgw}Let $(\hat\M,[\hat\xi],\hat g)$ be a gravitational wave. 

For any representative $\hat\xi\in[\hat\xi]$ and dual 1-form $\hat\psi:=\hat g(\hat\xi)$, the following statements hold:
\bi
\item Locally, there exists a nowhere vanishing function $\Om\in\foncmhn$, referred to as the \textbf{scaling factor} of $\hat\xi$ with respect to $\hat g$, such that: \bea \hat d\hat\psi=\hat d\ln\Om\w\hat\psi.\label{eqfrob}\eea
\item Letting $\hat{\bar\xi}:=\Om\un\hat\xi$, the 1-form $\hat{\bar\psi}\in\ff$ defined as $\hat{\bar\psi}:=\hat g(\hat{\bar\xi})=\Om\un\, \hat\psi$ is closed, \ie $\hat d\hat{\bar\psi}=0$. 

The representative $\hat{\bar\xi}$ (unique up to constant rescaling)  will be called the \textbf{special} vector field of $[\hat\xi]$. 
\item The space of gravitational waves is invariant under the action \eqref{groupaction2}. Under such an action, the scaling factor $\Om$ in \eqref{eqfrob} is rescaled according to $\Om\overset{\Lambda}{\mapsto}\Om\, \Lambda^{2}$. 
\ei
}
In the following, we will make use of two different parameterisations regarding the local expression of gravitational waves. These two classes of coordinate systems will prove useful in order to discuss the embedding of Galilean and Carrollian manifolds, respectively. 

Let $(\hat\M,[\hat\xi],\hat g)$ be a gravitational wave:
\bi
\item \textbf{Galilean}

The first parameterisation makes explicit use of Galilean variables\footnote{Note however that the variables appearing in the present section are \textit{a priori} defined over the whole ambient spacetime $\hat\M$.}, \cf \eqref{Trumpervariables}:
\bea
x^{\hat\mu}\in\pset{u,x^\mu}\quad,\quad\hat g\un:=\begin{pmatrix}
-\phi&Z^\nu\\
Z^\mu&h^{\mu\nu}
\end{pmatrix}\quad,\quad
\hat g:=\begin{pmatrix}
0&\psi_\nu\\
\psi_\mu&g_{\mu\nu}
\end{pmatrix}\label{hatgNC}
\eea
where all variables are functions of $x^{\hat\mu}\in\pset{u,x^\mu}$, where $\hat\mu\in\pset{0,\ldots,d+1}$ and $\mu\in\pset{0,\ldots,d}$.

The associated line element reads more explicitly 
\bea
\hat{ds}{}^2=2\, \psi_\mu\, dx^\mu\, du+g_{\mu \nu}\, dx^\mu\, dx^\nu.\label{lineelement}
\eea
As seen in \eqref{lineelement}, the $u$ direction is lightlike, allowing to define the lightlike vector field $\hat\xi:=\p_u$.  

The dual 1-form $\hat\psi:=\hat g(\hat \xi)$ reads $\hat\psi=\psi_\mu dx^\mu$ so that $\hat\psi(\hat\xi)=0$. 

The orbit $[\hat\xi]$ is obtained by acting on $\hat\xi$ with \eqref{groupaction1}. 

We further assume the following relations:
\bi
 \item\textbf{Algebraic}
\bea
\psi_\mu Z^\mu=1\quad,\quad g_{\mu\nu}Z^\nu=\phi\, \psi_\mu\quad,\quad h^{\mu\nu}\psi_\nu=0 \quad,\quad h^{\mu\rho}g_{\rho\nu}=\delta^{\mu}_\nu-Z^\mu \psi_\nu.
\eea 
 \item\textbf{Differential}
 \bea
h^{\mu\nu}\p_u\psi_\nu=0\quad\text{and}\quad d \psi\w\psi=0.
\eea 
\ei
\bremi{
\item Note that the above algebraic conditions formally identify with the defining relations \eqref{reltrump} of the Galilean variables $(\psi,h,\phi,Z,g)$. These algebraic relations guarantee that $\hat g^{\hat\mu \hat\lambda}\hat g_{\hat \lambda \hat \nu}=\delta^{\hat\mu}_{\hat \nu}$ while the differential ones ensure that $\hat \psi$ is Frobenius \ie $\hat d\hat\psi\w\hat\psi=0$.

\item Alternatively, one can choose to assume the stronger differential constraints: 
\bea
\p_u\psi_\nu=0\quad\text{and}\quad d \psi=0\label{condspec}
\eea 
which ensures that $\hat d\hat\psi=0$, so that $\hat\xi$ identifies in this case with the special vector field of $[\hat\xi]$, \cf \Prop{propgw}.

\item The line element \eqref{lineelement} is left invariant by the following shift\footnote{\textit{Cf}. \cite{Bekaert:2013fta,Bekaert:2014bwa} for an interpretation of the transformation \eqref{eqtransU12} both in a purely nonrelativistic and ambient context.} of the coordinate $u$ with respect to the invariant function $f(x)\in\foncmhinv$:
\bea
\begin{split}
&u\overset{f}{\mapsto}u-f\quad,\quad
g_{\mu\nu}\overset{f}{\mapsto}g_{\mu\nu}+2\, \psi _{(\mu}\p_{\nu)}f\\
&Z^\mu\overset{f}{\mapsto}Z^\mu-h^{\mu\nu}\p_\nu f\quad,\quad
\phi\overset{f}{\mapsto}\phi+2\, Z^\mu\p_\mu f-h^{\mu\nu}\p_\mu f\p_\nu f.\label{eqtransU12}
\end{split}
\eea
In other words, a shift of the lightlike coordinate $u$ amounts to a Maxwell-gauge transformation \eqref{eqtransU1} of the Galilean variables. 
\item For further use, we note that the Christoffel coefficients associated with \eqref{hatgNC} read:
\bea
\hat\Gamma^{u}_{uu}&=&Z^\lambda\p_u\psi_\lambda\nn\\
\hat\Gamma^{u}_{u\nu}&=&\half Z^\alpha\big( \p_u g_{\alpha\nu}+2\, \p_{[\nu}\psi_{\alpha]}\big)\nn\\
\hat\Gamma^{u}_{\mu\nu}&=&-\half\Lag_Z g_{\mu\nu}+\p_{(\mu}\phi\, \psi_{\nu)}+\half\phi\, \p_u g_{\mu\nu}\nn\\
\hat\Gamma^{\lambda}_{uu}&=&h^{\lambda\alpha}\p_u\psi_{\alpha}\nn\\
\hat\Gamma^{\lambda}_{u\nu}&=&\half h^{\lambda\alpha}\big( \p_u g_{\alpha\nu}+2\p_{[\nu}\psi_{\alpha]}\big)\nn\\
\hat\Gamma^{\lambda}_{\mu\nu}&=&Z^\lambda\p_{(\mu}\psi_{\nu)}+\half h^{\lambda\alpha}\big( \p_\mu g_{\alpha\nu}+\p_\nu g_{\alpha\mu}-\p_\alpha g_{\mu\nu}\big)-\half Z^\lambda\p_u g_{\mu\nu}\nn.
\eea
}
\item \textbf{Carrollian}

As an alternative to the Galilean choice, any gravitational wave can be parameterised  in terms of Carrollian variables \cite{Bekaert:2015xua,Hartong:2015xda}:
\bea
x^{\hat\mu}\in\pset{t,x^\mu}\quad,\quad\hat g\un:=\begin{pmatrix}
0&\xi^\nu\\
\xi^\mu&g^{\mu\nu}
\end{pmatrix}\quad,\quad
\hat g:=\begin{pmatrix}
-\lambda&A_\nu\\
A_\mu&\gamma_{\mu\nu}\label{hatgCarr1}
\end{pmatrix}
\eea
where all variables are functions of $x^{\hat\mu}\in\pset{t,x^\mu}$ and the line element takes the following form:
\bea
ds^2=-\lambda\, dt^2+2\, A_\mu dx^\mu\, dt+\gamma_{\mu \nu}dx^\mu\, dx^\nu.
\eea

Similarly to the Galilean case, we need to assume the following algebraic relations:
\bea
A_\mu\xi^\mu=1\quad,\quad \gamma_{\mu\nu} \xi^\nu=0\quad,\quad g^{\mu\nu}A_\nu=\lambda\, \xi^\mu\quad,\quad g^{\mu\rho}\gamma_{\rho\nu}=\delta^{\mu}_\nu-\xi^\mu A_\nu.\label{algrelcar}
\eea 
Acting on $\hat{\bar\xi}:=\xi^\mu\p_\mu$ with \eqref{groupaction1} defines the orbit $[\hat\xi]$. We note that the dual 1-form $\hat{\bar\psi}:=\hat g(\hat{\bar\xi})$ reads $\hat{\bar\psi}=dt$, so that $\hat d\hat{\bar\psi}=0$ and $\hat{\bar\xi}$ is thus the special vector field of $[\hat\xi]$. Contrarily to the Galilean case, no differential condition is necessary and any vector field $\hat\xi\in[\hat\xi]$ is automatically hypersurface-orthogonal.

For further applications, we introduce an alternative Carrollian parameterisation for which the representative vector field is not necessarily special, as:
\bea
\hat g\un:=\begin{pmatrix}
0&\Om\un\xi^\nu\\
\Om\un\xi^\mu&g^{\mu\nu}
\end{pmatrix}\quad,\quad
\hat g:=\begin{pmatrix}
-\Om^2\lambda&\Om\, A_\nu\\
\Om\, A_\mu&\gamma_{\mu\nu}\label{hatgCarr2}
\end{pmatrix}
\eea
where $\pset{\xi,\gamma, A,g}$ satisfy the algebraic relations \eqref{algrelcar} and $\Om\in\foncmhn$ is nowhere vanishing. 

Defining $\hat\xi:=\xi^\mu\p_\mu$ leads to  $\hat\psi:=\hat g(\hat\xi)=\Om\, dt$ so that $\hat d\hat\psi=\hat d\ln\Om\w\hat\psi$ and $\hat\psi$ is Frobenius. 

The Christoffel coefficients associated with \eqref{hatgCarr2} read:
\bea
\begin{split}
\hat\Gamma^{t}_{tt}&=\half\Om\,  \Lag_\xi\lambda+\xi^\alpha\p_t A_\alpha+\lambda\Lag_\xi\Om+\p_t\ln\Om\\
\hat\Gamma^{t}_{t\nu}&=\half\, (\Om\un\xi^\alpha\p_t\gamma_{\alpha\nu}-\Lag_\xi A_\nu+\p_\nu\ln\Om-\Lag_\xi\ln\Om\, A_\nu)
\\
\hat\Gamma^{t}_{\mu\nu}&=-\half\Om\un\Lag_\xi\gamma_{\mu\nu}\label{eqChrisCarr2}\\
\hat\Gamma^{\lambda}_{tt}&=-\half\, \Om\, \xi^\lambda\p_t\lambda+\Om\,  g^{\lambda\alpha}\p_tA_\alpha+\half \Om^2g^{\lambda\alpha}\p_\alpha\lambda+g^{\lambda\rho}\lambda\, \Om\, \p_\rho\Om\\
\hat\Gamma^{\lambda}_{t\nu}&=-\half\Om\, \xi^\lambda\p_\nu\lambda-\half\lambda\xi^\lambda\p_\nu\Om+\half g^{\lambda\alpha}\big( \p_t\gamma_{\alpha\nu}+2\Om\p_{[\nu}A_{\alpha]}-\p_\alpha\Om A_\nu\big)\\
\hat\Gamma^{\lambda}_{\mu\nu}&=\xi^\lambda\p_{(\mu}A_{\nu)}+\half g^{\lambda\alpha}\big( \p_\mu\gamma_{\alpha\nu}+\p_\nu\gamma_{\alpha\mu}-\p_\alpha\gamma_{\mu\nu}\big)-\half\Om\un\xi^\lambda\p_t\gamma_{\mu\nu}+\xi^\lambda\p_{(\mu}{\ln\Om}\, A_{\nu)}.
\end{split}
\eea
\ei
 \subsection{Kundt waves}
 \label{Kundt waves}
 We now focus on a subclass of gravitational waves, called \textit{Kundt waves}, \cf \cite{Kundt1961,Podolsky:2008ec,Coley:2009ut}. This class of spacetimes is of particular importance within the context of string theory as they constitute promising candidates for classical string vacua, generalising the Ricci flat pp-waves \cite{Amati:1988sa,Horowitz:1989bv}. 
 More precisely, it has been shown recently that all four-dimensional universal\footnote{Recall that a metric is called \textbf{universal} \cite{Coley:2008th} if the following condition is satisfied: all conserved symmetric rank-2 tensors constructed from the metric, the Riemann tensor (of the associated Levi--Civita connection) and its covariant derivatives are themselves multiples of the metric. Hence, universal metrics provide vacuum solutions for all gravitational theories whose Lagrangian is a diffeomorphism invariant density constructed from the metric, the Riemann tensor and its covariant derivatives. In other words, universal spacetimes are solutions to Einstein equations while being immune to any higher-order -- or ``quantum'' -- corrections. } metrics of Lorentzian signature are of the Kundt type \cite{Hervik:2017sdr}. Furthermore, all  known examples of Lorentzian universal spacetimes in higher-dimensions are Kundt, \cf \cite{Coley:2008th,Hervik:2015mja,Hervik:2013cla}. 
 
 Kundt waves can be geometrically defined as:
\bdefi{Kundt wave \cite{Coley:2009ut}}{\label{defiKundt}A Kundt wave is a gravitational wave $(\hat\M,[\hat\xi],\hat g)$ such that any representative $\hat\xi\in[\hat\xi]$ is geodesic, expansionless, shearless and twistless.
}
A more convenient definition for our purpose is given by the following lemma (\cf \eg \cite{Morand:2016rrt}):

\blem{}{\label{lemgravKundt}A gravitational wave $(\hat\M,[\hat\xi],\hat g)$ is a Kundt wave if and only if the following equivalent conditions hold:
\begin{enumerate}
\item For all $V,W\in\Milnegh$, $\hat\nabla_VW\in\Milnegh$.
\item For all $V,W\in\Milnegh$ and ${\hat\xi}\in[{\hat\xi}]$, $(\Lag_{\hat\xi} {\hat g})(V,W)=0$.
\item{For all $V,W\in\Milnegh$ and ${\hat\xi}\in[{\hat\xi}]$, $(\hat\nabla_V\hat\psi)(W)=0$. }
\item For all ${\hat\xi}\in[{\hat\xi}]$, there exist $\eta\in\Gamma\big({\Ann\hat\xi}\big)$ and $\om\in \ffh$
such that \bea\hat\nabla\hat\psi=\hat\psi\otimes \eta+\om\otimes\hat\psi.\label{eqKundt}\eea\label{item4}
\vspace{-0.8cm}

\end{enumerate}
Here $\Ker\hat\psi$ denotes the canonical distribution induced by $(\hat\M,[{\hat\xi}],{\hat g})$ and $\hat\nabla$ is the Levi--Civita connection associated with $\hat g$. 
}
\bremi{\label{remKundtauto}
\item The previous lemma can be used in order to characterise Kundt waves using the previously introduced parameterisations:
\bi
\item \textbf{Galilean:} $\p_ug|_{\Ker\psi}=0$ or equivalently $\p_u h^{\mu \nu}=0$.\itemnum\label{condhinv}
\item \textbf{Carrollian:} $\Lag_{\xi}\gamma=0$.\itemnum\label{condgammainv}
\ei
\item Equation \eqref{eqKundt} is invariant under the following action of the direct product of the abelian multiplicative groups of nowhere vanishing functions and nowhere vanishing invariant functions $\foncmhn\times\foncmhinvn$:
\bea\label{actionKundt}
\hat\xi\overset{\Om,\Lambda}{\mapsto}\Om\, \hat\xi\quad,\quad
\hat g\overset{\Om,\Lambda}{\mapsto} \Lambda^2\hat g\quad,\quad
\hat\psi\overset{\Om,\Lambda}{\mapsto}\Lambda^2\Om\, \hat\psi\quad,\quad
\eta\overset{\Om,\Lambda}{\mapsto}\eta-\hat d\ln\Lambda\quad,\quad
\om\overset{\Om,\Lambda}{\mapsto}\om+\hat d\ln\Lambda+\hat d\ln\Om
\eea
where $\Om\in\foncmhn$ and $\Lambda\in\foncmhinvn$. 
\item The first item of Lemma \ref{lemgravKundt} entails that the wavefront worldvolumes foliating a gravitational wave are autoparallel submanifolds\footnote{\label{ftauto}Recall that a distribution $\mathscr D\subset T\M$ is said to be \textbf{autoparallel} with respect to the connection $\nabla$ if $\nabla_XY\in\field{\mathscr D}$ for all $X,Y\in\field{\mathscr D}$. If $\nabla$ is torsionfree, then the autoparallel distribution $\mathscr D$ is involutive \big(\ie $\br{X}{Y}\in\field{\mathscr D}$ for all $X,Y\in\field{\mathscr D}$\big) hence integrable and the leaves of the induced foliation are said to be \textbf{autoparallel submanifolds}. } with respect to the associated Levi--Civita connection if and only if the gravitational wave is Kundt. This property will impart Kundt waves a particular rôle in the context of the embedding of Carrollian manifolds, \cf Sections \ref{secDuvalCarr} and \ref{secCarrembed}. 

}
\subsection{Bargmann--Eisenhart waves}
We pursue our typology of gravitational waves by reviewing two subclasses of Kundt waves that proved particularly relevant regarding the ambient approach to non-Riemannian structures. 
In the present section, we start by discussing a class first introduced by Brinkmann in \cite{Brinkmann} before being brought in the context of the embedding of nonrelativistic physics by Eisenhart in \cite{Eisenhart1928}. This class is at the center of the ambient approach of Duval \etal \cite{Duval:1984cj,Duval:1990hj} who in particular emphasised the deep connection between this class of gravitational waves and the Bargmann algebra\footnote{Recall that the Bargmann algebra is the central extension of the Galilei algebra, \cf footnote \ref{footnoteBarg}. }, \cf below. For these reasons, this important class was dubbed {\it Bargmann--Eisenhart waves} in   \cite{Bekaert:2013fta}.
\bdefi{Bargmann--Eisenhart wave}{\label{defBEW}A Bargmann--Eisenhart wave $(\hat\M,\hat\xi,\hat g)$ is a gravitational wave $(\hat\M,[\hat\xi],\hat g)$ such that there exists a distinguished representative $\hat\xi\in[\hat\xi]$ parallelised by the Levi--Civita connection $\hat \nabla$ associated with $\hat g$, \ie $\hat\nabla\hat\xi=0$.

Equivalently, the two following conditions are satisfied:
\begin{enumerate}
\item The vector field $\hat\xi$ is Killing for the metric $\hat g$, \ie $\Lag_{\hat\xi} \hat g=0$. 
\item The 1-form $\hat\psi$ is closed, \ie $\hat d\hat\psi=0$. 
\end{enumerate}
}
\bremi{\label{remBE2}
\item In other words, a Bargmann--Eisenhart wave is a gravitational wave for which the special vector field is Killing. 
\item Bargmann--Eisenhart waves are Kundt as the 1-form $\hat\psi$ dual to $\hat\xi$ satisfies $\hat\nabla\hat\psi=0$, \cf \eqref{eqKundt}.
\item As mentioned in Section \ref{secBargstruc}, Bargmann--Eisenhart waves are the only gravitational waves inducing a Bargmannian manifold $(\hat\M,\hat\xi,\hat g,\hat\nabla)$ (\cf Remark \ref{remBM}). More precisely, Bargmann--Eisenhart waves can be equivalently characterised as torsionfree Bargmannian manifolds. Put differently, Bargmann--Eisenhart waves can be seen as torsionfree Cartan geometries for the Bargmann algebra $\barg(d+1,1)$, \cf \cite{Andringa:2010it,Morand:2016rrt,Morand2018b}. 
\item The space of Bargmann--Eisenhart waves is preserved by the following action of the abelian multiplicative group of {invariant} functions $\foncmhinv$:
\bea
\hat\xi\overset{{\hat\phi}}{\mapsto}\hat\xi\quad,\quad
\hat g\overset{{\hat\phi}}{\mapsto}\hat g+{\hat\phi}\, \hat \psi\otimes\hat\psi\quad,\quad
\hat\psi\overset{{\hat\phi}}{\mapsto}\hat\psi \quad,\quad \text{with }{\hat\phi}\in\foncmhinv.\label{eqorbBE}
\eea
}

The crucial r\^ole played by Bargmann--Eisenhart waves within the context of the ambient approach (both of nonrelativistic and ultrarelativistic structures) will be reviewed in Section \ref{secDuval}.

\subsection{Platonic waves} 
The second subclass of Kundt waves we would like to discuss has first been introduced in the context of lightlike reduction in the works \cite{Lichnerowicz1955,Julia:1994bs} and further studied in \cite{Bekaert:2013fta} where it was dubbed \textit{Platonic waves} (based on an analogy of \cite{Minguzzi:2006wz,Minguzzi:2006gq}).  It can be seen as a generalisation of the class of Bargmann--Eisenhart waves for which the Killing vector field is not necessarily special. 
\bdefi{Platonic wave}{\label{defiPlato}A Platonic wave $(\hat\M,\hat\xi,\hat g)$ is a gravitational wave $(\hat\M,[\hat\xi],\hat g)$ such that there exists a distinguished representative $\hat\xi\in[\hat\xi]$ which is Killing with respect to the metric $\hat g$ \ie $\Lag_{\hat\xi} \hat g=0$.}
\bremi{
\item Platonic waves have been shown to be Kundt\footnote{More precisely, it has been shown in \cite{Bekaert:2013fta} that Platonic waves belong to the more restrictive class of \textbf{degenerate Kundt} spacetimes, \cf \cite{Coley:2009ut} for a geometric definition. } in \cite{Bekaert:2013fta}.
\item The space of Platonic waves is preserved by the following action of the abelian multiplicative group of nowhere vanishing {invariant} functions $\foncmhinvn$:
\bea
\hat\xi\overset{\Lambda}{\mapsto}\hat\xi\quad,\quad
\hat g\overset{\Lambda}{\mapsto}\Lambda^2\, \hat g\quad,\quad
\hat\psi\overset{\Lambda}{\mapsto}\Lambda^2\, \hat\psi. \label{eqorbPlato}
\eea
 \item Making use of the Galilean parameterisation \eqref{hatgNC}, a gravitational wave is Platonic if and only if $\p_u\psi_\mu=0$ and $\p_u g_{\mu \nu}=0$. 
}
The following statement was proved in \cite{Bekaert:2013fta}  \big(\cf the group action \eqref{eqorbPlato}\big): 
\bprop{\label{proporbplatBE}Any orbit of Platonic waves contains a Bargmann--Eisenhart wave. }
In other words, Platonic waves can be characterised as conformal Bargmann--Eisenhart waves with preserved lightlike Killing vector field. 

The rôle of Platonic waves regarding the embedding of Galilean structures will be discussed in Section \ref{secELl}. 
\subsection{Walker waves}
The fourth item of Lemma \ref{lemgravKundt} suggests to introduce two new subclasses of Kundt waves, namely when one or the other 1-form $\eta$ or $\om$ vanishes. Whenever the 1-form $\eta$ is assumed to vanish, the subclass of Kundt waves described is known as \textit{Walker waves} in the literature, \cf \cite{Walker1950,Gibbons:2007zu}.

\bdefi{Walker wave}{\label{defiWalker}A Walker wave $(\hat\M,[\hat\xi],\hat g)$ is a gravitational wave such that for any representative $\hat\xi\in[\hat\xi]$, there exists a 1-form $\om\in\ffh$ such that:
\bea
\hat\nabla\hat\xi=\om\otimes\hat\xi. \label{recrel}
\eea
}
\bremi{
\item Walker waves are Kundt waves, as is obvious from comparing condition \eqref{recrel} and the fourth item of Lemma \ref{lemgravKundt}. 
\item A vector field satisfying \eqref{recrel} is said \textbf{recurrent} and $\om$ is referred to as the recurrent 1-form \cite{Gibbons:2007zu}. 
\item Condition \eqref{recrel} is invariant under the following action of the abelian multiplicative group of nowhere vanishing functions $\foncmhn$:
\bea
\hat\xi\overset{\Om}{\mapsto}\Om\, \hat\xi\quad,\quad
\hat\psi\overset{\Om}{\mapsto}\Om\, \hat\psi\quad,\quad
\om\overset{\Om}{\mapsto} \om+\hat d\ln\Om.\label{Walkergroupaction}
\eea
This last property ensures that the definition of a Walker wave -- involving an orbit $[\hat\xi]$ of lightlike vector fields -- is consistent.
\item The subclass of Walker waves with exact recurrent 1-form identifies with the class of Bargmann--Eisenhart waves.
}
\subsection{Dodgson waves}
\label{Dodgson waves}
Dually to the Walker case, we now consider the subclass of Kundt waves defined by imposing that the 1-form $\om$ in \eqref{eqKundt} vanishes. 
We dubbed this subclass {\it Dodgson waves} in order to emphasise its connection with Carrollian geometry, as discussed in Section \ref{secCarrembed}.  
\bdefi{Dodgson wave}{\label{defiDodgsonwave}A Dodgson wave $(\hat\M,\hat\xi,\hat g)$ is a gravitational wave $(\hat\M,[\hat\xi],\hat g)$ such that there exists a distinguished representative $\hat\xi\in[\hat\xi]$ satisfying the condition:
\bi
\item For all $V\in\Milnegh$, $\hat\nabla_V\hat\xi=0$. \itemnum \label{Carrcond}
\ei
where $\Ker\hat\psi$ is the canonical distribution induced by $(\hat\M,[\hat\xi],\hat g)$. 

Equivalently, there exists a 1-form $\eta\in\Gamma\big({\Ann\hat\xi}\big)$, called the \textbf{Dodgson 1-form}, such that 
\bea
\hat\nabla\hat\psi=\hat\psi\otimes \eta.\label{defCareq}
\eea

}
\bremi{\label{remDodgsonKundt}
\item Dodgson waves are Kundt waves, as follows straightforwardly from comparing the defining condition \eqref{defCareq} with the fourth item of Lemma \ref{lemgravKundt}. 
\item Despite their seemingly symmetric definitions as subclasses of Kundt waves, Walker and Dodgson waves differ from the fact that \big(contrarily to \eqref{recrel}\big) the defining condition \eqref{defCareq} is {\it not} invariant under a rescaling\footnote{More precisely, a rescaling of $\hat\xi$ cannot be compensated by a mere shift of $\eta$, as was the case in \eqref{Walkergroupaction}.} of the lightlike vector field $\hat\xi$. Consequently, the definition of a Dodgson wave involves a distinguished representative $\hat\xi\in[\hat\xi]$, in contradistinction with \Defi{defiWalker}.
\item Furthermore, Walker waves can be shown to belong to the Dodgson class by writing the recurrency condition \eqref{recrel} with respect to the special vector field $\hat{\bar\xi}\in[\hat\xi]$ for which the recurrence form $\om$ is collinear to the closed dual 1-form $\hat{\bar\psi}:=\hat g(\hat{\bar\xi})$ \Big(\ie $\om=\kappa\, \hat{\bar\psi}$ for some function $\kappa\in\foncmh$\Big). 
\item A non-trivial group action on Dodgson waves can be defined, the latter involving a rescaling of the metric $\hat g$. Explicitly, condition \eqref{defCareq} is invariant under the following action of the abelian multiplicative group of nowhere vanishing {invariant} functions $\foncmhinvn$:
\bea
\hat\xi\overset{\Lambda}{\mapsto}\Lambda\un\, \hat\xi\quad,\quad
\hat g\overset{\Lambda}{\mapsto}\Lambda^2\, \hat g\quad,\quad
\hat\psi\overset{\Lambda}{\mapsto}\Lambda\, \hat\psi\quad,\quad
\eta\overset{\Lambda}{\mapsto} \eta-\hat d\ln\Lambda\label{eqactioncarr}
\eea
where $\Lambda\in\foncmhinvn$. 
}
We already noticed that Walker waves are Dodgson. Conversely -- in the same spirit as Platonic waves were characterised as conformal ``Bargmann--Eisenhart'', \cf \Prop{proporbplatBE} -- Dodgson waves can be characterised as ``conformal Walker'' in the following sense\footnote{We are grateful to S.~Hervik for a useful comment regarding this point. }:
\bprop{\label{proporbCar}Any orbit of Dodgson waves contains a Walker wave. }
\bproof{Skewsymmetrising \eqref{defCareq} leads to $\hat d\hat\psi=-\eta\w\hat\psi$ so that $\hat\psi$ is Frobenius and locally we have $\hat d\hat\psi=\hat d\ln\Om\w\hat\psi$ for some scaling factor $\Om\in\foncmhn$. Consequently, the 1-form $\eta$ can be written as $\eta=-\hat d\ln\Om-\kappa\, \hat\psi$ for some function $\kappa\in\foncmh$. Furthermore, the condition $\eta(\hat\xi)=0$ ensures that $\Lag_{\hat\xi}\Om=0$ \ie $\Om\in\foncmhinvn$. Acting with \eqref{eqactioncarr} and $\Lambda:=\Om\un$ leads to $\eta\overset{\Lambda}{\mapsto}-\kappa\, \hat\psi$ so that the resulting wave is Walker. 
}
The following proposition further relates Dodgson and Platonic waves:
\bprop{\label{propinplacar}Let $(\hat\M,\hat \xi,\hat g)$ be a Platonic wave with scaling factor $\Om$.
\be
\item The triplet $(\hat\M,\hat \xi,\hat g)$ is a Dodgson wave if and only if it is a Bargmann--Eisenhart wave. 
\item The triplet $(\hat\M,\hat{\xi}{'}{},\hat g)$ -- with $\hat{\xi}{'}{}=\Om^{-\half}\, \hat\xi$ -- is a Dodgson wave with Dodgson 1-form $\eta=-\half \hat d\ln\Om$.
\ee
}
\bproofe{
\item The defining properties of a Platonic wave impose $\eta=-\om$ in \eqref{eqKundt} while the defining condition of a Dodgson wave reads $\om=0$, so that $\eta=\om=0$ and the wave is Bargmann--Eisenhart. 
\item The defining relations of a Platonic wave can be restated in terms of $\hat\psi=\hat g(\hat\xi)$ as:
\bea
\hat\nabla_{(\mu}\hat\psi_{\nu)}=0\quad,\quad
\hat\nabla_{[\mu}\hat\psi_{\nu]}=\p_{[\mu}\ln\Om\,\hat \psi_{\nu]}\label{defPlato}
\eea
so that $\hat\nabla_\mu\hat\psi_\nu=\p_{[\mu}\ln\Om\, \hat\psi_{\nu]}$.
Introducing the vector field $\hat{\xi}{'}{}=\Om^{-\half}\hat\xi$ and the dual 1-form $\hat{\psi}{'}{}=\hat g(\hat{\xi}{'}{})$, the latter satisfies $\hat\nabla_\mu\hat{\psi}_\nu'=-\half\, \hat{\psi}_\mu'\, \p_\nu\ln\Om$ and thus $(\hat\M,\hat{\xi}{'}{},\hat g)$ is a Dodgson wave with Dodgson 1-form $\eta=-\half \hat d\ln\Om$. \footnote{Note that the invariance of the scaling factor of a Platonic wave (\ie $\Lag_{\hat\xi} \Om=0$) ensures that $\eta\in\Gamma\big({\Ann\hat\xi}\big)$. }
}
In other words, any Dodgson wave with exact Dodgson 1-form can be recast as a Platonic wave upon rescaling of the ambient vector field. 
\pagebreak
\bremi{
\item Using the Carroll parameterisation \eqref{hatgCarr2}, a Kundt wave is a Dodgson wave if and only if the following conditions hold:
\bea
\Lag_\xi\Om=0\quad,\quad\Lag_{\xi}A_\mu+\p_\mu\ln\Om-\Om\un\xi^\rho\, \p_t\gamma_{\rho\mu}=0.\label{eqcondCarr}
\eea
\item For further use, we display the Christoffel coefficients of a generic Dodgson wave -- using the Carrollian parameterisation \eqref{hatgCarr2} -- obtained by combining \eqref{eqChrisCarr2} with \eqref{condgammainv} and \eqref{eqcondCarr}:
\bea
\hat\Gamma^{t}_{tt}&=&\half\Om \Lag_\xi\lambda+\xi^\rho\p_t A_\rho+\p_t\ln\Om\nn\\
\hat\Gamma^{t}_{t\nu}&=&\p_\nu\ln\Om\nn\\
\hat\Gamma^{\lambda}_{tt}&=&-\half\, \Om\xi^\lambda\p_t\lambda+\Om g^{\lambda\rho}\p_tA_\rho+\half \Om^2g^{\lambda\rho}\p_\rho\lambda+g^{\lambda\rho}\lambda\Om\p_\rho\Om\nn\\
\hat\Gamma^{\lambda}_{t\nu}&=&-\half\Om\xi^\lambda\p_\nu\lambda-\half\lambda\xi^\lambda\p_\nu\Om+\half g^{\lambda\rho}\big( \p_t\gamma_{\rho\nu}+2\Om\p_{[\nu}A_{\rho]}-\p_\rho\Om A_\nu\big)\nn\\
\hat\Gamma^{\lambda}_{\mu\nu}&=&\xi^\lambda\p_{(\mu}A_{\nu)}+\half \ovA{h}{}^{\lambda\rho}\big( \p_\mu\gamma_{\rho\nu}+\p_\nu\gamma_{\rho\mu}-\p_\rho\gamma_{\mu\nu}\big)-\xi^\lambda  A_{(\mu}\Lag_{\xi} A_{\nu)}+ \xi^\lambda\overset{ A}{\Sigma}_{\mu\nu}\nn
\eea
where $\ovA{h}{}^{\mu\nu}:=g^{\mu\nu}-\lambda\, \xi^\mu\xi^\nu$ satisfies \eqref{eqdefhA} and $\overset{ A}{\Sigma}_{\mu\nu}:=-\half \Om\un\pl \p_t\gamma_{\mu\nu}-2\,  A_{(\mu}\p_t\gamma_{\nu)\rho}\, \xi^\rho\pr$ satisfies \eqref{eqdefSigma}. 
}

As suggested by the formal analogy between the coefficients $\hat\Gamma^{\lambda}_{\mu\nu}$ and the coefficients \eqref{Carrconn} of a generic Carrollian connection, Dodgson waves play an important r\^ole in the context of the embedding of Carrollian manifolds inside gravitational waves, as will be elaborated further on in Section \ref{secCarrembed}. 
~\\

We conclude this discussion on Dodgson waves by displaying three distinguished examples:

\bexa{Dodgson waves}{\label{Carrwaves}We introduce three Dodgson waves $(\hat\M,\hat\xi,\hat g)$ where:
\bi
\item {$\hat\M$} is a $d+2$-dimensional spacetime coordinatised by $(u,t,x^i)$ where $i\in\pset{1,\ldots,d}$.
\item $\hat\xi=\p_u$ is a lightlike vector field. 
\item The metric is defined by the following line elements:
\bi
\item \textbf{Minkowski wave:} The metric is the Minkowski metric written in lightlike coordinates:
\bea
d\hat s^2=2\, dt\, du+\delta_{ij} dx^idx^j\label{Minkspacetime}
\eea
where $\delta$ is the $d$-dimensional Euclidean metric.
\item \textbf{Newton-Hooke wave:} The metric is the Hpp-wave metric \cite{Gibbons:2003rv}:
\bea
d\hat s^2=\frac{|x|^2}{R^2}dt^2+2\, dt\, du+\delta_{ij} dx^idx^j\label{NHspacetime}
\eea
where $R$ stands for  the {Newton-Hooke radius}. 

\item \textbf{(Anti) de Sitter wave:} The metric is the (Anti) de Sitter metric written in lightlike coordinates:
\bea
d\hat s^2=-\frac{u^2}{R^2}dt^2+2\cosh\frac{|x|}{R}dt\big(du-\frac{u}{R}\frac{x_i}{|x|}\tanh\frac{|x|}{R} dx^i\big)+\gamma_{ij} dx^idx^j\label{AdSspacetime2}
\eea
where $\gamma$ is the $d$-dimensional (hyperbolic) spherical metric whenever ($R^2>0$) $R^2<0$ defined in \eqref{hypmet}.
\ei
\ei
}
\pagebreak
\bremi{\label{remdodgsonwaves}
\item 
Both the Minkowski and Newton-Hooke waves are:
\bi
\item  Bargmann--Eisenhart waves. 
\item maximally symmetric as  Bargmannian manifolds \ie their isometry algebras 
\bea
\alg:=\pset{X\in\vfh\ |\ \Lag_X\hat\xi=0 \text{ and } \Lag_X \hat g=0}\label{isobarg}
\eea
are of maximal dimension $\frac{(d+2)(d+1)}{2}+1$. 
\ei
\item Explicitly, the isometry algebra \eqref{isobarg} of the Minkowski wave (resp. Newton-Hooke wave) is isomorphic to the Bargmann algebra $\barg(d+1,1)$ \big(resp. Bargmann-Hooke algebra $\barg_\pm(d+1,1)$\big) \ie the central extension of the Galilei algebra $\gal(d,1)$ \big(resp. Newton-Hooke algebra $\mathfrak{nh}_\pm(d,1)$\big).
\item The Minkowski and Newton-Hooke waves belong to the same orbit of Bargmann--Eisenhart waves via the group action \eqref{eqorbBE} with ${\hat\phi}=\frac{|x|^2}{R^2}$. 
\item  Contrarily to the Minkowski and Newton-Hooke waves, the (Anti) de Sitter wave is a genuine Dodgson wave\footnote{\ie it is neither Walker, Platonic nor Bargmann--Eisenhart. Note however that the Anti de Sitter spacetime (but not the de Sitter spacetime) can define a Platonic wave \cite{Bekaert:2013fta}. Explicitly, using the Poincar\'e coordinates, the line element $\hat{ds}{}^2=\frac{1}{z^2}\big(2\, du\, dt+dz^2+\delta_{ab}\, dx^adx^b\big)$, with $a,b\in\pset{1,\ldots,d-1}$, admits the Killing hypersurface-orthogonal vector field $\hat\xi:=\p_u$. }. Denoting $\Om=\cosh\frac{|x|}{R}$ the scaling factor associated with the (Anti) de Sitter wave, the 1-form $\hat\psi=\Om\, dt$ dual to $\hat\xi$ can be checked to satisfy relation \eqref{defCareq} with Dodgson 1-form $\eta=-\hat d\ln\Om-\frac{u}{R^2\Om^2} \hat\psi$.
\item The line element \eqref{AdSspacetime2} can be put in the form \eqref{hatgCarr2} upon the identification:
\bea
\begin{split}
\lambda=\frac{u^2}{R^2}\cosh^{-2}\frac{|x|}{R}\quad,\quad A=du-\frac{u}{R} \frac{x_i}{|x|}\tanh\frac{|x|}{R}dx^i\quad,\quad \gamma=\gamma_{ij}\, dx^i\vee dx^j\\
\xi=\p_u\quad,\quad g\un=\frac{u^2}{R^2}\, \p_u\vee \p_u+2\, \frac{u}{R} \frac{x^i}{|x|}\tanh\frac{|x|}{R}\, \p_u\vee \p_i+\gamma^{ij}\, \p_i\vee \p_j
\end{split}
\eea
which can be checked to satisfy \eqref{algrelcar}, \eqref{condgammainv} and \eqref{eqcondCarr}.
}
\pagebreak
\subsection{Summary}
\begin{figure}[ht]
\centering
   \includegraphics[width=0.7\textwidth]{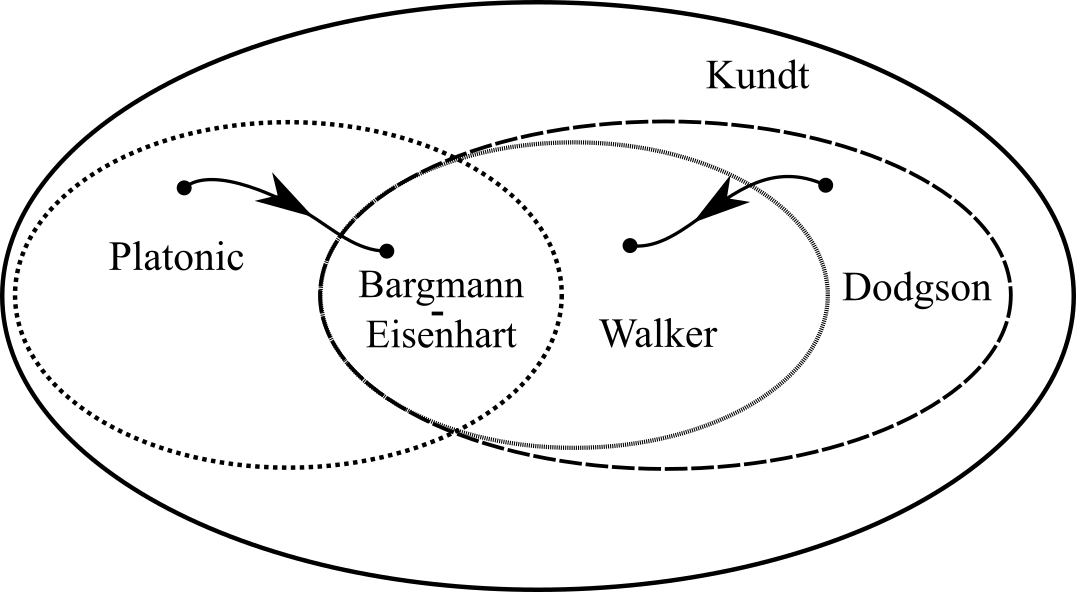}   
   
  \caption{Hierarchy of gravitational waves\label{diagboxes}}
\end{figure}

The various above definitions are summed up in Table \ref{diagboxes2}.
\begin{table}
\begin{center}
\begin{tabular}{|c|c|c|}
\hline
\rule[-0.2cm]{0cm}{0.6cm}\textbf{Bargmannian structures}&\textbf{Definition}&\textbf{Frobenius corollary}\\\hline\hline
\rule[-0.2cm]{0cm}{0.6cm}Gravitational waves&$\hat d\hat\psi\w\hat\psi=0$&$\hat d\hat\psi=\hat d\ln\Om\w\hat\psi$\\\hline
\rule[-0.2cm]{0cm}{0.6cm}Kundt waves&$\hat\nabla\hat\psi=\hat\psi\otimes \eta+\om\otimes\hat\psi$&$\om-\eta=\hat d\ln\Om+\kappa\, \hat\psi$\\\hline
\rule[-0.2cm]{0cm}{0.6cm}Dodgson waves&$\om=0$&$\eta=-\hat d\ln\Om-\kappa\, \hat\psi$\\\hline
\rule[-0.2cm]{0cm}{0.6cm}Walker waves&$\eta=0$&$\om=\hat d\ln\Om+\kappa\, \hat\psi$\\\hline
\rule[-0.2cm]{0cm}{0.6cm}Platonic waves&$\eta=-\om$&$\eta=-\om=-\half \hat d\ln\Om$\\\hline
\rule[-0.2cm]{0cm}{0.6cm}Bargmann--Eisenhart waves&$\eta=\om=0$&\\\hline
\end{tabular}
 \caption{Defining properties of Bargmannian structures\label{diagboxes2}}
\end{center}
\end{table}
The corresponding hierarchy of inclusions\footnote{Note that a non-trivial inclusion of Platonic waves within the class of Dodgson waves exists, modulo a rescaling of the vector field $\hat\xi$, \cf \Prop{propinplacar}. } 
is summarized in Figure \ref{diagboxes}, where the arrows refer to the group actions \eqref{eqorbPlato} and \eqref{eqactioncarr}.
 
\subsection{Brinkmann coordinates}
\label{appBrink}
We conclude this overview of gravitational waves by displaying local expressions using the so-called Brinkmann coordinates \cite{Brinkmann}, \cf also \cite{Blau2004}. 
\bi
\item \textbf{Gravitational wave:}
The most general form of a $(d+2)$-dimensional gravitational wave $(\hat\M,[\hat\xi],\hat g)$ reads locally:
\bi
\item $[\hat\xi]=\Omega(u,t,x^i)\, \p_u$\itemnum\label{Brinkeqxi}
\item $\hat{ds}{}^2=\Lambda(t,x^i)^2\crl2\, dt\big(du+\half\, \phi(u,t,x^i)\, dt+A_i(u,t,x^i)\, dx^i\big)+\gamma_{ij}(u,t,x^i)\, dx^i\, dx^j\crr$\itemnum\label{Brinkeq}
\ei
where $i\in\pset{1,\ldots,d}$ and $\gamma$ is a $d$-dimensional Riemannian metric. 

Different representatives in $[\hat\xi]$ differ by the choice of the nowhere vanishing function $\Om\in\foncmhn$. 

The associated equivalence class of dual 1-forms reads $[\hat\psi]=\Lambda(t,x^i)^2\, \Omega(u,t,x^i)dt$. 

Note that introducing Brinkmann coordinates singles out two distinguished representatives:
\bi
\item $\hat\xi:=\p_u$ whose associated dual 1-form $\hat\psi:=\hat g(\hat\xi)$ satisfies $\hat d\hat\psi=\hat d\ln\Lambda^2\w\hat\psi$. 
 \item $\hat{\bar{\xi}}:=\Lambda^{-2}\p_u$ whose associated dual 1-form $\hat{\bar{\psi}}:=\hat g(\hat{\bar{\xi}})$ is closed. 
 
 The vector field $\hat{\bar{\xi}}$ is thus the special vector field in $[\hat\xi]$.
 \ei
  \item \textbf{Kundt wave:} The Kundt condition imposes $\p_u\gamma_{ij}=0$ in \eqref{Brinkeq} so that $\gamma_{ij}(u,t,x^i)=\bar\gamma_{ij}(t,x^i)$. 
   \bi
\item $[\hat\xi]=\Om(u,t,x^i)\, \p_u$\itemnum\label{Kundteqxi}
\item $\hat{ds}{}^2=\Lambda(t,x^i)^2\crl2\, dt\big(du+\half\, \phi(u,t,x^i)\, dt+A_i(u,t,x^i)\, dx^i\big)+\bar\gamma_{ij}(t,x^i)\, dx^i\, dx^j\crr$.\itemnum\label{Kundteq}
\ei
Relation \eqref{eqKundt} can be checked to be satisfied for:
 \bi
  \item $\eta=\Lambda^{-2}\Om\un\kappa\, \hat\psi+\p_u\alpha-\hat d\ln\Lambda$
 \item $\om=-\Lambda^{-2}\Om\un\kappa\, \hat\psi+\p_u\alpha+\hat d\ln\Lambda+\hat d\ln\Om$
 \ei
 with $\alpha:=\frac{1}{4} \phi\, dt+\half A_i dx^i$ and $\kappa(u,t,x^i)$ an arbitrary function. 
 
 The 1-forms $\eta$ and $\om$ can be checked to scale according to \eqref{actionKundt}.

Rescaling $u\mapsto \tilde u=\Lambda^2 u$ allows to put the line element \eqref{Kundteq} in the usual Kundt form \cite{Podolsky:2008ec}:
\bea
\hat{ds}{}^2=2\, dt\big(d\tilde u+\half\, \tilde\phi(\tilde u,t,x^i)\, dt+\tilde A_i(\tilde u,t,x^i)\, dx^i\big)+\bar\gamma_{ij}(t,x^i)\, dx^i\, dx^j.
\eea 
 \item \textbf{Bargmann--Eisenhart wave:} Bargmann--Eisenhart waves can be locally defined as \cite{Brinkmann,Duval:1984cj,Duval:1990hj,Blau2004}:
   \bi
\item $\hat\xi=\p_u$\itemnum\label{BEeqxi}
\item $\hat{ds}{}^2=2\, dt\big(du+\half\, \bar\phi(t,x^i)\, dt+\bar A_i(t,x^i)\, dx^i\big)+\bar\gamma_{ij}(t,x^i)\, dx^i\, dx^j$.\itemnum\label{BEeq}
\ei
The ambient vector field $\hat\xi$ is readily seen to be Killing and the associated dual 1-form $\hat\psi=dt$ to be closed. 
Acting with the group action \eqref{eqorbBE} amounts to a shift of the potential $\bar\phi$.
 \item \textbf{Platonic wave:} The most general form of a Platonic wave is given by \cite{Lichnerowicz1955,Bekaert:2013fta}:
  \bi
\item $\hat\xi=\p_u$\itemnum\label{Platoneqxi}
\item $\hat{ds}{}^2=\Lambda(t,x^i)^2\crl2\, dt\big(du+\half\, \bar\phi(t,x^i)\, dt+\bar A_i(t,x^i)\, dx^i\big)+\bar\gamma_{ij}(t,x^i)\, dx^i\, dx^j\crr$.\itemnum\label{Platoneq}
\ei
The ambient vector field $\hat\xi$ is readily seen to be Killing and the associated dual 1-form $\hat\psi=\Lambda^2dt$ to be Frobenius. 
Acting with the group action  \eqref{eqorbPlato} is equivalent to a conformal rescaling of the invariant function $\Lambda$. 
Rescaling $\Lambda$ to 1 amounts to move along the Platonic orbit up to the Bargmann--Eisenhart  metric \eqref{BEeq}, \cf \Prop{proporbplatBE}. 

Note that upon rescaling the ambient vector field as $\hat{\xi}=\Lambda\un\p_u$, the Platonic wave \eqref{Platoneqxi}-\eqref{Platoneq} defines a Dodgson wave with Dodgson 1-form $\eta=-\hat d\ln\Lambda$, \cf \Prop{propinplacar} and \eqref{Dodgeqxi}-\eqref{Dodgeq} below.
  \item \textbf{Walker wave:} The Walker wave metric takes the form \cite{Gibbons:2007zu}:
 \bi
\item $[\hat\xi]=\Om(u,t,x^i)\, \p_u$\itemnum\label{Walkeqxi}
\item $\hat{ds}{}^2=2\, dt\big(du+\half\, \phi(u,t,x^i)\, dt+\bar A_i(t,x^i)\, dx^i\big)+\bar\gamma_{ij}(t,x^i)\, dx^i\, dx^j$.\itemnum\label{Walkeq}
\ei
The defining relation of a Walker wave \eqref{recrel} can be checked to hold with recurrent 1-form $\om=\hat d\ln\Om+\kappa\hat\psi$ where $\kappa=\half\, \Om\un \p_u\phi$. The recurrent 1-form can be checked to scale as \eqref{Walkergroupaction} under a change of representative $\hat\xi$.
\item \textbf{Dodgson wave:} We conclude by displaying the most general local form of a Dodgson wave:
 \bi
\item $\hat\xi=\Lambda(t,x^i)\un\, \p_u$\itemnum\label{Dodgeqxi}
\item $\hat{ds}{}^2=\Lambda(t,x^i)^2\crl2\, dt\big(du+\half\, \phi(u,t,x^i)\, dt+\bar A_i(t,x^i)\, dx^i\big)+\bar\gamma_{ij}(t,x^i)\, dx^i\, dx^j\crr$.\itemnum\label{Dodgeq}
\ei
Acting with the group action  \eqref{eqactioncarr} is equivalent to a rescaling of the invariant function $\Lambda$. 
Rescaling $\Lambda$ to 1 amounts to move along the Dodgson orbit up to the Walker metric \eqref{Walkeq}, \cf \Prop{proporbCar}. 
The defining relation of a Dodgson wave \eqref{defCareq} can be checked to hold with Dodgson 1-form $\eta=-\hat d\ln\Lambda+\kappa\hat\psi$ where $\kappa=\half\, \Lambda\un \p_u\phi$. The Dodgson 1-form $\eta$ scales according to \eqref{eqactioncarr}.
\ei

\pagebreak\section{Embedding intrinsic geometries \ala Duval \etal}
\label{secDuval}
In this section, we review the pioneering works of Duval and collaborators regarding the embedding of intrinsic non-Riemannian geometries inside Bargmann--Eisenhart waves\footnote{Or torsionfree Bargmannian manifolds, \cf Remarks \ref{remBM} and \ref{remBE2}.}, \cf \Defi{defBEW}. Explicitly, it has been shown in \cite{Duval:1984cj,Duval:1990hj} that the quotient manifold (or space of rays) of any Bargmann--Eisenhart wave was naturally a Newtonian manifold (\cf \Defi{defiNewtonman}) and furthermore that any Newtonian manifold could be obtained as a projection of a Bargmann--Eisenhart wave (\ie the ``projection map'' is surjective). This embedding approach can be understood as a geometric counterpart of the Eisenhart lift \cite{Eisenhart1928}, which relates nonrelativistic dynamical trajectories and relativistic geodesics. We first review how the metric structure of a Bargmann--Eisenhart wave induces a Galilean structure (\cf \Defi{defiGalstruc}) with closed clock $\psi$ on the quotient manifold and generalise this result to the larger class of Kundt waves, \cf \Defi{defiKundt}. Then we review the projection mechanism of the associated Levi--Civita connection onto a Newtonian connection compatible with the above Galilean structure. This projection of connection will be shown to generalise from Bargmann--Eisenhart to Platonic waves -- \cf \Defi{defiPlato} -- in Section \ref{secELl}.
\vspace{2mm}

More recently, the work \cite{Duval:2014uoa} showed that the lightlike hyperplanes (or wavefront worldvolumes) of any Bargmann--Eisenhart wave were naturally endowed with a structure of torsionfree Carrollian manifold, so that any Bargmann--Eisenhart wave admits a natural lightlike foliation by Carrollian manifolds. As a preliminary step, we review how the metric structure of a Bargmann--Eisenhart wave induces an invariant Carrollian structure on any wavefront worldvolumes and then generalise this result to Kundt waves. We then build on the result of \cite{Duval:2014uoa} and identify the necessary and sufficient condition a Carroll manifold must satisfy to be embedded using this scheme. 
In other words, we single out the relevant subset of torsionfree Carrollian manifolds -- called {\it invariant}\footnote{\Cf \Defi{deficarrinvar}.} -- for which exists an injective ``embedding map'' into Bargmann--Eisenhart waves. Such an embedding scheme will be generalised from Bargmann--Eisenhart to the larger class of Dodgson waves -- \cf \Defi{defiDodgsonwave} -- in Section \ref{secCarrembed}.
\subsection{Newtonian manifold as quotient}
\label{secnewtmanquot}
\subsubsection{Metric structure}
\label{MetstrucNewt}
This section addresses the procedure of embedding nonrelativistic structures inside (suitable classes of) gravitational waves, such that the former are obtained as projection or quotient\footnote{Or equivalently as (lightlike) Kaluza--Klein dimensional reduction.}.
The minimal structure needed to discuss projectability is called an \textbf{unparameterised ambient structure} and is defined as a pair $\big( \hat\M,[\hat\xi]\big)$ where:
\bi
\item $\hat\M$ is a manifold of dimension $d+2$.
\item $[\hat\xi]$ is a $\foncmhn$-orbit of nowhere vanishing vector fields $\hat\xi\in\vfh$ under the action \eqref{groupaction1}:
\bea
\hat\xi\overset{\Om}{\mapsto}\Om\, \hat\xi\quad,\quad \Om\in\foncmhn.\label{eqgroupaction}
\eea
\ei
Picking a representative $\hat\xi\in[\hat\xi]$, the pair $\big(\hat\M,\hat\xi\big)$ will be called an \textbf{ambient structure} \cite{Bekaert:2015xua}.
\vspace{2mm}

Given an unparameterised ambient structure $\big( \hat\M,[\hat\xi]\big)$, we will denote $\M:=\hat\M/\mR$ the d+1-dimensional \textbf{quotient manifold} of the ambient manifold $\hat\M$ by the $\mR$-action\footnote{Since any representative $\hat \xi$ is assumed to be nowhere vanishing, the induced flow action is free. We also assume that the latter is proper so that the quotient manifold theorem applies (\cf \eg Theorem 21.10 in \cite{Lee2003}) and $\M$ is hence a manifold.} induced by the flow of any representative $\hat\xi\in[\hat\xi]$. Equivalently, $\M$ can be seen as the space of integral curves of any representative $\hat\xi\in[\hat\xi]$. 
\bremi{
\item The construction is canonical \ie the quotient manifold $\M$ is independent of the choice of representative. 
\item The quotient manifold $\M$ is the base space of the following principal $\mR$-bundle:
  \bea
 \xymatrix{
\mR
 \ar@{^{(}->}[d]\\
\hat\M\ar@{->>}[d]^{\pi}\label{diagrammprincipalbundle}\\
\M
  }
  \eea
}

A Bargmannian structure on $\hat\M$ will be said \textbf{projectable} if it induces a well-defined Galilean structure on the quotient manifold $\M$.\footnote{\label{reminducedmet}Explicitly, given a projectable Bargmannian structure $\big(\hat\M,\hat\xi,\hat g\big)$, the 1-form $\hat\psi:=\hat g(\hat\xi)$ dual to $\hat\xi$ projects as an absolute clock $\psi$ on $\M$ whereas the contravariant metric $\hat g{}\un$ projects as a contravariant Galilean metric $h$ (or absolute rulers) compatible with $\psi$ on $\M$ (equivalently, the Leibnizian metric $\hat\gamma$ induced by $\hat g$ on $\Ker\hat\psi$ -- \cf Remark \ref{remBM} -- projects as the covariant Galilean metric $\gamma$ on $\M$ acting on $\Ker\psi$), \cf \cite{Bekaert:2015xua} for details. }
  The following lemma was shown in \cite{Bekaert:2015xua}:
\blem{Projection of Bargmannian structures}{\label{propinducedLeibniz}
A Bargmannian structure $\big(\hat\M,\hat\xi,\hat g\big)$ is projectable if and only if the two following conditions are met:
\bea
\hat\xi\in\Rad \Lag_{\hat\xi} \hat g\quad,\quad(\Lagxih\hat g)|_{\Ker\hat\psi}=0. \label{condprojBarg}
\eea
}
\bremi{
\item An obvious corollary is that the Killing condition $\Lagxih\hat g=0$ is a sufficient (but not necessary) condition for a Bargmannian structure to be projectable.
\item If one chooses to restrict to gravitational waves, the second condition of \eqref{condprojBarg} singles out the subclass of Kundt waves (\cf the second item of Lemma \ref{lemgravKundt}), as made precise by the following Proposition. 
}

\bprop{\label{propKundtproj}Let $\big(\hat\M,[\hat\xi],\hat g\big)$ be a Kundt wave with associated special vector field $\hat{\bar\xi}\in[\hat\xi]$. 

The quotient manifold of the Bargmannian structure $\big(\hat\M,\hat{\bar\xi},\hat g\big)$ is endowed with a Galilean structure $\big(\M,\psi,\gamma\big)$ such that the 1-form $\psi$ is closed.}
\bremi{\item Recall from \Prop{propgw} that the special vector field of a gravitational wave is characterised by the condition $\hat d \hat{\bar\psi}=0$ which ensures $\Lag_{\hat{\bar\xi}}\hat{\bar\psi}=0$, hence the first condition of  \eqref{condprojBarg} is satisfied. 
\item Using the Galilean parameterisation \eqref{hatgNC}, \Prop{propKundtproj} follows readily from \eqref{condspec} and \eqref{condhinv}. }

Since Bargmann--Eisenhart waves are Kundt, the following fact -- originally proved in \cite{Duval:1984cj} -- stems straightforwardly from \Prop{propKundtproj} (\cf also Remark \ref{remBE2}):
\bcor{}{\label{corLeibBarg}The quotient manifold of a  Bargmann--Eisenhart wave $\big(\hat\M,\hat\xi,\hat g\big)$ is endowed with a Galilean structure $\big(\M,\psi,\gamma\big)$ with closed clock $\psi$.}
The condition $d\psi=0$ satisfied by the induced Galilean structure is not incidental but is in fact necessary in order to ensure the existence of compatible torsionfree connections (\cf \Defi{deftorgal}). In the next section, we review how the Levi--Civita connection of a  Bargmann--Eisenhart wave projects on the quotient manifold as a Newtonian connection compatible with the induced Galilean structure.

\subsubsection{Connection}\label{seccon1}
Before describing the procedure of projection of connections on the quotient manifold of an unparameterised ambient structure, we recall the notion of lift of a vector field (\cf \eg \cite{Bekaert:2015xua}): 
\bdefi{Lift of a vector field}{Let $\big( \hat\M,[\hat\xi]\big)$ be an unparameterised ambient structure and $X\in\vf$ a vector field on the associated quotient manifold $\M$. 

The vector field $\hat X\in\vfh$ on $\hat\M$ will be called a lift of $X$ if:
\be
\item $\hat X$ is projectable \ie $\Lag_{\hat \xi} \hat X\sim \hat\xi$.
\item $\hat X$ projects on $X$ \ie $\pi_*\hat X=X$.
\ee
where $\hat\xi\in[\hat\xi]$ is any representative and $\pi_*:\Gamma\big(T\hat\M\big)\to\Gamma\big(T\M\big)$ is the pushforward of the projection map \eqref{diagrammprincipalbundle}. 
}
\bremi{\label{reminvlift}
\item Note that the first condition is indeed independent of the choice of representative. 
\item If $\big( \hat\M,\hat\xi\big)$ is an ambient structure, one can define the notion of \textbf{invariant} lift by trading the first condition with the stronger one $\Lag_{\hat \xi} \hat X=0$.
\item In the absence of further structure, the notion of lift of a vector field is not canonical. Rather, the space of lifts of a given vector field $X$ possesses the structure of an affine space modelled on the vector space $\Span\hat\xi$. 
\item A map assigning to each vector field on $\M$ a canonical lift on $\hat\M$ is known as an \textbf{Ehresmann connection} and can be represented as a 1-form $\hat A\in\ffh$ satisfying $\hat A(\hat\xi)\neq0$, for all $\hat\xi\in[\hat\xi]$. The canonical lift associated with $X$ by $\hat A$ is then the unique lift $\hat X$ satisfying $\hat A(\hat X)=0$.
}
\bdefi{Projection of connection}{\label{defiprojcon} Letting $\big( \hat\M,[\hat\xi]\big)$ be an unparameterised ambient structure and $\hat\nabla$ be a Koszul connection on $\hat\M$, a projected Koszul connection $\nabla$ can be canonically defined on the quotient manifold $\M:=\hat\M/\mR$ by making the following diagram commute:
\bea
\begin{split}
\xymatrix{
\big(\hat X,\hat Y \big)\ar@{->>}[d]_{\pi_*}\ar[r]^{\hat\nabla}&\hat\nabla_{\hat X}\hat Y\ar@{->>}[d]^{\pi_*}\\
\pl X, Y\pr\ar[r]^{\nabla}&\nabla_{X} Y \label{diagdefibarnabla}
}
\end{split}
\eea
where: 
\be
\item $(X,Y)$ is a pair of vector fields on $\M$.
\item $(\hat X,\hat Y \big)$ is a pair of lifts on $\hat\M$ of $X$ and $Y$, respectively.
\ee
provided that the following conditions are satisfied:
\begin{enumerate}[a)]
\item The ambient vector field $\hat\nabla_{\hat X}\hat Y\in\vfh$ is projectable.
\item The vector field $\nabla_{ X} Y:=\pi_*\pl\hat\nabla_{\hat X}\hat Y\pr$ is independent of the choice of lifts $\hat X$ and $\hat Y$. 
\item The derivative operator $\nabla$ satisfies the axioms of a Koszul connection\footnote{\Cf \eg footnote 7 of \cite{Bekaert:2014bwa}. }.
\end{enumerate}
}
\bremi{
\item A Koszul connection $\hat\nabla$ on $\hat\M$ such that conditions a)-c) are satisfied is referred to as \textbf{projectable}. 
\item The projection scheme \eqref{diagdefibarnabla} is geometric and canonical \ie it does not rely on any coordinate system nor choice of section $\sigma:\M\to\hat\M$ of the principal bundle \eqref{diagrammprincipalbundle}.
}

In \cite{Bekaert:2015xua}, we defined the class of \textbf{invariant} Koszul connections on ambient structures as follows:
\bdefi{Invariant connection}{\label{propinvKoszul}Let us consider the following structures:
\bi
\item $\big( \hat\M,\hat\xi\big)$ is an ambient structure.
\item $\hat\nabla$ is a Koszul connection on $\hat \M$.
\item $\hat T\in\Gamma\big({\w^2T^*\hat\M\otimes T\hat\M}\big)$ is the associated torsion tensor.
\ei
The connection $\hat\nabla$ will be said invariant if the following conditions are satisfied:
\begin{enumerate}
\item $\hat\nabla \hat\xi=0$
\item $\hat T\big(\hat \xi,\cdot\big)=0$
\item $\Lagxih \hat \nabla=0$. 
\end{enumerate}
}
\bremi{
\item The notion of invariant connection only stands for ambient structures (\ie with a fixed vector field $\hat\xi$), since Conditions 1. and 3. are not invariant under the group action \eqref{eqgroupaction}. 
}
Conditions 1.-3. were shown in \cite{Bekaert:2015xua} to provide a set of sufficient conditions ensuring projectability:

\bprop{\label{propprojinv}Any invariant Koszul connection is projectable. 
}

Now, let us particularise the discussion to the case at hand, namely the projection of the Levi--Civita connection $\hat\nabla$ associated with a  Bargmann--Eisenhart wave $\big(\hat\M,\hat\xi,\hat g\big)$. 
\blem{}{The Levi--Civita connection associated with a  Bargmann--Eisenhart wave is invariant. }
\bremi{
\item The proof is straightforward from \Defi{defBEW} and the fact that a Killing vector field for a Lorentzian metric is necessarily affine Killing for the associated Levi--Civita connection. 
}
By \Prop{propprojinv}, this result ensures that the Levi--Civita connection of a Bargmann--Eisenhart wave induces a well-defined connection $\nabla$ on the quotient manifold $\M$. 

Furthermore, the following Lemma can be shown:
\blem{}{\label{LemNewt}Let $\big(\hat\M,\hat\xi,\hat g\big)$ be a Bargmann--Eisenhart wave. 

The connection $\nabla$ induced on the quotient manifold $\M$ satisfies the following properties:
\be
\item $\nabla$ is torsionfree.
\item $\nabla$ is compatible with the Galilean structure $\big(\M,\psi,\gamma\big)$ induced by $\big(\hat\M,\hat\xi,\hat g\big)$ {\rm(}\cf \Prop{corLeibBarg}{\rm)}.
\item $\nabla$ satisfies the Duval--K\"unzle condition \eqref{eqDK}. 
\ee
}
A detailed geometric proof\footnote{Heuristically, the first two properties follow from the torsionfreeness of the ambient Levi--Civita connection together with its compatibility with both the ambient metric and vector field. The Duval--K\"unzle condition in turn follows from the symmetry of the associated Riemann tensor under exchange of the first and last pair of indices.} of Lemma \ref{LemNewt} can be found in \cite{Morand:2016rrt}. The above statements can be summarised in the following Theorem:
\bthm{Duval \etal \cite{Duval:1984cj}}{\label{thmDuval}The quotient manifold of a Bargmann--Eisenhart wave is a Newtonian manifold. }

A quick proof can be articulated using the Galilean parameterisation \eqref{hatgNC} for which a Bargmann--Eisenhart wave is characterised by the conditions:
\bea
\p_u\psi_\mu=0\quad,\quad\p_ug_{\mu \nu}=0\quad,\quad d\psi=0
\eea
so that the associated Christoffel coefficients read:
\bea
\hat\Gamma^{u}_{\mu\nu}&=&-\half\Lag_Z g_{\mu\nu}+\p_{(\mu}\phi\, \psi_{\nu)}\label{coeffBEwc1}\\
\hat\Gamma^{\lambda}_{\mu\nu}&=&Z^\lambda\p_{(\mu}\psi_{\nu)}+\half h^{\lambda\rho}\big( \p_\mu g_{\rho\nu}+\p_\nu g_{\rho\mu}-\p_\rho g_{\mu\nu}\big).\label{coeffBEwc2}
\eea
\bremi{
\item The Christoffel coefficients \eqref{coeffBEwc1}-\eqref{coeffBEwc2} are independent of the coordinate $u$ and the components $\Gamma^{\lambda}_{uu}$, $\Gamma^\lambda_{\mu u}$ and $\Gamma^\lambda_{u \mu}$ vanish, thus ensuring that the conditions a)-b) of \Prop{defiprojcon} are satisfied\footnote{We should emphasise that relaxing the closedness condition $d\psi=0$ does not lead to a projectable connection as the components $\Gamma^\lambda_{\mu u}$ do not vanish, hence Condition b) of \Prop{defiprojcon} would not be satisfied in this case. }. 
\item The projected connection $\nabla$ admits \eqref{coeffBEwc2} as Christoffel-like coefficients. These coincide with the coefficients \eqref{NewtMilne} of the most general Newtonian connection compatible with the Galilean structure $\big(\M,\psi,\gamma\big)$, as written in the variables \eqref{Trumpervariables}. 
Substituting 
\bea
Z^\mu=N^\mu-h^{\mu\nu} A_{\nu}\quad,\quad \phi=2\, {A}_\mu N^\mu-h^{\mu \nu} A_\mu A_\nu\quad,\quad g_{\mu \nu}=\gamma_{\mu \nu}+\psi_\mu  A_{\nu}+  A_{\mu}\psi_\nu
\eea
one recovers the familiar expression \eqref{Galconn} with $\N F_{\mu \nu}=2\, \p_{[\mu} A_{\nu]}$.
\item Projecting the group action \eqref{eqorbBE} on the set of Bargmann--Eisenhart waves to the quotient manifold amounts to perform a shift \eqref{eqshifttrumpvar}-\eqref{eqgroupactionNewtpotshift} of the Newtonian potential parameterised by $\bar\phi\in\fonc{\M}$, defined as $\pi^*\bar\phi=\hat\phi$. In particular, this projection of group action  allows to relate the Minkowski and Newton-Hooke waves of Example \ref{Carrwaves} to the maximally symmetric flat Galilei and Newton-Hooke manifolds of Example \ref{exaGal} as:
\bea
\begin{split}
\xymatrix{
\underset{\text{(Bargmann--Eisenhart)}}{\text{Minkowski wave}}\ar@{->>}[dd]^{\pi}_{\text{Theorem }\ref{thmDuval}}\ar[rr]^{\hat\phi}_{\eqref{eqorbBE}}&&\ar@{->>}[dd]_{\pi}^{\text{Theorem }\ref{thmDuval}}\underset{\text{(Bargmann--Eisenhart)}}{\text{Newton-Hooke wave}}\\\\
\underset{\text{(Newtonian)}}{\text{Flat Galilei}}\ar[rr]^{\bar\phi}_{\eqref{eqgroupactionNewtpotshift}}&&\underset{\text{(Newtonian)}}{\text{Newton-Hooke}}
}
\end{split}
\eea
where ${\hat\phi}=\frac{|x|^2}{R^2}$ and $\bar\phi=\pi^*{\hat\phi}=\frac{|x|^2}{R^2}$. 
\item The projection result of Theorem \ref{thmDuval} will be generalised to the larger class of Platonic waves in Theorem \ref{thmPlato}. 
}

The important Theorem \ref{thmDuval} of Duval \etal provides the natural framework necessary to a geometric understanding of the Eisenhart lift \cite{Eisenhart1928}. The latter relates the relativistic geodesics of a Bargmann--Eisenhart wave of dimension $d+2$ to nonrelativistic dynamical trajectories in $d+1$ spacetime dimension. A modern statement of the Eisenhart Theorem can be articulated as follows:
\bthm{Eisenhart lift \cite{Eisenhart1928,Duval:1984cj}}{\label{thmEisenhart}Let $\big(\hat\M,\hat\xi,\hat g\big)$ be a Bargmann--Eisenhart wave and denote $\hat\psi$ the closed 1-form $\hat\psi:=\hat g(\hat \xi)$. Let furthermore $\big(\M,\psi,\gamma,\nabla\big)$ be the induced Newtonian manifold on the quotient manifold. 

Solutions to the parameterised geodesic equation associated with the Bargmann--Eisenhart metric $\hat g$ are classified according to the two constant of motions:
\be
\item $M^2:=-\hat g(\dot x,\dot x)$
\item $m:=\hat \psi(\dot x)$
\ee
as follows:
\bi
\item $\mathbf{m=0}:$ These are the trajectories constrained on one hypersurface of the foliation induced by $\hat\psi$. 

The former are 
classified as\footnote{The condition $m=0$ ensures that $M^2=-\hat\gamma(\dot x,\dot x)\leqslant0$ with $\hat \gamma$ the Leibnizian metric induced by $\hat g$, \cf Remark \ref{remBM}. }:
\bi
\item $\mathbf{M^2=0:}$ These are the integral curves of the lightlike vector field $\hat\xi$ \ie $\dot x\sim\hat \xi$.
\item $\mathbf{M^2<0:}$ The projection of such trajectories on the quotient manifold are spacelike trajectories 

(\ie confined on one $d$-dimensional spacelike hypersurface or \textbf{absolute space}) satisfying the parameterised geodesic equation associated with the $d$-dimensional Riemannian metric $\gamma$. 
\ei
\item $\mathbf{m\neq0:}$ The projection of such trajectories\footnote{Irrespectively of the sign of $M^2$ \ie whether the relativistic trajectory is timelike, spacelike or lightlike.} on the quotient manifold are nonrelativistic timelike trajectories satisfying the parameterised geodesic equation associated with the Newtonian connection $\nabla$. 
\ei
}
\bproof{Denoting $\mathscr F^{\hat\lambda}:=\ddot x^{\hat\lambda}+\hat\Gamma^{\hat \lambda}_{\hat\mu\hat\nu}\, \dot x^{\hat\mu}\dot x^{\hat\nu}$ the parameterised geodesic equation -- with $\hat\Gamma^{\hat \lambda}_{\hat\mu\hat\nu}$ the Christoffel coefficients of the Levi--Civita connection -- associated with $\hat g$ allows to compute the following quantities:
\bi
\item $\hat g(\mathscr F,\dot x)=-\half \big(\frac{dM^2}{d\tau}+\hat\nabla_{\hat\lambda}\hat g_{\hat\mu\hat\nu}\, \dot x^{\hat\lambda}\dot x^{\hat\mu}\dot x^{\hat\nu}\big)$\itemnum\label{CM1}
\item $\hat\psi(\mathscr F)=\frac{dm}{d\tau}-\hat\nabla_{(\hat\mu}\hat\psi_{\hat\nu)}\, \dot x^{\hat\mu}\dot x^{\hat\nu}$.\itemnum\label{CM2}
\ei
Therefore, the metric compatibility and Killing condition\footnote{\label{footnoteKilBE}The fact that the proof requires the ambient vector field to be Killing allows to generalise the Eisenhart Theorem to Platonic waves -- \cf Theorem \ref{Lichnetheorem} -- but prevents further extension to larger classes of gravitational waves. } for $\hat\xi$ respectively ensure that $M^2$ and $m$ are constants of motion.

Let us analyse separately the following cases:
\bi
\item $\mathbf{m=0}$, $\mathbf{M^2=0:}$ Assuming $\hat\psi(\dot x)=0$ ensures that $M^2=-\hat\gamma(\dot x,\dot x)$ where $\hat\gamma$ is the Leibnizian metric induced by $\hat g$ on $\Ker\hat\psi$ (\cf Remark \ref{remBM}) satisfying $\Rad\hat\gamma=\Span\hat\xi$, so that $\dot x\sim \hat\xi$.
\item $\mathbf{m\neq0}$, $\mathbf{M^2<0:}$ Using the Galilean parameterisation \eqref{hatgNC}, the parameterised geodesic equation reads:
\be
\item $\mathscr F^u=\ddot u+\hat\Gamma^u_{\mu \nu}\dot x^\mu \dot x^\nu=0$
\item $\mathscr F^\lambda=\ddot x^\lambda+\hat\Gamma^\lambda_{\mu \nu}\dot x^\mu \dot x^\nu=0$
\ee
where the Christoffel coefficients are given by \eqref{coeffBEwc1}-\eqref{coeffBEwc2}. The relativistic parameterised geodesic equation thus projects downto the quotient manifold coordinatised by $x^\mu$ as the parameterised geodesic equation associated with the Newtonian connection $\nabla$ with coefficients \eqref{NewtMilne}. The case $m\neq0$ thus corresponds to nonrelativistic timelike geodesics for the Newtonian connection $\nabla$.
\item $\mathbf{m=0}$, $\mathbf{M^2<0:}$ Restricting to spacelike trajectories (\ie characterised by $m=0$), the Newtonian connection $\nabla$ reduces to the Levi--Civita connection $\nabla_\gamma$ associated with the Riemannian metric $\gamma$, that is $\nabla|_{\Ker\psi}=\nabla_\gamma$, \cf Remark \ref{remspacelikeabsspace}.
\ei
}
\bremi{
\item The (unprojected) trajectories characterised by $m=0$ will be given the interpretation of geodesics for a suitable Carrollian connection in Theorem \ref{thmCarroll}. 
\item As an illustration of the Eisenhart Theorem, it can be checked that the projection of the parameterised geodesic equation associated with the Newton-Hooke wave \eqref{NHspacetime} of Example \ref{Carrwaves} gives the harmonic (expanding) oscillator equation, \ie the geodesic equation of the Newton-Hooke manifold of Example \ref{exaGal},  \cf Remark \ref{remGalNH}. 
}
\subsection{Carrollian manifold as lightlike hypersurface}
\label{secDuvalCarr}
This section can be considered as the dual counterpart of Section \ref{secnewtmanquot}. We first review the result of \cite{Duval:2014uoa} showing that the wavefront worldvolumes of a Bargmann--Eisenhart wave are naturally endowed with a structure of torsionfree Carrollian manifolds. We then precise the above result by identifying the class of torsionfree Carrollian manifolds that can be embedded using this scheme. 
\subsubsection{Metric structure}
We let $\big(\hat\M,[\hat\xi],\hat g\big)$ be a gravitational wave and denote $\Ker \hat\psi$ the associated involutive distribution. The associated integral submanifolds are referred to as wavefront worldvolumes and denoted $\M_t$, \cf Section \ref{secGW}. To each wavefront worldvolume $\M_t$ one attaches an embedding map:
\bea
i_t:\M_t\hookrightarrow\hat\M.
\eea
The latter enjoys the following isomorphisms:
\bi
\item $i_{t*}(T\M_t)\cong\Ker\hat\psi$\itemnum\label{iso1}
\item $\Ker i_t^*\cong \Span\hat\psi$.\itemnum \label{iso2}
\ei

\bprop{\label{propCarrst}Let $\big(\M,[\hat\xi],\hat g\big)$ be a gravitational wave. 
\be
\item Each leaf $\M_t$ of the canonical foliation induced by $\Ker\hat\psi$ is endowed with an equivalence class of Carrollian structures $( \M_t,[\xi],\gamma)$.
\item The induced equivalence class $( \M_t,[\xi],\gamma)$ is invariant (\ie $\Lag_{\xi}\gamma=0$ for all $\xi\in[\xi]$) if and only if $\big(\M,[\hat\xi],\hat g\big)$ is a Kundt wave. 
\ee
}
\pagebreak
\bproofe{
\item Since $\hat\xi\in[\hat\xi]$ implies that $\hat\xi\in\Milnegh$, the isomorphism \eqref{iso1} ensures that there is a unique (nowhere vanishing) vector field $\xi\in\field{T\M_t}$ satisfying $i_{t*}\xi=\hat\xi|_{i_t(\M_t)}$. Secondly, the Carrollian metric is obtained as $\gamma:=i_t^*\hat g$. The expression $\gamma(X,V)=i_t^*\hat g(X,V)=\hat g(i_{t*}X,i_{t*}V)\circ i_t$ vanishes for all $V\in\field{T\M_t}$ if and only if $X\sim\xi$, so that the metric $\gamma$ is of rank $d$ and its radical is spanned by $[\xi]$. 
\item This point is readily seen from the second point of Lemma \ref{lemgravKundt}, or alternatively -- using the Carroll parameterisation \eqref{hatgCarr2} -- from \eqref{condgammainv}.  
}
\bremi{
\item As recalled in \Prop{proptorfreeCarr}, a Carrollian structure admits compatible torsionfree connections if and only if it is invariant. It follows from the second item of \Prop{propCarrst} that being Kundt is a necessary condition for a gravitational wave to induce a torsionfree Carrollian manifold on its leaves. 
\item The next section will show that being Kundt is also the necessary condition for a gravitational wave to induce a well-defined torsionfree connection on its leaves.
}
The following result -- originally proved in \cite{Duval:2014uoa} -- is immediate from \Prop{propCarrst}:
\bcor{}{\label{corDuvalBE}Any wavefront worldvolume of a Bargmann--Eisenhart wave is naturally endowed with a canonical invariant Carrollian structure. }
\subsubsection{Connection}
\label{sectionConnectionDuval}
Having seen how the wavefront worldvolumes of (suitable classes of) gravitational waves are naturally endowed with Carrollian metric structures, we now move to the next logical step and investigate under which conditions the ambient Levi--Civita connection endows the leaves of the foliation with compatible Carrollian connections.
\bprop{\label{projconnCarr}Let $\hat\M$ be a manifold and $\hat\nabla$ be a Koszul connection on $\hat\M$. Let furthermore $\mathscr D\subset T\hat\M$ be an involutive distribution and denote $\M_t$ the associated integral submanifolds foliating $\hat\M$.

\vspace{2mm}

Denoting  $i_t:\M_t\hookrightarrow\hat\M$ the embedding map, the following isomorphism holds:
\bi
\item ${i_t}_*(T\M_t)\cong\mathscr D|_{i_t(\M_t)}$.
\ei
The connection $\hat\nabla$ admits a well-defined projection $\nabla$ on any leaf $\M_t$ if and only if the distribution $\mathscr D$ is autoparallel with respect to $\hat \nabla$.

The projected connection $\nabla$ is then defined by making the following diagram commute:
\bea
\begin{split}
\xymatrix{
(X,Y)\ar@{^{(}->}[r]^{{i_t}_*}\ar[d]_{\nabla}&(\hat X,\hat Y)\ar[d]^{\hat \nabla}\\
\nabla_{X}Y\ar@{^{(}->}[r]^{{i_t}_*}&\hat\nabla_{\hat X}\hat Y \label{diagdefibarnablacar}
}
\end{split}
\eea
where:
\bi
\item $X,Y\in\field{T\M_t}$ are two vector fields on the leaf $\M_t$.
\item $\hat X,\hat Y\in\field{\mathscr D}|_{{i_t}(\M_t)}$ are the two vector fields defined as $\hat X:={i_t}_*X$ and $\hat Y:={i_t}_*Y$, respectively. 
\item $\nabla_{X}Y\in\field{T\M}$ is the vector field defined as $\nabla_{X}Y:={i_t}_*\un(\hat\nabla_{\hat X}\hat Y)$. 
\ei
}

Note that the last step of the procedure supposes that $\hat\nabla_{\hat X}\hat Y\in\field{\mathscr D}$ for all $\hat X,\hat Y\in\field{\mathscr D}$, \ie that $\mathscr D$ is an autoparallel distribution for $\hat \nabla$, \cf footnote \ref{ftauto}. This entails the following Proposition:
\bprope{\label{propKundtstructures}\item The wavefront worldvolumes of a Kundt wave are endowed with:
\be
\item  a set of invariant Carrollian structures $( \M_t,[\xi],\gamma)$.
\item a canonical connection $\nabla$ being
\bi
\item torsionfree
\item compatible with the Carrollian metric $\gamma$.
\ei
\ee
\item The wavefront worldvolumes of a Kundt wave are totally geodesic lightlike hypersurfaces\footnote{A lightlike hypersurface $i:\Sigma\hookrightarrow\hat\M$ of a Lorentzian manifold $(\hat\M,\hat g)$ is said to be \textbf{totally geodesic} if, for any geodesic $x(\tau)$ of the induced connection $\nabla$, the curve $i\circ x(\tau)$ is a geodesic of the Levi--Civita connection $\hat\nabla$ associated with $\hat g$, \cf \eg \cite{Duggal1996}. }. 

}

\bproofe{
\item 
\be
\item \cf \Prop{propCarrst}.
\item As noted in Remark \ref{remKundtauto}, the Kundt condition is necessary and sufficient for the canonical distribution $\Ker\hat\psi$ of a gravitational wave to be autoparallel. 
\bi
\item The torsionfreeness of the induced connection follows directly from the one of the ambient Levi--Civita connection $\hat\nabla$.
\item The compatibility condition $\nabla\gamma=0$ is a consequence of the ambient metric compatibility condition $\hat\nabla\hat g=0$. 
\ei
\ee
\item The totally geodesic property for lightlike hypersurfaces is equivalent to the autoparallel condition, \cf \eg Theorem 2.2 in \cite{Duggal1996}.
}

It follows from Proposition \ref{propKundtstructures} that any wavefront worldvolume $\M_t$ of a Kundt wave $\big(\hat\M,[\hat\xi],\hat g\big)$ is naturally endowed with a set of invariant Carrollian structures $( \M_t,[\xi],\gamma)$ as well as with a metric compatible torsionfree connection $\nabla$. However, let us emphasise that the connection $\nabla$ does {\it not} generically preserve any representative of the equivalence class $[\xi]$ so that the wavefront worldvolumes of a Kundt wave are not torsionfree Carrollian manifolds in general. 
However, restricting from Kundt to Bargmann--Eisenhart waves, the following result holds:
\bthm{Duval \etal \cite{Duval:2014uoa}}{\label{thmDuvcarcon}Any wavefront worldvolume of a Bargmann--Eisenhart wave is canonically a torsionfree Carrollian manifold.  }
\bproof{Denoting $\big(\M_t,\xi,\gamma\big)$ the induced invariant Carrollian structure (\cf Corollary \ref{corDuvalBE}) and $\nabla$ the induced torsionfree connection (\cf \Prop{projconnCarr}), the following compatibility relations hold:
\be
\item $\nabla\xi=0$ 
\item $\nabla \gamma=0$
\ee
as a consequence of $\hat\nabla\hat\xi=0$ and $\hat \nabla \hat g=0$. 
}
Before concluding the present section, we aim at completing Theorem \ref{thmDuvcarcon} by identifying the space of torsionfree Carrollian manifolds that can be embedded inside a Bargmann--Eisenhart wave via the above procedure. In other words, thinking of Theorem \ref{thmDuvcarcon} as a Carrollian dual of Theorem \ref{thmDuval}, we wish to identify the Carrollian avatar to the notion of Newtonian manifold. 
The following proposition singles out the category of {\it invariant} Carrollian manifolds -- \cf \Defi{deficarrinvar} -- as the natural counterpart of Newtonian manifolds within the Carrollian world: 
\bprop{\label{propCarrBE}The torsionfree Carrollian manifolds induced on the wavefront worldvolumes of a Bargmann--Eisenhart wave are invariant. Conversely, any invariant torsionfree Carrollian manifold can be embedded inside a Bargmann--Eisenhart wave. 
}
\bproof{Letting $\big(\hat\M,\hat\xi,\hat g\big)$ be a Bargmann--Eisenhart wave, we denote $\big(\M_t,\xi,\gamma\big)$ the torsionfree Carrollian manifold induced on the wavefront worldvolume $\M_t$. 
The first part of the proof is readily seen from the fact that $\Lag_{\hat \xi}\hat\nabla=0$ (as a consequence of the Killing condition $\Lagxih \hat g=0$) which in turn ensures that $\Lag_\xi \nabla=0$ so that $\big(\M_t,\xi,\gamma\big)$ is invariant (recall that the condition $\Lagxi\gamma=0$ follows from $\Lagxi g=0$). A proof of the converse statement will be given as a corollary of the proof of \Prop{propCarrconnamb}, \cf Remark \ref{rempostembedCar}.}

Combining Theorem \ref{thmDuvcarcon} and \Prop{propCarrBE} leads to the following result:

\bthm{}{\label{ThmCarrollBE}Any Bargmann--Eisenhart wave admits a lightlike foliation by invariant torsionfree Carrollian manifolds.}
A particular example of Bargmann--Eisenhart wave is the Minkowski wave of Example \ref{Carrwaves}. According to Theorem \ref{ThmCarrollBE}, each leaf $t=const$ is naturally endowed with a torsionfree Carrollian manifold. It can be checked that the Carrollian manifold induced on any leaf of the foliation identifies with the flat Carroll manifold of Example \ref{exaAdSCarr}. The Minkowski wave thus provides a natural embedding for the flat Carroll manifold ( \cf Remark \ref{rempostembedCar} below for more details).

However, as noted earlier, the Carrollian connection associated with the (A)dS-Carroll manifold (\cf Example \ref{exaAdSCarr}) is not invariant, and thus cannot be embedded inside a Bargmann--Eisenhart wave, as a consequence of \Prop{propCarrBE}. In Section \ref{secCarrembed}, we propose a generalisation of the ambient framework of Duval \etal from Bargmann--Eisenhart to Dodgson waves allowing the embedding of a larger class of Carrollian manifolds (dubbed {\it pseudo-invariant}) containing in particular the (A)dS-Carroll manifold. 
~\\

We conclude the present section by displaying an ultrarelativistic avatar of the Eisenhart Theorem \ref{thmEisenhart}. 
 \bthm{Carroll train}{\label{thmCarroll}Let $\big(\hat\M,\hat\xi,\hat g\big)$ be a Bargmann--Eisenhart wave and denote $\hat\psi$ the Frobenius 1-form $\hat\psi:=\hat g(\hat \xi)$ and  $\hat\nabla$ the Levi--Civita connection associated with the Bargmann--Eisenhart metric $\hat g$.  
 
 Let furthermore $i_t:\M_t\hookrightarrow \hat\M$ be a leaf of the foliation induced by $\Ker \hat\psi$ (or wavefront worldvolume). 
\be
\item Let $p\in i_t(\M_t)$ and $V_p\in\Ker\hat\psi_p$ be a tangent vector. Any geodesic $x(\tau)$ of $\hat\nabla$ such that $\dot x(0)=V_p$ stays on the wavefront worldvolume $\M_t$. 
\item The curve $x(\tau)$ is a geodesic for the Carrollian connection $\nabla$ induced by $\hat \nabla$ on $\M_t$.
\ee
 }
 \pagebreak
\bproofe{
\item Recall from Theorem \ref{thmEisenhart} that the quantity $m=\hat\psi(\dot x)$ is a constant of motion for any geodesic $x(\tau)$ of $\hat\nabla$, as ensured by the Killing property enjoyed by the ambient vector field $\hat \xi$. This ensures that $\dot x(\tau)\in\Ker \hat\psi_{x(\tau)}$ for any $\tau$, so that $\dot x$ is always tangent to $\M_t$. The ambient geodesic $x(\tau)$ is therefore constrained to stay\footnote{More precisely, $x(\tau)\in i_t(\M_t)$ for all $\tau$.} on the submanifold $\M_t$.
\item This proposition follows straightforwardly from Theorem \ref{thmDuvcarcon} and \Prop{projconnCarr}.
}
 \bremi{
 \item Theorem \ref{thmCarroll} complements the Eisenhart Theorem \ref{thmEisenhart} by providing a geometric interpretation for the class of (unprojected) geodesics characterised by $m=0$.
 \item A generalisation of the Carroll train Theorem \ref{thmCarroll} to the larger class of Dodgson waves will be provided in Section \ref{secCarrembed}.
 }
\pagebreak\section{Geometrising the Eisenhart--Lichnerowicz lift}
\label{secELl}
As reviewed in Section \ref{secDuval}, the ambient approach introduced by Duval \etal allows the embedding of Newtonian manifolds (on the quotient space) and invariant Carrollian manifolds (on leaves of the foliation) inside a single class of gravitational waves (Bargmann--Eisenhart waves). 
Building on these seminal results, our aim in the next two sections will be to generalise the ambient approach of intrinsic geometries to larger classes of gravitational waves. 
In the present section, we deal with the Galilean case and generalise the embedding of Newtonian manifolds by enlarging the class of embedding waves from Bargmann--Eisenhart to Platonic waves\footnote{\Cf \Defi {defiPlato}.}. 
This generalisation will require to relax some of the assumptions of the projection scheme detailed\footnote{\Cf Definitions \ref{defiprojcon} and \ref{propinvKoszul}.} in Section \ref{seccon1} -- the latter being too restrictive to account for Platonic connections -- thus requiring to introduce a more general projection scheme. 
The latter will be shown to provide a geometric understanding of the generalisation by Lichnerowicz \cite{Lichnerowicz1955} of the Eisenhart lift  of dynamical trajectories to relativistic geodesics. 

\subsection{Metric structure}Using the general framework of projection of Bargmannian structures, as developed in Section \ref{MetstrucNewt}, the following proposition is immediate:
\bprop{\label{corLeibParg}The quotient manifold of a  Platonic wave $\big(\hat\M,\hat\xi,\hat g\big)$ is endowed with a Galilean structure $\big(\M,\psi,\gamma\big)$ such that the 1-form $\psi$ is Frobenius.}
\bremi{
\item Platonic waves are projectable, in the sense of Lemma \ref{propinducedLeibniz}, as ensured by the Killing condition $\Lagxih\hat g=0$. 
\item Recall that the Frobenius condition takes the form $d\psi\w \psi=0$.
\item Since the 1-form $\psi$ is not closed in general, the Galilean structure $\big(\M,\psi,\gamma\big)$ does not generically admit torsionfree compatible connections, \cf \Defi{deftorgal}.
\item As ensured by \Prop{proporbplatBE}, any orbit of Platonic waves under the group action \eqref{eqorbPlato} contains a Bargmann--Eisenhart wave. Furthermore, any orbit of Galilean structures with Frobenius clock under the group action \eqref{groupactiongalconf} contains a Galilean structure with closed clock, \cf Remark \ref{remgalfrob}. These two statements can be related as follows: letting $\big(\hat\M,\hat\xi,\hat g\big)$ be a Platonic wave and denoting $\big(\hat\M,\hat\xi,\hat{\bar g}\big)$ the associated Bargmann--Eisenhart wave related to $\big(\hat\M,\hat\xi,\hat g\big)$ via the nowhere vanishing {invariant} function $\Lambda\in\foncmhinvn$ -- \ie $\hat g=\Lambda^2\hat{\bar g}$ -- the following diagram commutes:
\bea
\begin{split}
\xymatrix{
\underset{\text{\rm (Platonic)}}{\big(\hat\M,\hat\xi,\hat g\big)}\ar[rr]^{\Lambda\un}_{\eqref{eqorbPlato}}\ar@{->>}[dd]^{\pi}_{\text{\rm Proposition}\, \ref{corLeibParg}}&&\underset{\text{\rm (Bargmann--Eisenhart)}}{\big(\hat\M,\hat\xi,\hat{\bar g}\big)}\ar@{->>}[dd]_{\pi}^{\text{\rm Corollary}\, \ref{corLeibBarg}}\\\\
\underset{\text{\rm (Frobenius)}}{\big(\M,\psi,\gamma\big)}\ar[rr]^{\bar\Lambda\un}_{\eqref{groupactiongalconf}}&&\underset{\text{\rm (Closed)}}{\big(\M,\bar\psi,\bar\gamma\big)} \label{diagplatoBEprojmet}
}
\end{split}
\eea
where $\bar\Lambda\in\fonc{\M}$ denotes the projection of the invariant function $\Lambda\in\foncmhinvn$ on the quotient manifold $\M$  \big(\ie $\pi^*\bar\Lambda=\Lambda$\big) and $(\bar\psi,\bar\gamma)$ is the special Galilean structure -- \ie with closed absolute clock -- associated with $(\psi,\gamma)$, \cf Remark \ref{remgalfrob}.
}
\subsection{Connection} In Section \ref{seccon1}, we introduced a projection scheme (\cf \Defi{defiprojcon}) that proved suitable for Bargmann--Eisenhart waves (and more generally for Bargmannian manifolds, as will be detailed in \cite{Morand2018b}). The present section is devoted to the generalisation of this projection scheme to larger classes of gravitational waves. In Section \ref{Kundt connection on absolute spaces}, we show that the very same projection procedure can be applied to the larger class of Kundt waves, provided one restricts to spacelike vector fields (\ie the projected connection is only defined on absolute spaces). 
In Section \ref{Platonic connection}, we introduce a generalised scheme that will prove suitable to account for the Platonic case, thus providing the geometric background underlying the Lichnerowicz lift.
\subsubsection{Kundt connection on absolute spaces}
\label{Kundt connection on absolute spaces}
Kundt waves have been introduced in Section \ref{Kundt waves} and shown to be projectable (at the level of metric structures) in Section \ref{secnewtmanquot}. We now complete these previous statements by displaying the suitable restriction under which the Kundt connection admits a well-defined projection on the quotient manifold.
\bprop{\label{propKundtabs}Let $(\hat\M,[\hat\xi],\hat g)$ be a Kundt wave and denote $\big([\psi],\gamma\big)$ the class of Frobenius Galilean structures induced on the quotient manifold $\M$. The Levi--Civita connection $\hat\nabla$ associated with $\hat g$ admits a well-defined projection on the absolute spaces foliating the quotient manifold $\M$. Furthermore, the projected connection on each absolute space identifies with the Levi--Civita connection $\nabla$ associated with the Riemannian metric $\gamma$. 
}
\bproof{Let $V,W\in\Milneg$ be a pair of spacelike\footnote{We will denote $\Ker\hat\psi$ the canonical involutive distribution induced by $(\hat\M,[\hat\xi],\hat g)$ -- \cf Remark \ref{remBM} -- and $\Ker \psi$ its projection on $\M$. } vector fields on $\M$ and $\hat V,\hat W\in\Gamma\big({\Ker\hat\psi}\big)$ be a pair of lifts of $(V,W)$ on $\hat\M$. The following facts follow\footnote{\Cf \cite{Morand:2016rrt} for details.} straightforwardly from Lemma \ref{lemgravKundt}:
\be
\item  The vector field $\hat\nabla_{\hat V}\hat W$ is spacelike, \ie $\hat\nabla_{\hat V}\hat W\in\Gamma\big({\Ker\hat\psi}\big)$, \cf the first point of Lemma \ref{lemgravKundt}.
\item The vector field $\hat\nabla_{\hat V}\hat W$ is projectable, as follows from the projectability of $\hat V,\hat W$ and the second point of Lemma \ref{lemgravKundt}.
\item The pushforward $\pi_*\pl\hat\nabla_{\hat V}\hat W\pr$ is a spacelike vector field independent of the choice of lifts $\hat V$ and $\hat W$. 
\ee
The three above statements ensure that the Kundt connection $\hat\nabla$ induces a canonical projection $\nabla|_{\Ker\psi}$ on the absolute spaces of $\M$. The connection $\nabla|_{\Ker\psi}$ is torsionfree -- as follows from the torsionfreeness of $\hat\nabla$  -- and compatible with the Riemannian metric $\gamma$, as ensured\footnote{Recall that the Riemannian metric $\gamma$ is obtained as projection of the Leibnizian metric $\hat\gamma:=g|_{\Ker\hat\psi}$ , \cf footnote \ref{reminducedmet}.} by the compatibility between $\hat\nabla$ and $\hat g$. Consequently, the induced connection $\nabla|_{\Ker\psi}$ identifies with the Levi--Civita connection associated with $\gamma$.
}

\subsubsection{Platonic connection}
\label{Platonic connection}
We now turn to the class of Platonic waves and discuss a projection scheme suitable for this particular class. In the rest of this section, we will let $\big(\hat\M,\hat\xi,\hat g\big)$ be a Platonic wave and denote $\hat\nabla$ the associated Levi--Civita connection. We start by noting that the naive application of the projection scheme described in \Defi{defiprojcon} to the Platonic case faces the two following obstructions: 
\begin{enumerate}[a)]
\item The vector field $\hat\nabla_{\hat X}\hat Y$ in Diagram \ref{diagdefibarnabla} is not necessarily projectable whenever $\hat X$ and $\hat Y$ are projectable. 
\item The vector field $\nabla_{ X} Y:=\pi_*\pl\hat\nabla_{\hat X}\hat Y\pr$ depends on the choice of lifts $\hat X$ and $\hat Y$, thus preventing $\nabla$ to be canonically defined. 
\ee
The first drawback can be circumvented by restricting to invariant\footnote{The notion of invariant lifts (\cf Remark \ref{reminvlift}) is well-defined here since the definition of a Platonic wave involves a distinguished vector field $\hat \xi$, so that $(\hat\M,\hat\xi)$ is a (parameterised) ambient structure.} lifts (then, the Killing condition ensures that $\hat\nabla_{\hat X}\hat Y$ is invariant, hence projectable). This restriction is however not sufficient to evade the second obstruction, which will require further restriction of the class of lifts. This is the leitmotiv of the alternative scheme we now introduce. 
\bdefi{Weighted projection}{\label{defiweightedproj}Let $\big(\hat\M,\hat\xi,\hat g\big)$ be a projectable Bargmannian structure and denote $(\M,\psi,\gamma)$ the induced Galilean structure on the quotient manifold $\M:=\hat\M/\mR$. Let $\hat\nabla$ be a Koszul connection on $\hat\M$. A projected Koszul connection $\overset{w}{\nabla}$ of weight $w$ can be defined by making the following diagram commute:
\bea
\begin{split}
\xymatrix{
\big(\hat X,\hat Y \big)_{w}\ar@{->>}[d]_{\pi_*}\ar[r]^{\hat\nabla}&\hat\nabla_{\hat X}\hat Y\ar@{->>}[d]^{\pi_*}\\
\pl X, Y\pr\ar[r]^{\overset{w}{\nabla}}&\overset{w}{\nabla}_{X} Y \label{diagdefibarnabla2}
}
\end{split}
\eea
where 
\be
\item $(X,Y)$ is a pair of vector fields on $\M$ such that $X$ and $Y$ are not both spacelike\footnote{\ie $\psi(X)\neq0$ or $\psi(Y)\neq0$.}.
\item $(\hat X,\hat Y \big)_w$ is a pair of invariant lifts on $\hat\M$ of $X$ and $Y$, respectively, satisfying the additional constraint:
\bea
\hat g(\hat X,\hat Y)=w\, \hat \psi(\hat X)\, \hat\psi(\hat Y)\text{ where }\hat\psi:=\hat g(\hat \xi)\label{eqweightcond}
\eea
and referred to as a \textbf{pair of weighted lifts} of weight $w$, with $w$ a constant.
\ee
provided the following conditions are satisfied:
\begin{enumerate}[a)]
\item The ambient vector field $\hat\nabla_{\hat X}\hat Y\in\vfh$ is projectable.
\item The vector field $\wnab_{ X} Y:=\pi_*\pl\hat\nabla_{\hat X}\hat Y\pr$ is independent of the choice of pair of weighted lifts $\big(\hat X,\hat Y \big)_{w}$.
\item The derivative operator $\wnab$ satisfies the axioms of a Koszul connection.
\end{enumerate}
}
\bremi{\item The conditions for a Bargmannian structure to be projectable are given in Lemma \ref{propinducedLeibniz}. 
\item Note that the above projection scheme is {\it not} canonical but depends on the choice of a constant $w$ called the \textbf{weight}.
\item It can be checked that the previous prescription completely determines the induced connection $\wnab$. 
\item When acting on two spacelike vector fields $V,W\in\Milneg$, one can define the resulting action of $\wnab$ as $\wnab_VW=\wnab_VX+\wnab_VY$ where $X:=\half (W+N)$ and $Y:=\half (W-N)$ with $N\in\FO$ an arbitrary field of observers.
}

The following lemma ensures that this alternative scheme is suitable for Platonic connections:
\blem{}{\label{lemPlatoproj}\label{lemprojconPlato}Let $\big(\hat\M,\hat\xi,\hat g\big)$ be a Platonic wave. The Levi--Civita connection $\hat\nabla$ associated with $\hat g$ admits a well-defined weighted projection on the quotient manifold $\M:=\hat\M/\mR$. }
\bproof{Let $\Lambda\in\foncmhinvn$ be an invariant nowhere vanishing function such that $\Lambda^2$ is the scaling factor of the Platonic wave $\big(\hat\M,\hat\xi,\hat g\big)$, \ie $\hat d\hat \psi=2\, \hat d\ln\Lambda\w\hat\psi$. The present proof will make use of the Galilean parameterisation\footnote{\Cf  \cite{Morand:2016rrt} for a more detailed geometric proof. } \eqref{hatgNC} in which a Platonic wave is characterised by the conditions:
\bea
\p_u\psi_\mu=0\quad,\quad\p_ug_{\mu \nu}=0
\eea
which in turn ensure that $\p_uh^{\mu \nu}=0$ and $\p_u\phi=0$. All the Galilean objects parameterising the Platonic wave thus admit a well-defined projection on the quotient manifold $\M$, as a consequence of the Killing condition. 
The previous relations allow to write the Christoffel coefficients associated with $\hat\nabla$ as:
\bea
\begin{split}
\hat\Gamma^{u}_{u\nu}&=-\half \Lag_Z\psi_\nu\\
\hat\Gamma^{u}_{\mu\nu}&=-\half\Lag_Z g_{\mu\nu}+\p_{(\mu}\phi\, \psi_{\nu)}\\
\hat\Gamma^{\lambda}_{u\nu}&=h^{\lambda\rho}\p_{[\nu}\psi_{\rho]}\\
\hat\Gamma^{\lambda}_{\mu\nu}&=Z^\lambda\p_{(\mu}\psi_{\nu)}+\half h^{\lambda\rho}\big( \p_\mu g_{\rho\nu}+\p_\nu g_{\rho\mu}-\p_\rho g_{\mu\nu}\big).
\end{split}
\eea
Now, let $(X,Y)$ be a pair of vector fields on $\M$ (such that $X$ and $Y$ are not both spacelike) and $(\hat X,\hat Y \big)_w$ be an associated {pair of weighted lifts} of weight $w$. Condition \eqref{eqweightcond} translates as the following constraint:
\bea
\psi(X) Y^u+\psi(Y)X^u+g(X,Y)=w\, \psi(X)\psi(Y)\label{constraintproof}
\eea
 where $X^u,Y^u$ denote respectively the components of $\hat X,\hat Y$ along $\p_u$.
 
Computing the projectable components of $\hat\nabla_{\hat X}\hat Y$ leads to 
\bea
(\hat\nabla_{\hat X}\hat Y)^\lambda=X^\mu(\p_\mu Y^\lambda+\hat\Gamma\lmn Y^\nu)-h^{\lambda \rho}\p_\rho\ln\bar\Lambda\, \big(\psi(X)Y^u+\psi(Y) X^u\big)
\eea
where  we used $d\psi=2\, d\ln\bar\Lambda\w \psi$, where $\bar\Lambda\in\fonc{\M}$ denotes the projection of the invariant scaling factor $\Lambda\in\foncmhinvn$ on $\M$  \big(\ie $\pi^*\bar\Lambda=\Lambda$\big). Substituting \eqref{constraintproof} leads to $\pi_*(\hat\nabla_{\hat X}\hat Y)=\wnab_{X}Y$ where the components of the induced connection $\wnab$ read: 
  \bea
 \overset{w}{\Gamma}\lmn :=Z^\lambda\p_{(\mu}\psi_{\nu)}+\half h^{\lambda\rho}\big( \p_\mu g_{\rho\nu}+\p_\nu g_{\rho\mu}-\p_\rho g_{\mu\nu}\big) +  h^{\lambda \rho}\p_\rho\ln\bar\Lambda\, \big(g_{\mu\nu}-w\, \psi_\mu\psi_\nu\big).\label{coeffweightPlato}
 \eea
 We conclude that the weighted projection procedure is well-defined for Platonic waves and the induced torsionfree connection $\wnab$ is given by explicit formulas, \cf \eqref{coeffweightPlato}.
}
Combining \Prop{corLeibParg} and Lemma \ref{lemPlatoproj}, we conclude that the quotient manifold of a Platonic wave $\big(\hat\M,\hat\xi,\hat g\big)$ is endowed with a Galilean structure $(\M,\psi,\gamma)$ together with a torsionfree connection $\overset{w}{\nabla}$, with $w$ an arbitrary constant. It should be emphasised however that the quadruplet $(\M,\psi,\gamma,\overset{w}{\nabla})$ is {\it not} a torsionfree Galilean manifold in general, since the absolute clock $\psi$ -- being Frobenius but not necessarily closed -- does not generically admit torsionfree compatible connections.  
Therefore, in contradistinction with the Bargmann--Eisenhart case -- for which the projected connection was naturally compatible with the induced Galilean metric structure (\cf Lemma \ref{LemNewt}) -- additional work will be necessary in order to interpret the projected connection $\overset{w}{\nabla}$ \big(\cf coefficients \eqref{coeffweightPlato}\big) as a Galilean connection.
To provide such an interpretation, we will rely on the notion of projective equivalence, understood in the sense of the following classic theorem:
\bthm{\cf \eg \cite{Spivak1970,Eastwood2008}}{\label{thmEastwood}Two torsionfree Koszul connections $\nabla'$ and $\nabla$ on a manifold $\M$ share the same geodesics as unparameterised curves if and only if there exists a 1-form $\Upsilon\in\ff$ such that the following relation holds:
\bea
\nabla'_XY=\nabla_XY+\Upsilon\pl X\pr Y+\Upsilon\pl Y\pr X\quad\text{ for all }X,Y\in\vf. \label{eqthmEastwood}
\eea
Two Koszul connections $\nabla'$ and $\nabla$ satisfying eq.\eqref{eqthmEastwood} for some (exact) 1-form $\Upsilon$ are said to be (strongly) \textbf{projectively equivalent}. 
}
\bremi{
\item In components, relation \eqref{eqthmEastwood} can be expressed as 
\bea
\Gamma'{}^{\lambda}_{\mu\nu}=\Gamma^{\lambda}_{\mu\nu}+2\, \delta^\lambda_{(\mu}\Upsilon_{\nu)} \label{eqprojstructGamma}
\eea
such that $\nabla'_{\dx}\dx\sim \dot x\Leftrightarrow \nabla_{\dx}\dx\sim \dot x$.
}
The notion of projective equivalence allows us to articulate the following proposition:
\bprop{\label{propprojeqplato}Let $\big(\hat\M,\hat\xi,\hat g\big)$ be a Platonic wave inducing the Frobenius Galilean structure $(\M,\psi,\gamma)$ on $\M$. Let us furthermore denote $\overset{w}{\nabla}$ the weighted projection of the Levi--Civita connection $\hat\nabla$ associated with $\hat g$, with weight $w$. 
The connection $\overset{w}{\nabla}$ is strongly projectively equivalent to a Newtonian connection  $\overset{w}{\bar\nabla}$ on $\M$ compatible with the special Galilean structure $(\M,\bar\psi,\bar\gamma)$ associated with $(\M,\psi,\gamma)$.
}
\bproof{Recalling from the proof of Lemma \ref{lemPlatoproj} the expression \eqref{coeffweightPlato} for the components of the connection $\wnab{}$ obtained by weighted projection of weight $w$ of the Platonic connection $\hat\nabla{}$:
  \bea
 \overset{w}{\Gamma}\lmn :=Z^\lambda\p_{(\mu}\psi_{\nu)}+\half h^{\lambda\rho}\big( \p_\mu g_{\rho\nu}+\p_\nu g_{\rho\mu}-\p_\rho g_{\mu\nu}\big) +  h^{\lambda \rho}\p_\rho\ln\bar\Lambda\, \big(g_{\mu\nu}-w\, \psi_\mu\psi_\nu\big)
 \eea
 and performing the reparameterisation: 
 \bea
 \psi=\bar\Lambda^2\, \bar\psi\quad,\quad h=\bar\Lambda^{-2}\, \bar h\quad,\quad g=\bar\Lambda^{2}\, \bar g\quad,\quad Z=\bar\Lambda^{-2}\bar Z\quad,\quad \phi=\bar\Lambda^{-2}\,\bar \phi
 \eea
leads to 
     \bea
 \overset{w}{\Gamma}{}\lmn=\bar Z^\lambda\p_{(\mu}\bar \psi_{\nu)}+\half \bar h^{\lambda\rho}\big( \p_\mu {\bar g}_{\rho\nu}+\p_\nu {\bar g}_{\rho\mu}-\p_\rho {\bar g}_{\mu\nu}\big)+2\, \delta^\lambda_{(\mu}\p_{\nu)}\ln\bar\Lambda-w\, \bar\Lambda^2\bar h^{\lambda \rho}\p_\rho\ln\bar\Lambda\,  \bar \psi_\mu \bar \psi_\nu.\label{eqproofint}
  \eea
Introducing the shifted Lagrangian metric $\overset{w}{\bar g}_{\mu\nu}={\bar g}_{\mu\nu}+w\, \bar\Lambda^{2}\, \bar \psi_{\mu}\bar \psi_{\nu}$ and using\footnote{Recall that $d\psi=2\, d\ln\bar\Lambda\w \psi$, so that $d\bar\psi=0$, \ie $(\bar\psi,\bar \gamma)$ is the special Galilean structure associated with $(\psi,\gamma)$, \cf Remark \ref{remgalfrob}. } $d\bar \psi=0$, expression \eqref{eqproofint} can be reformulated as:
\bea
 \overset{w}{\Gamma}{}\lmn&=&\bar Z^\lambda\p_{(\mu}\bar \psi_{\nu)}+\half \bar h^{\lambda\rho}\big( \p_\mu \overset{w}{\bar g}_{\rho\nu}+\p_\nu \overset{w}{\bar g}_{\rho\mu}-\p_\rho \overset{w}{\bar g}_{\mu\nu}\big)+2\, \delta^\lambda_{(\mu}\p_{\nu)}\ln\bar\Lambda\label{eqwprojN}.
\eea
Comparing \eqref{eqwprojN} and \eqref{eqprojstructGamma} shows that $\wnab{}$ is strongly projectively equivalent \big(in the sense of Theorem \eqref{thmEastwood}\big) to the Newtonian connection $\overset{w}{\bar \nabla}$ defined by the coefficients:
\bea
\overset{w}{\bar \Gamma}{}^{\lambda}_{\mu\nu}=\bar Z^\lambda\p_{(\mu}\bar \psi_{\nu)}+\half \bar h^{\lambda\rho}\big( \p_\mu \overset{w}{\bar g}_{\rho\nu}+\p_\nu \overset{w}{\bar g}_{\rho\mu}-\p_\rho \overset{w}{\bar g}_{\mu\nu}\big)\label{eqcoeffbarwnabla}
\eea
with projective 1-form $\Upsilon=d\ln\bar\Lambda$. 
}
\pagebreak
\bremi{
\item Note that the Newtonian manifold $\big(\M,\bar\psi,\bar\gamma,\overset{w}{\bar\nabla}\big)$ -- whose Newtonian connection $\overset{w}{\bar \nabla}$ is defined by the coefficients \eqref{eqcoeffbarwnabla} -- can be directly obtained from the projection (\cf Theorem \ref{thmDuval}) of a Bargmann--Eisenhart wave, denoted $\big(\hat\M,\hat\xi,\overset{w}{\hat{\bar g}}\big)$. The latter is related to the Bargmann--Eisenhart wave $\big(\hat\M,\hat\xi,\hat{\bar g}\big)$ conformally equivalent to the Platonic wave $\big(\hat\M,\hat\xi,\hat g\big)$ (\cf \Prop{proporbplatBE}) via a shift \eqref{eqorbBE} parameterised by $\hat\phi=w\, \Lambda^2$.
}
We sum up the previous discussion by the following theorem:
\bthm{}{\label{thmPlato}Let $\big(\hat\M,\hat\xi,\hat g\big)$ be a Platonic wave. The following diagram commutes:
\bea
\begin{split}
\xymatrix{
&\underset{\text{\rm (Bargmann--Eisenhart)}}{\big(\hat\M,\hat\xi,\overset{w}{\hat{\bar g}}\big)}
\ar@{->>}[dd]
&\\
\underset{\text{\rm (Platonic)}}{\big(\hat\M,\hat\xi,\hat g\big)}\ar[rr]^{\quad\quad\Lambda\un}_{\quad\quad\eqref{eqorbPlato}}|!{[r];[r]}\hole\ar@{->>}[dd]^{\pi}_{\text{\rm $w$-projection (Lemma}\, \ref{lemprojconPlato})}&&\underset{\text{\rm (Bargmann--Eisenhart)}}{\big(\hat\M,\hat\xi,\hat{\bar g}\big)}\ar@{->>}[dd]_{\pi}^{\text{\rm Theorem }\ref{thmDuval}}\ar[lu]_{w\, \Lambda^2}^{\eqref{eqorbBE}}\\
&\underset{\text{\rm (Newtonian)}}{\big(\M,\bar\psi,\bar\gamma,\overset{w}{\bar\nabla}\big)} &\\
\underset{\text{\rm (Not Galilean)}}{\big(\M,\psi,\gamma,\overset{w}{\nabla}\big)}
\ar@{<->}[ru]
_{\eqref{eqprojstructGamma}}
^{\text{\rm Proj. equiv.\, }}&&\underset{\text{\rm (Newtonian)}}{\big(\M,\bar\psi,\bar\gamma,\bar\nabla\big)}\ar[lu]_{w\, \bar\Lambda^2}^{\eqref{eqshifttrumpvar}}
}
\end{split}
\eea
 }
 
 The most salient feature of the previous construction is that a conformal transformation at the (relativistic) ambient spacetime level translates into a shift of the (nonrelativistic) Newtonian potential on the quotient manifold. Such a property was first observed by Lichnerowicz \cite{Lichnerowicz1955} -- at the level of unparameterised geodesics -- in his generalisation of the Eisenhart lift to Platonic waves. In Section \ref{seccon1}, Theorem \ref{thmDuval} was argued to provide the geometric rationale behind the Eisenhart lift (\cf Theorem \ref{thmEisenhart}). In the same fashion, we argue that Theorem \ref{thmPlato} provides the geometric background underlying Lichnerowicz's generalisation:
\bthm{Eisenhart--Lichnerowicz lift \cite{Lichnerowicz1955}}{\label{Lichnetheorem}Let $\big(\hat\M,\hat\xi,\hat g\big)$ be a Platonic wave and denote $\big(\hat\M,\hat\xi,\hat{\bar g}\big)$ the conformally related Bargmann--Eisenhart wave with conformal factor $\Lambda^2\in\foncmhinv$ -- \ie $\hat g=\Lambda^2\, \hat{\bar g}$ -- and ${\hat\psi}$ the Frobenius 1-form ${\hat\psi}:=\hat g(\hat \xi)$. Let furthermore $\big(\M,{\psi}, \gamma\big)$ \Big(resp. $\big(\M,{\bar\psi}, {\bar\gamma}\big)$\Big) be the Galilean structure induced by $\big(\hat\M,\hat\xi,\hat g\big)$ \Big(resp. $\big(\hat\M,\hat\xi,\hat{\bar g}\big)$\Big). 

Solutions to the parameterised geodesic equation associated with the Platonic metric $\hat g$ are classified according to the two constant of motions:
\be
\item $M^2:=-\hat g(\dot x,\dot x)$
\item $m:=\hat \psi(\dot x)$
\ee
as follows:
\bi
\item $\mathbf{m=0}:$ These are the trajectories constrained on one hypersurface of the foliation induced by ${\hat\psi}$. 

The former are 
classified as\footnote{The condition $m=0$ ensures that $M^2=-\hat\gamma(\dot x,\dot x)\leqslant0$ with $\hat{\gamma}$ the Leibnizian metric induced by $\hat g$. }:

\bi
\item $\mathbf{M^2=0:}$ These are the integral curves of the lightlike vector field $\hat\xi$ \ie $\dot x\sim\hat \xi$.
\item $\mathbf{M^2<0:}$ The projection of such trajectories on the quotient manifold are spacelike trajectories 

(\ie confined on one $d$-dimensional spacelike hypersurface or \textbf{absolute space}) satisfying the parameterised geodesic equation associated with the $d$-dimensional Riemannian metric $\gamma$. 
\ei
\item $\mathbf{m\neq0:}$ The projection of such trajectories\footnote{Irrespectively of the sign of $M^2$ \ie whether the relativistic trajectory is timelike, spacelike or lightlike.} on the quotient manifold are nonrelativistic timelike trajectories satisfying the unparameterised geodesic equation\footnote{Performing a redefinition of $\tau$ allows to put equation \eqref{equnparamLich} in the parameterised form $\overset{w}{{\bar \nabla}}_{\dot x}\dot x=0$.}:
\bea
\overset{w}{{\bar \nabla}}_{\dot x}\dot x=-\frac{d\ln\bar\Lambda^2}{d\tau}\dot x\label{equnparamLich}
\eea
where $\overset{w}{{\bar \nabla}}$ is the Newtonian connection compatible with the Galilean structure $\big(\M,{\bar\psi}, {\bar\gamma}\big)$ related to the Newtonian connection $\bar\nabla$ {\rm(}obtained as projection of the Bargmann--Eisenhart connection $\hat{\bar \nabla}${\rm)} via a shift \eqref{eqgroupactionNewtpotshift} of the Newtonian potential parameterised by $\bar\phi=w\, \bar\Lambda^2$, where $w=-\frac{M^2}{m^2}$ is a constant of motion.
\ei

}
\bproof{The proof follows the one of Theorem \ref{thmEisenhart} with the exception of the case $m\neq0$, which requires additional care. 
Denoting $\hat\nabla$ (resp. $\hat {\bar\nabla}$) the Levi--Civita connections associated with $\hat g$ (resp. $\hat{\bar g}$), the following well-known relation between Levi--Civita connections associated with conformally related (pseudo)-Riemannian metrics holds:
\bea
{\hat\Gamma}^{\hat \lambda}_{{\hat \mu}{\hat \nu}}=\hat{\bar \Gamma}^{\hat \lambda}_{{\hat \mu}{\hat \nu}}+2\, \delta^{\hat \lambda}_{({\hat \mu}}\p_{{\hat \nu})}\ln\Lambda- \hat{\bar g}^{{\hat \lambda}{\hat \rho}}\p_{\hat \rho}\ln\Lambda\, \hat{\bar g}_{{\hat \mu}{\hat \nu}}.\label{eqprojcon}
\eea
Plugging \eqref{eqprojcon} into the parameterised geodesic equation leads to:
\bea
\ddot x^{\hat\lambda}+\hat{\bar \Gamma}^{\hat \lambda}_{\hat\mu\hat\nu}\, \dot x^{\hat\mu}\dot x^{\hat\nu}+2\, \frac{d\ln\Lambda}{d\tau}\dot x^{\hat\lambda}+M^2\Lambda^{-2}\hat{\bar g}^{\hat\lambda \hat\rho}\p_{\hat \rho}\ln\Lambda\, =0.
\eea
Projecting on the quotient manifold $\M$ by means of the Galilean parameterisation \eqref{hatgNC}, one obtains:
\bea
\ddot x^{\lambda}+{\bar \Gamma}^{ \lambda}_{\mu\nu}\, \dot x^{\mu}\dot x^{\nu}+2\, \frac{d\ln\bar\Lambda}{d\tau}\dot x^{\lambda}+M^2\bar\Lambda^{-2}\bar h^{\lambda \rho}\p_{ \rho}\ln\bar\Lambda=0\label{eqproof1}
\eea
where we used that $\p_u\Lambda=0$ and denoted $\bar\Lambda$ the corresponding projection. The coefficients ${\bar \Gamma}$ stand for the components of the Newtonian connection $\bar\nabla$ obtained as projection of the Bargmann--Eisenhart connection $\hat{\bar\nabla}$ and read explicitly:
\bea
{{\bar \Gamma}}{}^{\lambda}_{\mu\nu}=\bar Z^\lambda\p_{(\mu}{\bar\psi}_{\nu)}+\half \bar h^{\lambda\rho}\big( \p_\mu {\bar g}_{\rho\nu}+\p_\nu {\bar g}_{\rho\mu}-\p_\rho {\bar g}_{\mu\nu}\big).
\eea
Now, defining the constant of motion $w:=-\frac{M^2}{m^2}$ allows to reexpress $M^2$ as $M^2=-w\, \bar\Lambda^4{\bar\psi}_\mu {\bar\psi}_\nu \dot x^\mu \dot x^\nu$ such that eq.\eqref{eqproof1} takes the form:
\bea
\ddot x^{\lambda}+\big({\bar \Gamma}^{ \lambda}_{\mu\nu}-\frac{w}{2}\bar h^{\lambda \rho}\p_\rho \bar\Lambda^2\, {\bar\psi}_\mu {\bar\psi}_\nu\big)\, \dot x^{\mu}\dot x^{\nu}+\frac{d\ln\bar\Lambda^2}{d\tau}\dot x^{\lambda}=0\label{eqproof2}.
\eea
Defining the shifted Lagrangian metric $\overset{w}{\bar g}_{\mu\nu}=\bar{g}_{\mu\nu}+w\, \bar\Lambda^2\, {\bar\psi}_{\mu}{\bar\psi}_{\nu}$ allows to interpret the term between brackets as the components \eqref{eqcoeffbarwnabla} of the Newtonian connection $\overset{w}{\bar \nabla}$ associated with $\overset{w}{\bar g}$. 
}
\pagebreak
\bremi{
\item The geometric rationale behind the case $m=0$, ${M^2<0}$ of spacelike (both in the relativistic and nonrelativistic sense) trajectories is provided by \Prop{propKundtabs} stating that the Levi--Civita connection associated with a Kundt wave (and hence to a Platonic wave) admits a canonical projection\footnote{Note that the induced Newtonian connection $\overset{w}{{\bar \nabla}}$ for timelike trajectories also defines a connection on the absolute spaces. The latter coincides with the Levi--Civita connection $\nabla_{\bar\gamma}$ associated with the Riemannian metric $\bar\gamma$, \cf Remark \ref{remspacelikeabsspace}. However, in contrast to the Eisenhart lift, the spacelike trajectories are controlled by a different connection, namely the Levi--Civita connection $\nabla_\gamma$ associated with $\gamma$.} on the absolute spaces of the quotient manifold $\M$ as the Levi--Civita connection $\nabla_\gamma$ associated with the Riemannian metric $\gamma$. 
\item Theorem \ref{thmPlato} provides the apparatus necessary to a geometric understanding of the case $m\neq0$.
\item As an illustration of the Eisenhart--Lichnerowicz Theorem, we introduce the Platonic wave characterised by the ambient vector field $\hat\xi=\p_u$ and the metric
\bea
d\hat s^2=\frac{|x|^2}{R^2}\big(2\, dt\, du+\delta_{ij} dx^idx^j\big)\label{eqPlatoNH}
\eea
conformally related to the Minkowski wave \eqref{Minkspacetime} with conformal factor $\Lambda=\frac{|x|}{R}$. It can be checked that the parameterised geodesic equation associated with \eqref{eqPlatoNH} projects onto the harmonic (expanding) oscillator equation\footnote{Upon setting $\om=1$.}, \ie the geodesic equation of the Newton-Hooke manifold of Example \ref{exaGal},  \cf Remark \ref{remGalNH}. Note that performing a  conformal transformation thus amounts in turning on the Newtonian potential.
}
\pagebreak\section{From Dodgson to Carroll}
\label{secCarrembed}
As emphasised in the previous section, the ambient formalism of Duval \etal allowing to embed Newtonian manifolds into Bargmann--Eisenhart waves admits a suitable generalisation to the larger class of Platonic waves. 
In the present section, we address the dual counterpart of this statement by generalising the embedding procedure of Carrollian manifolds from Bargmann--Eisenhart waves to a larger subclass of gravitational waves, namely the one of Dodgson waves introduced in Section \ref{Dodgson waves}. Explicitly, we show that Dodgson waves are the largest subclass of gravitational waves admitting a canonical lightlike foliation by codimension one Carrollian manifolds. This result is then refined by identifying the subclass of Carrollian manifolds admitting such an embedding, leading us to single out the class of  torsionfree \textit{pseudo-invariant} Carrollian manifolds, \cf \Defi{propinvKoszulCarrpseudo}. As an application of the general formalism, we provide an explicit embedding of the (A)dS-Carroll manifold of Example \ref{exaAdSCarr} inside the (Anti) de Sitter wave of Example \ref{Carrwaves}. We conclude the section by a generalisation of the Carroll train (\cf Theorem \ref{thmCarroll}) applicable to the larger class of Dodgson waves.
~\\

Let us start the discussion by recalling some facts established in Section \ref{secDuval}, more precisely from Proposition \ref{propKundtstructures} where it was shown that the wavefront worldvolumes of a Kundt wave were naturally endowed with:
\be
\item  a set of invariant Carrollian structures.
\item a canonical connection being
\bi
\item torsionfree
\item compatible with the induced Carrollian metric.
\ei
\ee
As emphasised earlier, for a generic Kundt wave, the induced geometry is {\it not} a (torsionfree) Carrollian geometry. This is due to the fact that the induced connection generically fails to preserve any of the induced Carrollian vector fields. This drawback was then circumvented by focusing on the subclass of Bargmann--Eisenhart waves, \cf Theorem \ref{thmDuvcarcon} where it was shown that the parallel condition on the privileged ambient vector field was shown to be sufficient to ensure compatibility between the induced Carrollian vector field and the induced metric connection -- the latter becoming Carrollian. However, such a restriction to the class of Bargmann--Eisenhart waves was shown to strongly constrain the space of Carrollian geometries arising in this way, namely to invariant Carrollian manifolds, \cf \Prop{propCarrBE}. 

In the present section, we relax the ambient parallel condition of Bargmann--Eisenhart waves and argue that the defining condition \eqref{Carrcond} of a Dodgson wave is precisely the necessary and sufficient condition a Kundt wave should satisfy in order to be foliated by codimension one Carrollian manifolds. This extension of the ambient approach allows to embed a more general class of Carrollian geometries inside gravitational waves. To identify this class more precisely, we prove the following technical lemma:

\blem{}{\label{lemCarrpseudo}Let $(\hat\M,\hat\xi,\hat{g})$ be a Dodgson wave with scaling factor $\Om\in\foncmhinv$.
\be
\item The following relation holds for all $V,W\in\Milnegh$:
\bea
(\Lag_{\hat\xi}\hat\nabla)_VW=
-\half \Om\un\Big((\hat\nabla_V\hat d \Om)(W)+(\hat\nabla_W\hat d \Om)(V)\Big)\, \otimes\, \hat\xi\label{eqrelCarrinv}.
\eea
In components:
\bea
V^\mu W^\nu\crl(\Lag_{\hat\xi}\hat\nabla)\lmn+\Om\un\hat\nabla_\mu\hat\nabla_\nu\Om\ \hat\xi^\lambda\, \crr=0.\label{eqrelCarrinv2}
\eea
\item Defining $\hat{\bar\xi}:=\Om\, \hat\xi$, the following relation holds for all $V,W\in\Milnegh$:
\bea
(\Lag_{\hat{\bar\xi}}\hat\nabla)_VW=0.
\eea
\ee
}
\bproofe{

\item Let $(\hat\M,\hat g)$ be a (pseudo)-Riemannian manifold with associated Levi--Civita connection $\hat\nabla$. 

The following relation holds for all $X,V,W,Z\in\vf$:
\bea
\hat g\big((\Lag_X\hat\nabla)_VW,Z\big)+\Lag_X \hat g(\hat\nabla_VW,Z)
&=&\half\Big(V[\Lag_X \hat g(W,Z)]+W[\Lag_X \hat g(V,Z)]-Z[\Lag_X \hat g(V,W)]\nn\\
&&+\Lag_X \hat g(\br{Z}{V},W)+\Lag_X \hat g(V,\br{Z}{W})+\Lag_X \hat g(\br{V}{W},Z)\Big)\nn.
\eea
Assuming that $(\hat\M,[\hat\xi],\hat g)$ is a Kundt wave, any representative $\hat\xi\in[\hat\xi]$ satisfies the relation $\hat\nabla{\hat\psi}={\hat\psi}\otimes \eta+\om\otimes {\hat\psi}$ where ${\hat\psi}:=\hat g(\hat\xi)$. 
Defining $\alpha:=\eta+\om$ and assuming $V,W\in\Milnegh$, we have:
\bea
(\Lag_{\hat\xi}\hat\nabla)_VW&=&\half\Big(\alpha(V)\, \hat d\ln\Om(W)+\alpha(W)\, \hat d\ln\Om(V)+\hat\nabla_V\alpha(W)+\hat\nabla_W\alpha(V)\Big)\otimes\hat\xi
\eea
where we used $\hat d{\hat\psi}=\hat d\ln\Om\, \w\, {\hat\psi}$. Assuming $(\hat\M,\hat\xi,\hat g)$ is a Dodgson wave ensures $\om=0$ and $\alpha=\eta=-\hat d\ln\Om-\kappa{\hat\psi}$. Substituting leads to \eqref{eqrelCarrinv}.
\item The second point is straightforward from substituting $\hat\xi=\Om\un\hat{\bar\xi}$ inside \eqref{eqrelCarrinv}.
}
Comparing constraint \eqref{eqrelCarrinv2} with \eqref{pseudonabxicond} already suggests a relation between Dodgson waves and pseudo-invariant Carrollian manifolds. This relation is made explicit in the following proposition, which constitutes the main result of the present section:
\bprop{\label{propCarrconnamb}Any Dodgson wave admits a canonical lightlike foliation by pseudo-invariant torsionfree Carrollian manifolds of codimension one.
Conversely, any pseudo-invariant torsionfree Carrollian manifold can be embedded inside a Dodgson wave. }
\bproof{Regarding the first statement, the only missing element compared to Proposition \ref{propKundtstructures} lies in the fact that the induced connection preserves the Carrollian vector field $\xi$ (\ie $\nabla\xi=0$) if and only if condition  \eqref{Carrcond} is satisfied, as follows straightforwardly from the definition of the induced connection as displayed in \Prop{projconnCarr}. This condition being the defining condition of a Dodgson wave, we conclude that Dodgson waves are the most general subclass of gravitational waves such that any wavefront worldvolume thereof is canonically a torsionfree Carrollian manifold $\big(\M_t,\xi,\gamma,\nabla\big)$. Furthermore, it follows from Lemma \ref{lemCarrpseudo} that the induced Carrollian manifold is pseudo-invariant, in the sense of \Defi{propinvKoszulCarrpseudo}. The converse statement will be proved constructively in the following Proposition. 
 }
In order to make the previous statement more concrete, we now introduce an explicit embedding procedure allowing to cast any pseudo-invariant torsionfree Carrollian manifold as a wavefront worldvolume of a particular Dodgson wave:
\bpropp{Embedding procedure}{\label{embedproc}Let $\big(\M,\xi,\gamma,\nabla\big)$ be a $d+1$-dimensional pseudo-invariant torsionfree Carrollian manifold with scaling factor\footnote{Recall that a pseudo-invariant connection satisfies by definition the following relation $\Lag_\xi\nabla\lmn=-\xi^\lambda\, \Om\un\nabla_\mu\nabla_\nu\, \Om$ with $\Om$ a nowhere vanishing invariant function, \cf Remark \ref{rempseudocar}. } $\Om\in\foncinvn{\M}$. 

Let $A\in\EC$ be an Ehresmann connection satisfying the following conditions:
\bea
A(\xi)=1\quad,\quad \Lagxi A=-d\ln\Om\label{condA}
\eea
and denote $\overset{A}\Sigma$ the invariant\footnote{The invariance of $\ovA\Sigma$ (\ie $\Lag_\xi\ovA\Sigma$=0) follows from \eqref{pseudoSigmacond} and \eqref{condA}, \cf Remark \ref{rempseudocar}. } tensor defined as 
\bea
\overset{A}\Sigma_{\mu\nu}:=-\nabla_{(\mu}A_{\nu)}+A_{(\mu}\, \Lag_\xi A_{\nu)}.
\eea
The Dodgson wave $(\hat\M,\hat\xi,\hat g)$ where:
\bi
\item $\hat\M$ is a $d+2$-dimensional manifold coordinatised by $x^{\hat\mu}:=\pset{t,x^\mu}$ such that the hypersurface $t=t_0$ coincides with $\M$. 
\item $\hat\xi:=\xi^\mu\p_\mu$ is a nowhere vanishing vector field. 
\item $\hat g$ is the Lorentzian metric defined by the line element
\bea
\hat{ds}{}^2:=-\Om^2(x)\lambda(x)\, dt^2+2\, \Om(x)\, A_\mu(x)\, dx^\mu dt+\big(\gamma_{\mu\nu}(x)-2\, (t-t_0)\, \Om(x)\, \overset{A}\Sigma_{\mu\nu}(x)\big)\, dx^\mu dx^\nu\label{metembpseudo}
\eea
where $\lambda\in\foncm$ is an arbitrary function.
\ei
induces the pseudo-invariant torsionfree Carrollian manifold $\big(\M,\xi,\gamma,\nabla\big)$ on $\M_{t_0}=\M$. 
}
\bproof{
Defining $\hat\psi:=\hat g(\hat\xi)$, the latter reads $\hat\psi=\Om\, dt$ and can be checked to satisfy $\hat\psi(\hat\xi)=0$ and $\hat\nabla\hat\psi=-\hat\psi\otimes \big(\hat d\ln\Om+\half \Lag_\xi\lambda\, \hat\psi\big)$ where $\hat\nabla$ is the Levi--Civita connection associated with $\hat g$, so that the triplet $(\hat\M,\hat\xi,\hat g)$ is indeed a Dodgson wave with Dodgson 1-form $\eta=- \hat d\ln\Om-\half \Lag_\xi\lambda\, \hat\psi$.
In particular, $(\hat\M,\hat\xi,\hat g)$ is a Kundt wave so that the wavefront worldvolume $\M_{t_0}$ characterised by $t=t_0$ is naturally endowed with a (canonical) invariant Carrollian metric structure -- \cf Proposition \ref{propCarrst} -- which naturally identifies with $\big(\M,\xi,\gamma\big)$. Furthermore, \Prop{propKundtstructures} ensures that the Levi--Civita connection $\hat\nabla$ admits a well-defined projection on $\M_{t_0}$. 
Explicitly, restricting the Christoffel coefficients associated with $\hat\nabla$ to the hypersurface $\M_{t_0}$ leads to:
\bea
\hat\Gamma^{t}_{tt}|_{t=t_0}&=&\half\Om\,  \Lag_\xi\lambda\nn\\
\hat\Gamma^{t}_{t\nu}|_{t=t_0}&=&\p_\nu\ln\Om\nn\\
\hat\Gamma^{\lambda}_{tt}|_{t=t_0}&=&\half\, \Om^2\, \big(\ovA h{}^{\lambda\rho}+\lambda\, \xi^\lambda\xi^\rho\big)\p_\rho\lambda+\lambda\, \Om\, \ovA h{}^{\lambda\rho}\p_\rho\Om\nn\\
\hat\Gamma^{\lambda}_{t\nu}|_{t=t_0}&=&-\half\Om\, \xi^\lambda\p_\nu\lambda-\Om\, \ovA h{}^{\lambda\rho}\big(\overset{A}\Sigma_{\rho\nu}+\p_{[\rho}A_{\nu]}+\half\p_\rho\ln\Om\, A_\nu\big)\nn\\
\hat\Gamma^{\lambda}_{\mu\nu}|_{t=t_0}&=&\xi^\lambda\p_{(\mu}A_{\nu)}+\half \ovA h{}^{\lambda\rho}\big( \p_\mu\gamma_{\rho\nu}+\p_\nu\gamma_{\rho\mu}-\p_\rho\gamma_{\mu\nu}\big)-\xi^\lambda  A_{(\mu}\Lag_{\xi} A_{\nu)}+ \xi^\lambda\overset{ A}{\Sigma}_{\mu\nu}\nn
\eea
where $\ovA h\in\field{\vee^2T\M}$ is the transverse cometric associated with the Ehresmann connection $A$ \big(\cf \eqref{eqdefhA}\big). Using the projection scheme of Proposition \ref{projconnCarr}, the induced connection on $\M_{t_0}$ has thus coefficients $\hat\Gamma^{\lambda}_{\mu\nu}|_{t=t_0}$. Comparison with the classification result of Proposition \ref{proptorfreeCarr} -- specifically expression \eqref{Carrconn} -- shows that the induced connection is the unique Carrollian connection compatible with the invariant Carrollian structure $\big(\M,\xi,\gamma\big)$ characterised by the couple $(A,\ovA\Sigma)$ and thus identifies with the original Carrollian connection $\nabla$. 
}
\bremi{\label{rempostembedCar}
\item Since the function $\lambda\in\fonc{\M}$ is arbitrary, one can always choose it to be invariant (\ie $\Lag_\xi\lambda=0$). In this case, the line element \eqref{metembpseudo} defines a Platonic wave upon a rescaling of the ambient vector field $\hat\xi\overset{\Om}{\mapsto}\Om\, \hat\xi$, \cf Proposition \ref{propinplacar}.
\item Heuristically, the crux of the previous construction lies in the fact that the pseudo-invariance of $\nabla$ guarantees the existence of an Ehresmann connection $A$ such that $\ovA\Sigma$ is invariant (\ie $\Lag_\xi\ovA\Sigma$=0). The invariance of $\ovA\Sigma$ thus allows to make it part of the ambient Leibnizian metric $\gamma$, the latter being invariant due to the Kundt condition \eqref{condgammainv}. 
\item The embedding procedure of Proposition \ref{embedproc} provides a constructive existence proof of the second assertion of \Prop{propCarrconnamb}. Note however that the Dodgson wave defined by the line element \eqref{metembpseudo} is {\it not} the most general Dodgson wave embedding  $\big(\M,\xi,\gamma,\nabla\big)$.
\item The previous procedure can in particular be applied to invariant Carrollian manifolds. In this case, the invariance of the Carrollian connection $\nabla$ allows to set the factor $\Om$  to $1$. The resulting wave \eqref{metembpseudo} is then Walker, \cf \Defi{defiWalker}. Further restricting $\lambda$ to be invariant makes the embedding wave to be Bargmann--Eisenhart. In other words, any invariant torsionfree Carrollian manifold can be embedded into a Bargmann--Eisenhart wave. This fact provides an \textit{a posteriori} proof of the second statement of \Prop{propCarrBE}. 
 \item The Minkowski wave of Example \ref{Carrwaves} can be seen to fit into the framework of Proposition \ref{embedproc} upon the following identification:
 \bea
 \lambda=0\quad,\quad\Om=1\quad,\quad \xi=\p_u\quad,\quad A=du\quad,\quad\gamma=\delta_{ij}\, dx^i\vee  dx^j\quad,\quad\ovA h=\delta^{ij}\, \p_i\vee  \p_j\quad,\quad\ovA\Sigma=0.
\eea
The latter thus admits a lightlike foliation by flat Carroll manifolds -- \cf Example \ref{exaAdSCarr} -- characterised by
\bea
\xi=\p_u\quad,\quad\gamma=\delta_{ij}\, dx^i\vee  dx^j\quad,\quad\Gamma=0.
\eea
\item Note that the Newton-Hooke wave of Example \ref{Carrwaves} can also be cast in the form \eqref{metembpseudo}, where it only differs from the Minkowski wave by the value of $\lambda=-\frac{|x|^2}{R^2}$. Since the function $\lambda$ has no Carrollian counterpart, we conclude that the Newton-Hooke wave is also foliated by flat Carroll manifolds.
 }
 We now apply the general machinery developed previously to the maximally symmetric (A)dS-Carroll manifold of Example \ref{exaAdSCarr}. 
As discussed in Remark \ref{remCarrinv}, the latter is not invariant (\ie $\Lag_\xi\nabla\neq0$) and as such cannot be embedded into a Bargmann--Eisenhart wave (\cf \Prop{propCarrBE}). However, as shown in \Prop{propAdSCarr}, the (A)dS-Carroll manifold is pseudo-invariant and as such can be embedded inside a Dodgson wave\footnote{That is, if one trusts the \textit{modus ponens}, \cf \cite{Carroll1995}} (\cf \Prop{embedproc}). A natural candidate is provided by the (Anti) de Sitter wave introduced in Example \ref{Carrwaves}. 
The following proposition identifies the Carrollian manifolds foliating the (Anti) de Sitter wave as (A)dS-Carroll manifolds:
 \bprop{The (Anti) de Sitter wave admits a lightlike foliation by (A)dS Carroll manifolds of equal radii. }
 \bproof{
The line element \eqref{AdSspacetime2} of the (Anti) de Sitter wave can be put in the form \eqref{metembpseudo} upon the identification:
\bea
\begin{split}
&\lambda=\frac{u^2}{R^2}\cosh^{-2}\frac{|x|}{R}\quad,\quad\Om=\cosh\frac{|x|}{R}\quad,\quad\xi=\p_u\quad,\quad A=du-\frac{u}{R} \frac{x_i}{|x|}\tanh\frac{|x|}{R}dx^i
\\
&\gamma=\gamma_{ij}\, dx^i\vee dx^j \quad,\quad \ovA h=\frac{u^2}{R^2}\tanh^2 \frac{|x|}{R}\p_u\vee \p_u+2\, \frac{u}{R} \frac{x^i}{|x|}\tanh\frac{|x|}{R}\, \p_u\vee \p_i+\gamma^{ij}\, \p_i\vee \p_j\quad,\quad\ovA\Sigma=0
\end{split}
\eea
which can be checked to satisfy \eqref{algrelcar}, \eqref{condgammainv} and \eqref{eqcondCarr}.
The geometry induced on any leaf of the lightlike foliation reproduces precisely the formulation of the (A)dS Carroll manifold given by \eqref{eqadscarrdesc}. Thus, the (Anti) de Sitter wave is naturally foliated by (A)dS Carroll manifolds of equal radii $R$.
}
We conclude our analysis of the embedding of Carrollian structures into Dodgson waves with a generalisation of Theorem \ref{thmCarroll}:

 \bthm{Carroll train}{\label{thmCarroll2}Let $\big(\hat\M,\hat\xi,\hat g\big)$ be a Dodgson wave and denote $\hat\psi$ the Frobenius 1-form $\hat\psi:=\hat g(\hat \xi)$ and  $\hat\nabla$ the Levi--Civita connection associated with the Dodgson metric $\hat g$.  Let furthermore $i_t:\M_t\hookrightarrow \hat\M$ be a leaf of the foliation induced by $\Ker \hat\psi$ (or wavefront worldvolume). 
\be
\item Let $p\in i_t(\M_t)$ and $V_p\in\Ker\hat\psi_p$ be a tangent vector. Any geodesic $x(\tau)$ of $\hat\nabla$ such that $\dot x(0)=V_p$ {locally} stays on the wavefront worldvolume $\M_t$. 
\item On the small interval\footnote{Note that, even for a generic Dodgson wave, the solution $m=0$ for all $\tau$ is compatible with equation \eqref{CM2} and thus defines a subclass of geodesic curves for which the second statement of Theorem \ref{thmCarroll2} holds globally.} for which the geodesic $x(\tau)$ stays on $\M_t$, the latter is a geodesic for the Carrollian connection $\nabla$ induced by $\hat \nabla$ on $\M_t$.
\ee
 }
\bproofe{
\item We note that, for a generic Dodgson wave and geodesic $x(\tau)$ of $\hat\nabla$, the quantity $m=\hat\psi(\dot x)$ is \textit{not} a constant of motion since the ambient vector field $\hat \xi$ is not Killing in general\footnote{Whenever $\hat\xi$ is Killing, the underlying wave is Bargmann--Eisenhart -- \cf \Prop{propinplacar} -- so that one recovers the case covered by Theorem \ref{thmCarroll}.}. However, recall from \Prop{propKundtstructures} that the wavefront worldvolumes induced by a Dodgson wave (and more generally by a Kundt wave) are totally geodesic. This property ensures that any geodesic $x(\tau)$ of $\hat\nabla$ defined on a small interval $(-\epsilon,\epsilon)$ and such that $\dot x(0)=V_p$ stays on the wavefront worldvolume $\M_t$.
\item The statement follows straightforwardly from Propositions \ref{propCarrconnamb} and \ref{projconnCarr}.
}
\section{Summary}
The embedding of torsionfree Galilean and Carrollian manifolds into gravitational waves discussed in Sections \ref{secELl} and \ref{secCarrembed} respectively is summarised in Figure \ref{diagboxemb} (thus completing the hierarchy of gravitational waves displayed in Figure \ref{diagboxes}) where the diagonal arrow stands for lightlike projection and the horizontal arrows stand for embedding (or foliation). 
 
\begin{figure}[ht]
\centering
   \includegraphics[width=1\textwidth]{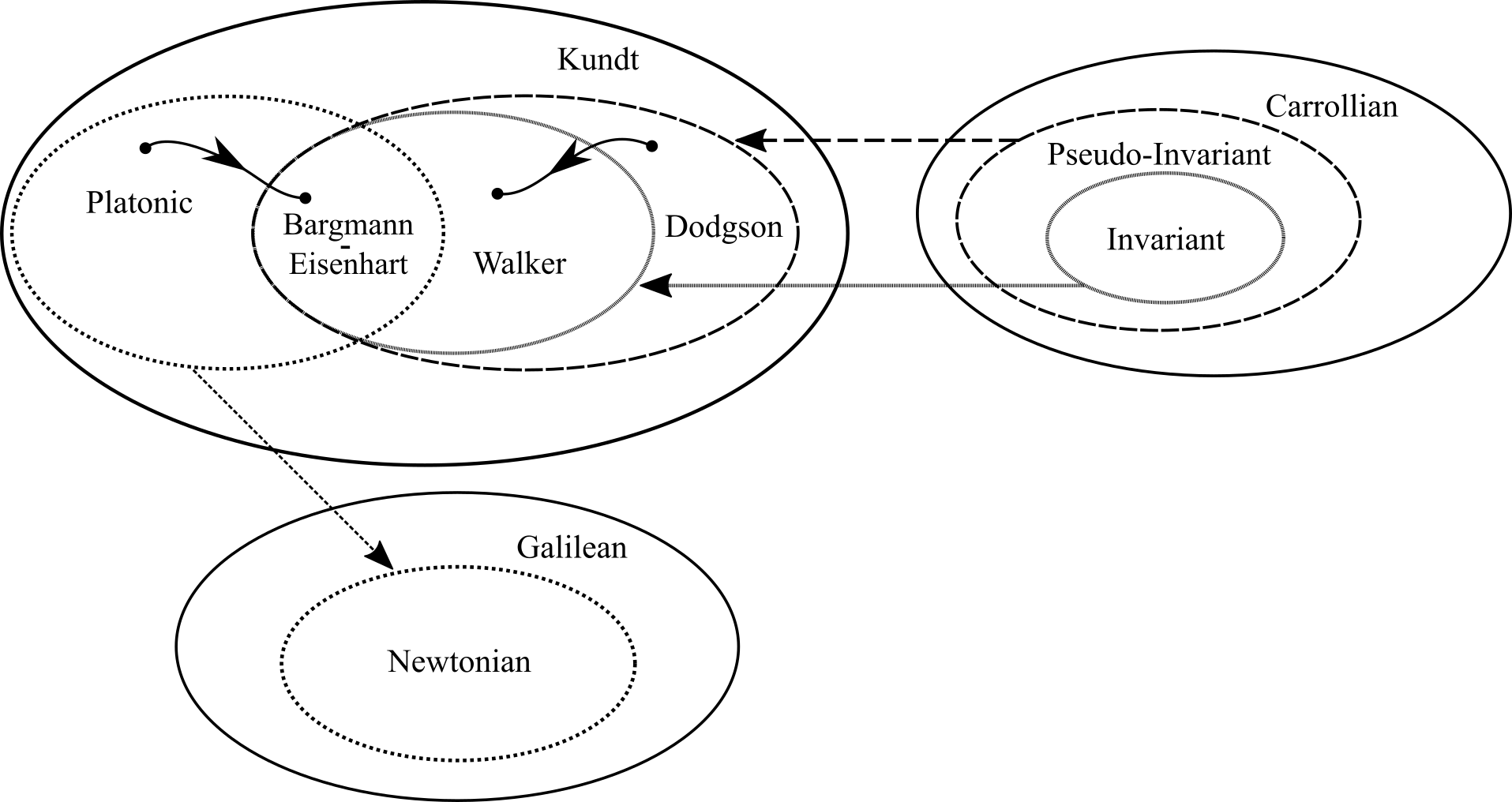}
  \caption{Summary of the embedding of torsionfree Galilean and Carrollian manifolds into gravitational waves\label{diagboxemb}}
\end{figure}
\pagebreak
\section*{Dedication}
\noindent This series of papers is dedicated to the memory of Christian Duval, for the continuous source of inspiration that constituted his work on non-Riemannian structures. 

\section*{Acknowledgements}
\noindent We are indebted to Xavier Bekaert for numerous discussions and constant support as well as for fruitful collaboration at the early stage of this project. We would like to thank Jeong--Hyuck Park for useful exchanges about non-Riemannian structures. We are also grateful to S.~Hervik and M.~Ortaggio for correspondence and helpful comments regarding the geometry of Kundt waves. We thank Lebedev Physical Institute and Centro Cient\'ifico-Tecnol\'ogico de Valpara\'iso
for hospitality. This work was supported by the Chilean Fondecyt Postdoc Project $\text{N}^\circ$3160325 until June 2018 and by the Korean Research Fellowship Grant $\text{N}^\circ$2018H1D3A1A01030137 afterward.


\newpage

\begin{thebibliography}{100}

	\bibitem{Chebyshev} \textsc{G.~ B.~ Halsted}
  \btitle{Biography: Paenutij Lvovitsch Tchebychev}
\newblock \href{http://www.jstor.org/stable/10.2307/2969930}{The American Mathematical Monthly  {\bf 2} No. 3, 61-63} (1895)

	\bibitem{Mink} \textsc{H. Minkowski}
    \btitle{Raum und Zeit}
    \newblock \href{https://en.wikisource.org/wiki/Translation:Space_and_Time}{Jahresberichte der Deutschen Mathematiker-Vereinigung. B.G. Teubner 1-14} (1909)

	\bibitem{Bacry:1968zf} \textsc{H.~Bacry and J.~M.~L\'evy--Leblond}
  \btitle{Possible kinematics}
  \newblock \href{http://dx.doi.org/10.1063/1.1664490}{J.\ Math.\ Phys.\  {\bf 9} 1605} (1968)

	\bibitem{Figueroa-OFarrill:2017sfs} \textsc{J.~Figueroa-O'Farrill}
    \btitle{Classification of kinematical Lie algebras} (2017)
  \barxiv{1711.05676}
    \btitle{Kinematical Lie algebras via deformation theory}
          \newblock \href{http://dx.doi.org/10.1063/1.5016288}{J.\ Math.\ Phys.\  {\bf 59} no.6,  061701} (2018) 
\barxiv{1711.06111}
    \btitle{Higher-dimensional kinematical Lie algebras via deformation theory}
          \newblock \href{http://dx.doi.org/10.1063/1.5016616}{  J.\ Math.\ Phys.\  {\bf 59} no.6,  061702} (2018)
\barxiv{1711.07363}

	\bibitem{Andrzejewski:2018gmz} \textsc{T.~Andrzejewski and J.~M.~Figueroa-O'Farrill}
    \btitle{Kinematical lie algebras in 2 + 1 dimensions}~\\
           \newblock \href{http://dx.doi.org/10.1063/1.5025785}{ J.\ Math.\ Phys.\  {\bf 59} no.6,  061703} (2018)
\barxiv{1802.04048}

	\bibitem{Morand2018b} \textsc{{K}. {Morand}}
\btitle{Embedding Galilean and Carrollian geometries II. Bargmann vs. Leibniz}  
(To appear)

	\bibitem{LevyLeblond1965} \textsc{J.~M.~L\'evy--Leblond}
  \btitle{Une nouvelle limite non-relativiste du groupe de Poincar\'e}
  \newblock \href{http://eudml.org/doc/75509}{Annales de l'I.H.P. Physique th\'eorique\  {\bf 3} 1, 1-12} (1965)

	\bibitem{Carroll1865} \textsc{L.~Carroll}
  \btitle{Alice's Adventures in Wonderland}
\newblock \href{}{Macmillan} (1865)

	\bibitem{Dyson:1972sd} \textsc{F.~J.~Dyson}
  \btitle{Missed opportunities}
\newblock \href{http://dx.doi.org/10.1090/S0002-9904-1972-12971-9}{  Bull.\ Am.\ Math.\ Soc.\  {\bf 78} 635} (1972)

	\bibitem{Carroll1871} \textsc{L.~Carroll}
  \btitle{Through the Looking-Glass, and What Alice Found There}
\newblock \href{}{Macmillan} (1871)

	\bibitem{Duval:2014uoa} \textsc{C.~Duval, G.~W.~Gibbons, P.~A.~Horvathy and P.~M.~Zhang}
  \btitle{Carroll versus Newton and Galilei: two dual non-Einsteinian concepts of time}
  \newblock \href{http://dx.doi.org/10.1088/0264-9381/31/8/085016}{Class.\ Quant.\ Grav.\  {\bf 31}, 085016} (2014)
\barxiv{1402.0657}

	\bibitem{Henneaux1979} \textsc{M.~Henneaux}
    \btitle{Zero hamiltonian signature spacetimes}
\newblock \href{}{Bull. Soc. Math. de Belgique {XXXI} {\bf31} 47} (1979)

	\bibitem{Son:2013rqa} \textsc{{D}.{T}. {S}on}
\btitle{{{N}ewton-{C}artan {G}eometry and the {Q}uantum {H}all {E}ffect}} (2013) 
\barxiv{1306.0638}

	\textsc{{M}. {G}eracie, {D}.{T}. {S}on, {C}. {W}u, and {S}.{F} {W}u}
\btitle{{{S}pacetime {S}ymmetries of the {Q}uantum {H}all {E}ffect}}
  \newblock \href{http://dx.doi.org/10.1103/PhysRevD.91.045030}{
Phys. Rev. D \textbf{91}:045030} (2015) 
\barxiv{1407.1252}

	\bibitem{Geracie:2014zha} \textsc{{M}. {G}eracie, {D}.{T}. {S}on}
\btitle{{H}ydrodynamics on the lowest {L}andau level}
  \newblock \href{http://dx.doi.org/10.1007/JHEP06(2015)044}{JHEP {\bf 1506} 044} (2015)
\barxiv{1408.6843}

	\bibitem{Jensen:2014aia} \textsc{K.~Jensen}
  \btitle{On the coupling of Galilean-invariant field theories to curved spacetime}
    \newblock \href{http://dx.doi.org/10.21468/SciPostPhys.5.1.011}{SciPost Phys.\  {\bf 5} no.1,  011} (2018)
\barxiv{1408.6855}

	\bibitem{Geracie:2014mta} \textsc{{M}. {G}eracie, {S}. {G}olkar, and {M}.{M}.~{R}oberts}
\btitle{{H}all viscosity, spin density, and torsion} (2014)
\barxiv{1410.2574}

	\bibitem{condmatt} \textsc{{B}. {Carter}, {I}.~{M}. {Khalatnikov}}
\btitle{{Canonically Covariant Formulation of Landau's Newtonian Superfluid
  Dynamics}}
    \newblock \href{http://dx.doi.org/10.1142/S0129055X94000134}{{R}eviews in {M}athematical {P}hysics \textbf{6}:277-304} (1994)
\textsc{{A}. {G}romov, {A}.{G}.~{A}banov}
\btitle{{{T}hermal {H}all {E}ffect and {G}eometry with {T}orsion}}
  \newblock \href{http://dx.doi.org/10.1103/PhysRevLett.114.016802}{Phys. Rev. Lett. \textbf{114}:016802} (2014) 
\barxiv{1407.2908}
\textsc{{B}. {Bradlyn}, {N}. {Read}}
\btitle{{{Low-energy effective theory in the bulk for transport in a
  topological phase}}}
  \newblock \href{http://dx.doi.org/10.1103/PhysRevB.91.125303}{Phys. Rev. B \textbf{91}:125303} (2015)
\barxiv{1407.2911}
\textsc{{T}. {B}rauner, {S}. {E}ndlich, {A}. {M}onin and {R}. {P}enco}
\btitle{{{G}eneral coordinate invariance in quantum many-body systems}}
  \newblock \href{http://dx.doi.org/10.1103/PhysRevD.90.105016}{Phys.Rev. D \textbf{90}:105016} (2014)
\barxiv{1407.7730}
\textsc{{S}. {M}oroz, {C}. {H}oyos}
\btitle{{E}ffective theory of two-dimensional chiral superfluids: gauge
  duality and {N}ewton-{C}artan formulation.}
  \newblock \href{http://dx.doi.org/10.1103/PhysRevB.91.064508}{Phys. Rev. \textbf{B} 91:064508} (2015)
\barxiv{1408.5911}

	\bibitem{Hall} \textsc{S.~Golkar, D.~X.~Nguyen and D.~T.~Son}
  \btitle{Spectral Sum Rules and Magneto-Roton as Emergent Graviton in Fractional Quantum Hall Effect}
      \newblock \href{http://dx.doi.org/10.1007/JHEP01(2016)021}{JHEP {\bf 1601} 021} (2016) 
\barxiv{1309.2638}
  \textsc{R.~Banerjee, A.~Mitra and P.~Mukherjee}
  \btitle{A new formulation of non-relativistic diffeomorphism invariance}
    \newblock \href{http://dx.doi.org/10.1016/j.physletb.2014.09.004}{Phys.\ Lett.\ B {\bf 737} 369} (2014)
\barxiv{1404.4491}

\bibitem{Banerjee:2015rca}
  \textsc{R.~Banerjee and P.~Mukherjee}
  \btitle{New approach to nonrelativistic diffeomorphism invariance and its applications}
 \newblock \href{http://dx.doi.org/10.1103/PhysRevD.93.085020}{Phys.\ Rev.\ D {\bf 93} no.8,  085020} (2016)
\barxiv{1509.05622}

	\bibitem{fluid} \textsc{M.~Geracie, K.~Prabhu and M.~M.~Roberts}
  \btitle{Fields and fluids on curved non-relativistic spacetimes}
    \newblock \href{http://dx.doi.org/10.1007/JHEP08(2015)042}{JHEP {\bf 1508} 042} (2015)
\barxiv{1503.02680}
  \textsc{L.~Ciambelli, C.~Marteau, A.~C.~Petkou, P.~M.~Petropoulos and K.~Siampos}
  \btitle{Covariant Galilean versus Carrollian hydrodynamics from relativistic fluids}
          \newblock \href{http://dx.doi.org/10.1088/1361-6382/aacf1a}{Class.\ Quant.\ Grav.\  {\bf 35} no.16,  165001} (2018)
\barxiv{1802.05286}

	\bibitem{flatholography} \textsc{C.~Duval, G.~W.~Gibbons and P.~A.~Horvathy}
  \btitle{Conformal Carroll groups and BMS symmetry}
  \newblock \href{http://dx.doi.org/10.1088/0264-9381/31/9/092001}{Class.\ Quant.\ Grav.\  {\bf 31}, 092001} (2014)
\barxiv{1402.5894}
  \textsc{J.~Hartong}
  \btitle{Holographic Reconstruction of 3D Flat Space-Time}
          \newblock \href{http://dx.doi.org/10.1007/JHEP10(2016)104}{JHEP {\bf 1610} 104} (2016)
\barxiv{1511.01387}
  \textsc{A.~Bagchi, R.~Basu, A.~Kakkar and A.~Mehra}
  \btitle{Flat Holography: Aspects of the dual field theory}
      \newblock \href{http://dx.doi.org/10.1007/JHEP12(2016)147}{JHEP {\bf 1612} 147} (2016)
\barxiv{1609.06203}
  \textsc{D.~Grumiller, W.~Merbis and M.~Riegler}
  \btitle{Most general flat space boundary conditions in three-dimensional Einstein gravity}
        \newblock \href{http://dx.doi.org/10.1088/1361-6382/aa8004}{Class.\ Quant.\ Grav.\  {\bf 34} no.18,  184001} (2017)
\barxiv{1704.07419}
    \textsc{L.~Ciambelli, C.~Marteau, A.~C.~Petkou, P.~M.~Petropoulos and K.~Siampos}
    \btitle{Flat holography and Carrollian fluids}
          \newblock \href{http://dx.doi.org/10.1007/JHEP07(2018)165}{JHEP {\bf 1807} 165} (2018)
\barxiv{1802.06809}

	\bibitem{Christensen:2013lma} \textsc{{M}.{H}. {Christensen}, {J}. {Hartong}, {N}.{A}. {Obers}, and {B}. {Rollier}}
\btitle{Torsional Newton-Cartan geometry and Lifshitz holography}
      \newblock \href{http://dx.doi.org/10.1103/PhysRevD.89.061901}{{Physical Review D} \textbf{89}(6):061901} (2014)
      \barxiv{1311.4794 }

	\bibitem{Christensen:2013rfa} \textsc{{M}.{H}. {C}hristensen, {J}. {H}artong, {N}.{A}.~{O}bers, and {B}. {R}ollier}
\btitle{{B}oundary {S}tress-{E}nergy {T}ensor and {N}ewton-{C}artan
  {G}eometry in {L}ifshitz {H}olography} 
    \newblock \href{http://dx.doi.org/10.1007/JHEP01(2014)057}{  JHEP {\bf 1401} 057} (2014)
\barxiv{1311.6471}

	\bibitem{Bergshoeff:2014uea} \textsc{{E}.{A}.~{B}ergshoeff, {J}. {H}artong, and {J}. {R}osseel}
\btitle{{T}orsional {N}ewton-{C}artan {G}eometry and the {S}chr\"odinger
  {A}lgebra} 
      \newblock \href{http://dx.doi.org/10.1088/0264-9381/32/13/135017}{Class.\ Quant.\ Grav.\  {\bf 32}, no. 13, 135017} (2015)
\barxiv{1409.5555}

	\bibitem{lifshitz} \textsc{{J}. {H}artong, {E}. {K}iritsis and {N}.{A}.~{O}bers}
\btitle{{L}ifshitz {S}pace-{T}imes for {S}chr\"odinger {H}olography} 
      \newblock \href{http://dx.doi.org/10.1016/j.physletb.2015.05.010}{Phys.\ Lett.\ B {\bf 746}, 318} (2015)
\barxiv{1409.1519}
\textsc{{J}. {H}artong, {E}. {K}iritsis and {N}.{A}.~{O}bers}
\btitle{{S}chr\"odinger {I}nvariance from {L}ifshitz {I}sometries in
  {H}olography and {F}ield {T}heory}
        \newblock \href{http://dx.doi.org/10.1103/PhysRevD.92.066003}{Phys.\ Rev.\ D {\bf 92}, 066003} (2015)
\barxiv{1409.1522}
\textsc{{J}. {H}artong, {E}. {K}iritsis and {N}.{A}.~{O}bers}
\btitle{{F}ield {T}heory on {N}ewton-{C}artan {B}ackgrounds and {S}ymmetries
  of the {L}ifshitz {V}acuum}
        \newblock \href{http://dx.doi.org/10.1007/JHEP08(2015)006}{JHEP {\bf 1508} 006} (2015)
\barxiv{1502.00228}
  \textsc{J.~Hartong, N.~A.~Obers and M.~Sanchioni}
  \btitle{Lifshitz Hydrodynamics from Lifshitz Black Branes with Linear Momentum}
  \newblock \href{http://dx.doi.org/10.1007/JHEP10(2016)120}{JHEP {\bf 1610} 120} (2016)
\barxiv{1606.09543}

	\bibitem{horava} 
\textsc{{J}. {H}artong, {N}.{A}.~{O}bers}
\btitle{Ho{\v{r}}ava-Lifshitz gravity from dynamical Newton-Cartan geometry}
  \newblock \href{http://dx.doi.org/10.1007/JHEP07(2015)155}{JHEP {\bf 1507} 155} (2015)
\barxiv{1504.07461}
  \textsc{H.~R.~Afshar, E.~A.~Bergshoeff, A.~Mehra, P.~Parekh and B.~Rollier}
  \btitle{A Schr\"odinger approach to Newton-Cartan and Ho{\v{r}}ava-Lifshitz gravities}
  \newblock \href{http://dx.doi.org/10.1007/JHEP04(2016)145}{JHEP {\bf 1604} 145} (2016)
\barxiv{1512.06277}
\textsc{J.~Hartong, Y.~Lei and N.~A.~Obers}
  \btitle{Nonrelativistic Chern-Simons theories and three-dimensional Ho{\v{r}}ava-Lifshitz gravity}
  \newblock \href{http://dx.doi.org/10.1103/PhysRevD.94.065027}{Phys.\ Rev.\ D {\bf 94}, no. 6, 065027} (2016)
\barxiv{1604.08054}
  \textsc{D.~O.~Devecioglu, N.~Ozdemir, M.~Ozkan and U.~Zorba}
  \btitle{Scale Invariance in Newton-Cartan and Ho\v{r}ava-Lifshitz Gravity}
    \newblock \href{http://dx.doi.org/10.1088/1361-6382/aac07e}{Class.\ Quant.\ Grav.\  {\bf 35} no.11,  115016} (2018)
\barxiv{1801.08726}

	\bibitem{galstring} \textsc{C.~Batlle, J.~Gomis and D.~Not}
  \btitle{Extended Galilean symmetries of non-relativistic strings}
        \newblock \href{http://dx.doi.org/10.1007/JHEP02(2017)049}{JHEP {\bf 1702} 049} (2017)
\barxiv{1611.00026}
  \textsc{J.~Gomis and P.~K.~Townsend}
  \btitle{The Galilean Superstring}
      \newblock \href{http://dx.doi.org/10.1007/JHEP02(2017)105}{JHEP {\bf 1702} 105} (2017)
\barxiv{1612.02759}
  \textsc{T.~Harmark, J.~Hartong and N.~A.~Obers}
  \btitle{Nonrelativistic strings and limits of the AdS/CFT correspondence}
    \newblock \href{http://dx.doi.org/10.1103/PhysRevD.96.086019}{Phys.\ Rev.\ D {\bf 96} no.8,  086019} (2017)
\barxiv{1705.03535}
    \textsc{T.~Harmark, J.~Hartong, L.~Menculini, N.~A.~Obers and Z.~Yan}
  \btitle{Strings with Non-Relativistic Conformal Symmetry and Limits of the AdS/CFT Correspondence}
\barxiv{1810.05560}


	\bibitem{Cardona:2016ytk} \textsc{B.~Cardona, J.~Gomis and J.~M.~Pons}
  \btitle{Dynamics of Carroll Strings}
    \newblock \href{http://dx.doi.org/10.1007/JHEP07(2016)050}{JHEP {\bf 1607} 050} (2016)
\barxiv{1605.05483}

	\bibitem{stringygravity} \textsc{R.~Andringa, E.~Bergshoeff, J.~Gomis and M.~de Roo}
  \btitle{`Stringy' Newton-Cartan Gravity}
          \newblock \href{http://dx.doi.org/10.1088/0264-9381/29/23/235020}{Class.\ Quant.\ Grav.\  {\bf 29} 235020} (2012) 
\barxiv{1206.5176}
  \textsc{S.~M.~Ko, C.~Melby-Thompson, R.~Meyer and J.~H.~Park}
  \btitle{Dynamics of Perturbations in Double Field Theory \& Non-Relativistic String Theory}
  \newblock \href{http://dx.doi.org/10.1007/JHEP12(2015)144}{JHEP {\bf 1512} 144} (2015)
\barxiv{1508.01121}
  \textsc{K.~Morand and J.~H.~Park}
  \btitle{Classification of non-Riemannian doubled-yet-gauged spacetime.}
\newblock \href{http://dx.doi.org/10.1140/epjc/s10052-017-5257-z}{  Eur.\ Phys.\ J.\ C {\bf 77}, no. 10, 685} (2017)
  \barxiv{1707.03713}

	\bibitem{Eisenhart1928} \textsc{L.~P.~{E}isenhart}
\btitle{{D}ynamical {T}rajectories and {G}eodesics}
  \newblock \href{https://www.jstor.org/stable/1968307}{{A}nn. {M}ath. {\bfseries 30} 591} (1928)

	\bibitem{ambientgroup} \textsc{J.~Gomis and J.~M.~Pons}
  \btitle{Poincare Transformations and Galilei Transformations}
      \newblock \href{http://dx.doi.org/10.1016/0375-9601(78)90397-3}{Phys.\ Lett.\ A {\bf 66} 463} (1978)
  \textsc{E.~Elizalde and J.~Gomis}
    \btitle{The Groups of Poincare and Galilei in Arbitrary Dimensional Spaces}
        \newblock \href{http://dx.doi.org/10.1063/1.523877}{J.\ Math.\ Phys.\  {\bf 19} 1790} (1978)
      \textsc{J.~Gomis and J.~M.~Pons}
  \btitle{The Centralizer Subalgebras of Poincare $IO(3,1)$}
            \newblock \href{http://dx.doi.org/10.1007/BF02778049}{Nuovo Cim.\ A {\bf 47}, 166} (1978)
    \textsc{J.~Gomis and J.~M.~Pons}
    \btitle{Coordinate Transformations and Centralizer Subalgebras of Poincare $IO(3,1)$}
          \newblock \href{http://dx.doi.org/10.1007/BF02778050}{Nuovo Cim.\ A {\bf 47} 175} (1978)
      \textsc{J.~Gomis, A.~Poch and J.~M.~Pons}
      \btitle{Poincare Wave Equations As Fourier Transforms Of Galilei Wave Equations}
            \newblock \href{http://dx.doi.org/10.1063/1.524369}{J.\ Math.\ Phys.\  {\bf 21} 2682} (1980)

	\bibitem{Duval:1984cj} \textsc{C.~Duval, G.~Burdet, H.~P.~K\"unzle and M.~Perrin}
  \btitle{Bargmann Structures and Newton-Cartan Theory}
  \newblock \href{http://dx.doi.org/10.1103/PhysRevD.31.1841}{Phys.\ Rev.\ D {\bf 31}, 1841} (1985)

	\bibitem{Duval:1990hj} \textsc{C.~Duval, G.~W.~Gibbons and P.~Horvathy}
  \btitle{Celestial mechanics, conformal structures and gravitational waves}
  \newblock \href{http://dx.doi.org/10.1103/PhysRevD.43.3907}{Phys.\ Rev.\ D {\bf 43}, 3907} (1991)
\barxiv{hep-th/0512188}

	\bibitem{Balachandran:1986hv} \textsc{A.~P.~Balachandran, H.~Gomm and R.~D.~Sorkin}
  \btitle{Quantum Symmetries From Quantum Phases: Fermions From Bosons, a $\mathbb Z_2$ Anomaly and Galilean Invariance}
     \newblock \href{http://dx.doi.org/10.1016/0550-3213(87)90420-2}{Nucl.\ Phys.\ B {\bf 281} 573} (1987)

	\bibitem{nullclassicalandquantummechanicalsystems} \textsc{M.~Cariglia, C.~Duval, G.~W.~Gibbons and P.~A.~Horvathy}
  \btitle{Eisenhart lifts and symmetries of time-dependent systems}
        \newblock \href{http://dx.doi.org/10.1016/j.aop.2016.07.033}{Annals Phys.\  {\bf 373} 631} (2016)
\barxiv{1605.01932}
 \textsc{P. M.~Zhang, M.~Cariglia, C.~Duval, M.~Elbistan, G.~W.~Gibbons and P.~A.~Horvathy}
    \btitle{Ion Traps and the Memory Effect for Periodic Gravitational Waves}
         \newblock \href{http://dx.doi.org/10.1103/PhysRevD.98.044037}{ Phys.\ Rev.\ D {\bf 98} no.4,  044037} (2018)
\barxiv{1807.00765}

	\bibitem{nullfluids} \textsc{M.~Hassaine and P.~A.~Horvathy}
  \btitle{Field dependent symmetries of a nonrelativistic fluid model}
       \newblock \href{http://dx.doi.org/10.1006/aphy.1999.6002}{Annals Phys.\  {\bf 282} 218} (2000)
\barxiv{math-ph/9904022}
  \textsc{M.~Hassaine and P.~A.~Horvathy}
  \btitle{Symmetries of fluid dynamics with polytropic exponent}
       \newblock \href{http://dx.doi.org/10.1016/S0375-9601(00)00834-3}{  Phys.\ Lett.\ A {\bf 279} 215} (2001)
  \barxiv{hep-th/0009092}
  \textsc{M.~Rangamani, S.~F.~Ross, D.~T.~Son and E.~G.~Thompson}
  \btitle{Conformal non-relativistic hydrodynamics from gravity}
      \newblock \href{http://dx.doi.org/10.1088/1126-6708/2009/01/075}{JHEP {\bf 0901} 075} (2009)
\barxiv{0811.2049}
  \textsc{M.~d.~Montigny, F.~C.~Khanna and A.~E.~Santana}
  \btitle{Lorentz-like covariant equations of non-relativistic fluids}
       \newblock \href{http://dx.doi.org/10.1088/0305-4470/36/8/301}{  J.\ Phys.\ A {\bf 36} 2009} (2003)
  \textsc{K.~Jensen}
  \btitle{Aspects of hot Galilean field theory}
         \newblock \href{http://dx.doi.org/10.1007/JHEP04(2015)123}{JHEP {\bf 1504} 123} (2015)
\barxiv{1411.7024}
  \textsc{M.~Geracie, K.~Prabhu and M.~M.~Roberts}
  \btitle{Curved non-relativistic spacetimes, Newtonian gravitation and massive matter}
           \newblock \href{http://dx.doi.org/10.1063/1.4932967}{J.\ Math.\ Phys.\  {\bf 56} no.10,  103505} (2015)
\barxiv{1503.02682}
  \textsc{N.~Banerjee, S.~Dutta and A.~Jain}
  \btitle{Equilibrium partition function for nonrelativistic fluids}
           \newblock \href{http://dx.doi.org/10.1103/PhysRevD.92.081701}{Phys.\ Rev.\ D {\bf 92} 081701} (2015)
\barxiv{1505.05677}
  \textsc{N.~Banerjee, S.~Dutta and A.~Jain}
  \btitle{Null Fluids - A New Viewpoint of Galilean Fluids}
    \newblock \href{http://dx.doi.org/10.1103/PhysRevD.93.105020}{Phys.\ Rev.\ D {\bf 93} no.10,  105020} (2016)
\barxiv{1509.04718}
  \textsc{A.~Jain}
  \btitle{Galilean Anomalies and Their Effect on Hydrodynamics}
    \newblock \href{http://dx.doi.org/10.1103/PhysRevD.93.065007}{  Phys.\ Rev.\ D {\bf 93} no.6,  065007} (2016)
\barxiv{1509.05777}
  \textsc{J.~Armas, J.~Bhattacharya, A.~Jain and N.~Kundu}
  \btitle{On the surface of superfluids}
       \newblock \href{http://dx.doi.org/10.1007/JHEP06(2017)090}{  JHEP {\bf 1706} 090} (2017)
\barxiv{1612.08088}
  \textsc{S.~Dutta and H.~Krishna}
  \btitle{Light-Cone Reduction vs. TsT transformations : A Fluid Dynamics Perspective}
    \newblock \href{http://dx.doi.org/10.1007/JHEP05(2018)029}{JHEP {\bf 1805} 029} (2018)
\barxiv{1803.03948}

	\bibitem{nullcondensedmatter} \textsc{M.~Cariglia, R.~Giamb\`o and A.~Perali}
  \btitle{Curvature-tuned electronic properties of bilayer graphene in an effective four-dimensional spacetime}
          \newblock \href{http://dx.doi.org/10.1103/PhysRevB.95.245426}{Phys.\ Rev.\ B {\bf 95} no.24,  245426} (2017)
\barxiv{1611.06254}

	\bibitem{nullcosmology} \textsc{C.~Duval, G.~Gibbons and P.~Horvathy}
  \btitle{Conformal and projective symmetries in Newtonian cosmology}
        \newblock \href{http://dx.doi.org/10.1016/j.geomphys.2016.11.012}{J.\ Geom.\ Phys.\  {\bf 112} 197} (2017)
\barxiv{1605.00231}
  \textsc{M.~Cariglia, A.~Galajinsky, G.~W.~Gibbons and P.~A.~Horvathy}
  \btitle{Cosmological aspects of the Eisenhart-Duval lift}
      \newblock \href{http://dx.doi.org/10.1140/epjc/s10052-018-5789-x}{Eur.\ Phys.\ J.\ C {\bf 78} no.4,  314} (2018)
\barxiv{1802.03370}

	\bibitem{null3D} \textsc{E.~A.~Bergshoeff, J.~Rosseel and P.~K.~Townsend}
  \btitle{Gravity and the Spin-2 Planar Schr\"odinger Equation}
          \newblock \href{http://dx.doi.org/10.1103/PhysRevLett.120.141601}{Phys.\ Rev.\ Lett.\  {\bf 120} no.14,  141601} (2018)
\barxiv{1712.10071}
  \btitle{On nonrelativistic 3D Spin-1 theories}
            \newblock \href{http://dx.doi.org/10.1134/S1063779618050064}{Phys. Part. Nuclei {\bf 49} 813} (2018)
\barxiv{1801.02527}

	\bibitem{nullholography} \textsc{W.~D.~Goldberger}
  \btitle{AdS/CFT duality for non-relativistic field theory}
  \newblock \href{http://dx.doi.org/10.1088/1126-6708/2009/03/069}{JHEP {\bf 0903} 069} (2009)
\barxiv{0806.2867}
  \textsc{J.~L.~F.~Barbon and C.~A.~Fuertes}
  \btitle{On the spectrum of nonrelativistic AdS/CFT}
  \newblock \href{http://dx.doi.org/10.1088/1126-6708/2008/09/030}{JHEP {\bf 0809} 030} (2008)
\barxiv{0806.3244}
  \textsc{F.~L.~Lin and S.~Y.~Wu}
  \btitle{Non-relativistic Holography and Singular Black Hole}
  \newblock \href{http://dx.doi.org/10.1016/j.physletb.2009.07.002}{Phys.\ Lett.\ B {\bf 679}, 65} (2009)
\barxiv{0810.0227}
  \textsc{J.~Hartong, Y.~Lei, N.~A.~Obers and G.~Oling}
  \btitle{Zooming in on AdS$_{3}$/CFT$_{2}$ near a BPS bound}
    \newblock \href{http://dx.doi.org/10.1007/JHEP05(2018)016}{JHEP {\bf 1805} 016} (2018)
\barxiv{1712.05794}


\bibitem{Cariglia:2018hyr}
  \textsc{M.~Cariglia}
  \btitle{General theory of Galilean gravity}
  \newblock \href{http://dx.doi.org/10.1103/PhysRevD.98.084057}{Phys.\ Rev.\ D {\bf 98} no.8,  084057} (2018)
\barxiv{1811.03446}

	\bibitem{Duval1977} \textsc{C.~{D}uval, H.~P.~{K}\"unzle}
\btitle{{S}ur les connexions newtoniennes et l'extension non triviale du groupe de {G}alil\'ee}
\newblock \href{https://www.researchgate.net/publication/267065597_Sur_les_connexions_newtoniennes_et_l%27extension_non_triviale_du_groupe_de_Galilee}{{C}. {R}. {A}cad. {S}c. {P}aris
  {\bfseries 285} 813} (1977) [in French].

	\bibitem{Duval:1976ht} \textsc{C.~Duval and H.~P.~K\"unzle}
 \btitle{Dynamics of Continua and Particles from General Covariance of Newtonian Gravitation Theory,}
  \newblock \href{http://dx.doi.org/10.1016/0034-4877(78)90063-0}{Rept.\ Math.\ Phys.\  {\bf 13} 351} (1978)

	\bibitem{Andringa:2010it} \textsc{R.~Andringa, E.~Bergshoeff, S.~Panda and M.~de Roo}
    \btitle{Newtonian Gravity and the Bargmann Algebra}
      \newblock \href{http://dx.doi.org/10.1088/0264-9381/28/10/105011}{Class.\ Quant.\ Grav.\  {\bf 28} 105011} (2011)
\barxiv{1011.1145}

	\bibitem{stringvacua} \textsc{A.~A.~Tseytlin}
  \btitle{A Class of finite two-dimensional sigma models and string vacua}
\newblock \href{http://dx.doi.org/10.1016/0370-2693(92)91104-H}{  Phys.\ Lett.\ B {\bf 288} 279} (1992)
    \barxiv{hep-th/9205058}
  \textsc{A.~A.~Tseytlin}
  \btitle{String vacuum backgrounds with covariantly constant null Killing vector and 2-d quantum gravity}
\newblock \href{http://dx.doi.org/10.1016/0550-3213(93)90389-7}{  Nucl.\ Phys.\ B {\bf 390} 153} (1993)
  \barxiv{hep-th/9209023}
  \textsc{C.~Duval, Z.~Horvath and P.~A.~Horvathy}
    \btitle{Strings in plane fronted gravitational waves}
\newblock \href{http://dx.doi.org/10.1142/S0217732393003482}{Mod.\ Phys.\ Lett.\ A {\bf 8} 3749} (1993)
\barxiv{hep-th/0602128}


	\bibitem{Amati:1988sa} \textsc{D.~Amati and C.~Klimcik}
  \btitle{Nonperturbative Computation of the Weyl Anomaly for a Class of Nontrivial Backgrounds}
        \newblock \href{http://dx.doi.org/10.1016/0370-2693(89)91092-7}{  Phys.\ Lett.\ B {\bf 219} 443} (1989)

	\bibitem{Horowitz:1989bv} \textsc{G.~T.~Horowitz and A.~R.~Steif}
  \btitle{Space-Time Singularities in String Theory}
       \newblock \href{http://dx.doi.org/10.1103/PhysRevLett.64.260}{Phys.\ Rev.\ Lett.\  {\bf 64} 260} (1990)

	\bibitem{Bekaert:2013fta} \textsc{{X}. {Bekaert}, {K}. {Morand}}
\btitle{{E}mbedding nonrelativistic physics inside a gravitational wave.}
\newblock \href{http://journals.aps.org/prd/abstract/10.1103/PhysRevD.88.063008}{{Physical Review D} \textbf{88}(6):063008} (2013)
\barxiv{1307.6263}

	\bibitem{Bergshoeff:2015wma} \textsc{E.~Bergshoeff, J.~Gomis and L.~Parra}
    \btitle{The Symmetries of the Carroll Superparticle}
    \newblock \href{http://dx.doi.org/10.1088/1751-8113/49/18/185402}{J.\ Phys.\ A {\bf 49} no.18,  185402} (2016) 
  \barxiv{1503.06083}

	\bibitem{Gibbons:2003rv} \textsc{G.~W.~Gibbons and C.~E.~Patricot}
    \btitle{Newton-Hooke space-times, Hpp waves and the cosmological constant}
      \newblock \href{http://dx.doi.org/10.1088/0264-9381/20/23/016}{Class.\ Quant.\ Grav.\  {\bf 20} 5225} (2003)
\barxiv{hep-th/0308200}

	\bibitem{Bekaert:2015xua} \textsc{{X}. {Bekaert}, {K}. {Morand}}
\btitle{Connections and dynamical trajectories in generalised Newton-Cartan
  gravity II. An ambient perspective.}  
  \newblock \href{http://dx.doi.org/10.1063/1.5030328}{J.\ Math.\ Phys. \  {\bf 59}, no. 7, 072503} (2018)
\barxiv{1505.03739}

	\bibitem{Brinkmann} \textsc{M.~W.~Brinkmann}
  \btitle{On Riemann spaces conformal to Euclidean space}
  \newblock \href{https://www.jstor.org/stable/83978}{Proc.\ Natl.\ Acad.\ Sci.\ U.S. {\bf 9} 1} (1923)
\btitle{Einstein spaces which are mapped conformally on each other}
 \newblock \href{https://link.springer.com/article/10.1007/BF01208647}{Math.\ Ann.\ {\bf 94} 119} (1925)

	\bibitem{Blau2004} \textsc{M.~Blau}
  \btitle{Plane waves and Penrose limit} 
   \newblock \href{http://www.blau.itp.unibe.ch/lecturesPP.pdf}{Online lecture notes}

	\bibitem{Lichnerowicz1955} \textsc{{A}. {L}ichnerowicz}
\btitle{{T}h\'eories relativistes de la gravitation et de
  l'\'electromagn\'etisme} {L}ivre {II}, {S}ection {I}-11, {M}asson (1955)

	\bibitem{Morand:2016rrt} \textsc{K.~Morand}
  \btitle{Nonrelativistic Symmetries and Newton-Cartan Gravity} \href{http://inspirehep.net/record/1411872/files/kevin.morand_4532.pdf}{PhD thesis} (2014)

	\bibitem{Kuenzle:1972zw} \textsc{H.~P.~K\"unzle}
  \btitle{Galilei and Lorentz structures on space-time: comparison of the corresponding geometry and physics}
  \newblock \href{http://www.numdam.org/item?id=AIHPA_1972__17_4_337_0}{Ann.\ Inst.\ H.\ Poincare Phys.\ Theor.\  {\bf 17} 337} (1972)

	\bibitem{Bernal:2002ph} \textsc{A.~N.~Bernal and M.~Sanchez}
  \btitle{Leibnizian, Galilean and Newtonian structures of space-time}
  \newblock \href{http://dx.doi.org/10.1063/1.1541120}{J.\ Math.\ Phys.\  {\bf 44} 1129} (2003)
    \barxiv{gr-qc/0211030}

	\bibitem{Bekaert:2014bwa} \textsc{{X}. {Bekaert}, {K}. {Morand}}
  \btitle{Connections and dynamical trajectories in generalised Newton-Cartan gravity I. An intrinsic view.}
\newblock \href{http://dx.doi.org/10.1063/1.4937445}{J.\ Math.\ Phys. \  {\bf 57}, no. 2, 022507} (2016)
\barxiv{1412.8212}

	\bibitem{Duval:1993pe} \textsc{C.~Duval}
  \btitle{On Galilean isometries}
  \newblock \href{http://dx.doi.org/10.1088/0264-9381/10/11/006}{Class.\ Quant.\ Grav.\  {\bf 10}, 2217} (1993)
\barxiv{0903.1641}

	\bibitem{Trumper1983} \textsc{M.~{Tr{\"u}mper}}
\btitle{Lagrangian mechanics and the geometry of configuration spacetime}
   \newblock\href{http://dx.doi.org/10.1016/0003-4916(83)90305-6}{{{A}nn. {P}hys.} {\bfseries 149} 203} (1983)

	\bibitem{Hartong:2015xda} \textsc{J.~Hartong}
    \btitle{Gauging the Carroll Algebra and Ultra-Relativistic Gravity}
    \newblock \href{http://dx.doi.org/10.1007/JHEP08(2015)069}{JHEP {\bf 1508} 069} (2015)
\barxiv{1505.05011}

	\bibitem{Gibbons:2007zu} \textsc{G.~W.~Gibbons and C.~N.~Pope}
  \btitle{Time-dependent multi-centre solutions from new metrics with holonomy SIM$(n-2)$}
    \newblock \href{http://dx.doi.org/10.1088/0264-9381/25/12/125015}{Class.\ Quant.\ Grav.\  {\bf 25} 125015} (2008)
\barxiv{0709.2440}

	\bibitem{Kundt1961} \textsc{W.~Kundt}
   \btitle{The plane-fronted gravitational waves}
    \newblock \href{http://dx.doi.org/10.1007/BF01328918}{Zeitshrift f\"{u}r Physik \textbf{163}, 77} (1961)

	\bibitem{Podolsky:2008ec} \textsc{J.~Podolsk\'y and M.~\v{Z}ofka}
  \btitle{General Kundt spacetimes in higher dimensions}
      \newblock \href{http://dx.doi.org/10.1088/0264-9381/26/10/105008}{Class.\ Quant.\ Grav.\  {\bf 26} 105008} (2009)
  \barxiv{0812.4928}

	\bibitem{Coley:2009ut} \textsc{A.~Coley, S.~Hervik, G.~O.~Papadopoulos and N.~Pelavas}
  \btitle{Kundt Spacetimes}
        \newblock \href{http://dx.doi.org/10.1088/0264-9381/26/10/105016}{Class.\ Quant.\ Grav.\  {\bf 26} 105016} (2009)
\barxiv{0901.0394}

	\bibitem{Coley:2008th} \textsc{A.~A.~Coley, G.~W.~Gibbons, S.~Hervik and C.~N.~Pope}
  \btitle{Metrics With Vanishing Quantum Corrections}
  \newblock \href{http://dx.doi.org/10.1088/0264-9381/25/14/145017}{Class.\ Quant.\ Grav.\  {\bf 25} 145017} (2008)
\barxiv{0803.2438}

	\bibitem{Hervik:2017sdr} \textsc{S.~Hervik, V.~Pravda and A.~Pravdov\'a}
  \btitle{Universal spacetimes in four dimensions}
  \newblock \href{http://dx.doi.org/10.1007/JHEP10(2017)028}{JHEP {\bf 1710} 028} (2017)
\barxiv{1707.00264}

	\bibitem{Hervik:2015mja} \textsc{S.~Hervik, T.~M\'alek, V.~Pravda and A.~Pravdov\'a}
  \btitle{Type II universal spacetimes}
    \newblock \href{http://dx.doi.org/10.1088/0264-9381/32/24/245012}{Class.\ Quant.\ Grav.\  {\bf 32} no.24,  245012} (2015)
\barxiv{1503.08448}

	\bibitem{Hervik:2013cla} \textsc{S.~Hervik, V.~Pravda and A.~Pravdova}
  \btitle{Type III and N universal spacetimes}
      \newblock \href{http://dx.doi.org/10.1088/0264-9381/31/21/215005}{  Class.\ Quant.\ Grav.\  {\bf 31} no.21,  215005} (2014)
  \barxiv{1311.0234}

	\bibitem{Julia:1994bs} \textsc{B.~Julia and H.~Nicolai}
  \btitle{Null Killing vector dimensional reduction and Galilean geometrodynamics}
  \newblock \href{http://dx.doi.org/10.1016/0550-3213(94)00584-2}{Nucl.\ Phys.\ B {\bf 439}, 291} (1995)
\barxiv{hep-th/9412002}

	\bibitem{Minguzzi:2006wz} \textsc{E.~Minguzzi}
  \btitle{Classical aspects of lightlike dimensional reduction}
      \newblock \href{http://dx.doi.org/10.1088/0264-9381/23/23/029}{Class.\ Quant.\ Grav.\  {\bf 23} 7085} (2006) 
\barxiv{gr-qc/0610011}

	\bibitem{Minguzzi:2006gq} \textsc{E.~Minguzzi}
    \btitle{Eisenhart's theorem and the causal simplicity of Eisenhart's spacetime}
        \newblock \href{http://dx.doi.org/10.1088/0264-9381/24/11/002}{Class.\ Quant.\ Grav.\  {\bf 24} 2781} (2007)
\barxiv{gr-qc/0612014}

	\bibitem{Walker1950} \textsc{A.~G.~Walker}
\btitle{Canonical form for a Riemannian space with a parallel field of null planes}
  \newblock \href{http://dx.doi.org/10.1093/qmath/1.1.69}{The Quarterly Journal of Mathematics \  {\bf 1}, Issue 1, 69-79} (1950)

	\bibitem{Lee2003} \textsc{{J}.~{M}.~{L}ee}
  \btitle{{I}ntroduction to {S}mooth {M}anifolds}
\newblock \href{https://link.springer.com/book/10.1007/978-1-4419-9982-5?page=1}{Graduate Texts in Mathematics {\bf 218}, Springer} (2003)

	\bibitem{Duggal1996} \textsc{{K}.~{L}.~Duggal, {A}.~Bejancu}
  \btitle{Lightlike Submanifolds of Semi-Riemannian Manifolds and Applications}
\newblock \href{http://dx.doi.org/10.1007/978-94-017-2089-2}{Kluwer Academic {\bf 364}} (1996)

	\bibitem{Spivak1970} \textsc{M.~D.~Spivak}
\btitle{A comprehensive introduction to differential geometry}
       \newblock \href{http://cds.cern.ch/record/872235}{ (Publish or Perish) Volume II, Addendum 1 to Chapter 6} (1999)

	\bibitem{Eastwood2008} \textsc{{M}.~{E}astwood}
\btitle{{N}otes on {P}rojective {D}ifferential {G}eometry}
in M.~{E}astwood and W.~Miller (eds), \textit{Symmetries and Overdetermined Systems of Partial Differential Equations},
       \newblock \href{https://link.springer.com/chapter/10.1007/978-0-387-73831-4_3}{IMA Volumes in Mathematics and its Applications {\bf 144} 41} (2008)

	\bibitem{Carroll1995} \textsc{L.~Carroll}
  \btitle{What the Tortoise Said to Achilles}
  \newblock \href{https://academic.oup.com/mind/article-abstract/104/416/691/1096551?redirectedFrom=fulltext}{Mind New Series, {\bf104} No. 416, 691-693} (1995)








\end{thebibliography}
    \end{document}